\documentclass[twocolumn]{aastex62}

\usepackage{amsmath}
\usepackage{color}
\graphicspath{{./}{figures/}}
\usepackage{natbib}
\usepackage{enumitem}
\usepackage{float}
\usepackage{url}
\usepackage{color,soul}
\usepackage{graphicx}

\def\be{\begin{eqnarray}}
\def\ee{\end{eqnarray}}

\newcommand{\refeq}[1]{Eq.~(\ref{eq:#1})}          
          
\newcommand{\reffig}[1]{Fig.~\ref{fig:#1}}          
\newcommand{\refsec}[1]{Sec.~\ref{sec:#1}}
\newcommand{\refapp}[1]{App.~\ref{app:#1}}

\newcommand{\Vvec}{\mathbf{V}}

\def\stars{\textsf{Stars}}

\def\NFW{\textsf{NFW}}
\def\Burk{\textsf{Burkert}}

\newcommand{\UDG}{UDG1}
\newcommand{\MGC}{m_\star}
\newcommand{\LGC}{{\rm L}_\star}

\newcommand{\sersic}{S\'ersic}

\shorttitle{Dynamics in NGC5846-UDG1}
\shortauthors{Bar, Danieli \& Blum}

\begin{document}

\title{Dynamical friction in globular cluster-rich ultra-diffuse galaxies: the case of NGC5846-UDG1}

%\correspondingauthor{Nitsan Bar}

\author[0000-0002-3724-5082]{Nitsan Bar}
\affiliation{Department of Particle Physics and Astrophysics,
	Weizmann Institute of Science, Rehovot 7610001, Israel}
\email{nitsan.bar@weizmann.ac.il}
  
\author[0000-0002-1841-2252]{Shany Danieli}
\altaffiliation{NASA Hubble Fellow}
\affil{Department of Astrophysical Sciences, 4 Ivy Lane, Princeton University, Princeton, NJ 08544, USA }
\affil{Institute for Advanced Study, 1 Einstein Drive, Princeton, NJ 08540, USA }
\email{sdanieli@astro.princeton.edu}

\author[0000-0001-8978-5155]{Kfir Blum} 
\affiliation{Department of Particle Physics and Astrophysics,
 	Weizmann Institute of Science, Rehovot 7610001, Israel}
\email{kfir.blum@weizmann.ac.il}

\begin{abstract}
Ultra-diffuse galaxies that contain a large sample of globular clusters (GCs) offer an opportunity to test the predictions of galactic dynamics theory. NGC5846-UDG1 is an excellent example, with a high-quality sample of dozens of GC candidates. We show that the observed distribution of GCs in NGC5846-UDG1 is suggestive of mass segregation induced by gravitational dynamical friction. We present simple analytic calculations, backed by a series of numerical simulations, that naturally explain the observed present-day pattern of GC masses and radial positions. Subject to some assumptions on the GC population at birth, the analysis supports the possibility that NGC5846-UDG1 resides in a massive dark matter halo. This is an example for the use of GC-rich systems as dynamical (in addition to kinematical) tracers of dark matter.
\end{abstract}

\keywords{.}%{galaxies: photometry -- galaxies: dwarf -- galaxies: star clusters: general -- galaxies: formation -- galaxies: dynamics -- galaxies: individual (NGC5846-UDG1)}

%\tableofcontents
%%%%%%%%%%%%%%%%%%%%%%%%%%%%%%% Section 1 %%%%%%%%%%%%%%%%%%%%%%%%%%%%%%% 
\section{Introduction}\label{sec:introduction}

\begin{figure*}
	\centering% trim={<left> <lower> <right> <upper>}
	\includegraphics[trim={0.5cm 3.4cm 0.5cm 3.2cm},width=\textwidth]{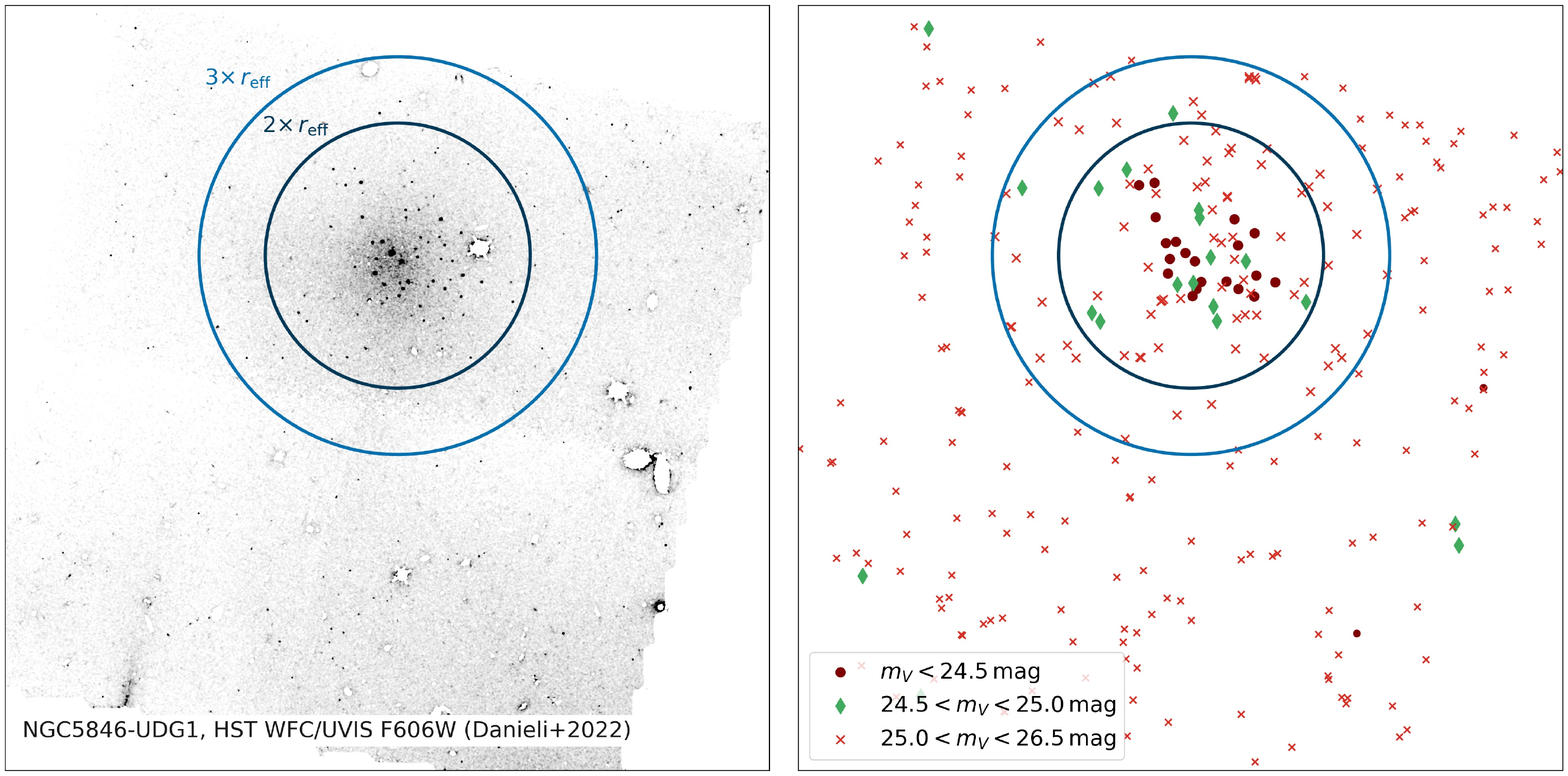}	
	\caption{\textbf{Left:} Reproduction of V-band data obtained in \cite{Danieli2021} using \textit{Hubble Space Telescope} WFC3/UVIS camera, containing \UDG\ and a nearby field (post-selection criteria described in \cite{Danieli2021}). Circles represent $2r_{\rm eff}$ and $3r_{\rm eff}$ of the stellar light profile, with $r_{\rm eff}$ the \sersic\ radius.  
	\textbf{Right:} A scatter plot of objects from the left panel, divided into magnitude bins. The magnitude bins for objects at $m_V < 25.0~$mag are relatively clean from background contamination. In comparison, contamination is significant for the bin $25.0<m_V < 26.5~$mag. In our main analysis, we primarily use the $r< 2r_{\rm eff}$ data of the $m_V < 25.0~$mag bins. We present a preliminary analysis of the $25.0<m_V < 26.5~$mag bin in \refapp{faintset}, showing that the faint objects also exhibit radial clustering above the background, comparable to the stellar body. %Properly accounting for this data sample would a require a more sophisticated statistical analysis, which we reserve for future work. 
	} \label{fig:udg1_hst}
\end{figure*}

Dynamical processes shape galaxies and may provide constraints on the nature of dark matter \citep{BinneyTremaine2}. In particular, dynamical friction \citep{Chandra43} can significantly impact the orbits of globular clusters (GCs) near the center of massive galaxies \citep{Tremaine1975} or in the halos of dwarf and ultra-diffuse galaxies \citep[e.g.][]{Tremaine1976a,SanchezSalcedo:2006fa,Nusser2018a, Dutta_Chowdhury_2020}. The Fornax dwarf satellite galaxy, hosting 5 or 6 GCs, is a well-studied test case where dynamical friction should have imprinted itself in the galaxy. Indeed, it was argued long ago that the lack of a nuclear star cluster in Fornax is surprising and perhaps poses a puzzle, because the dynamical friction time for  GC orbits appears to be short compared with the age of the system \citep{Tremaine1976a}. Several studies have revisited this ``Fornax globular cluster timing problem'', primarily focusing on the possibility that the dark matter halo in Fornax is cored \citep{ohlinricher2000,SanchezSalcedo:2006fa,Goerdt2006,Cole2012,boldrini2020embedding,Meadows20,Shao:2020tsl,Bar:2021jff}. However, the number of GCs in Fornax, although large relative to other Milky Way dwarf satellites, may be too small to allow  robust conclusions.

The potential to constrain dark matter via dynamical considerations motivates us to look for additional galaxies with a large population of GCs. This is timely in part due to recent studies of ultra-diffuse galaxies (UDGs) which often times host large numbers of GCs \citep{vanDokkum_2017,vanDokkum_2018, Lim_2018, Shen_2021}.
In this paper we consider NGC5846-UDG1 (\UDG\ for short; \citealt{Forbes2019,Forbes2021, muller2020spec,muller21udg,Danieli2021}), that recently attracted considerable attention.\footnote{See also NGC1052-DF2, with $10$ spectroscopically-confirmed GCs, which was studied in \cite{Nusser2018a} and in \cite{Dutta_Chowdhury_2019,Dutta_Chowdhury_2020}.}  At a distance of $\sim 25$~Mpc, stellar luminosity of $\sim 6\times 10^7~L_\odot$ and half light radius $r_{\rm eff}\sim 2$~kpc, UDG1 harbors some $\sim 50$ GC candidates, representing $\sim 10\%$ of the stellar mass at the preset day \citep{Danieli2021}. 

The left panel of Fig. \ref{fig:udg1_hst} shows the $V$-band Hubble Space Telescope (HST) WFC3/UVIS image of UDG1 and its nearby field, adapted from \cite{Danieli2021}. The right panel shows all compact sources that were selected as GC candidates based on the photometric selection criteria in \cite{Danieli2021}.
% We reproduce a V-band image of \UDG\ and its nearby field in the left panel of \reffig{udg1_hst}, along with a corresponding plot (right panel) of point sources defined using the selection criteria of \cite{Danieli2021}. 
%
In this work we focus on a low contamination sample of GC candidates, consisting of the 33  $m_V<25.0$~mag objects contained within $2r_{\rm eff}$ 
(twice the \sersic\ half-light radius of the stellar body; inner circle in Fig. \ref{fig:udg1_hst}), which has a background contamination of about $1$ object, estimated by comparison to the nearby field \citep{Danieli2021}. Spectrocscopic information is available for $11$ of these bright GCs \citep{muller2020spec}. 

\begin{figure}[htbp!]
	\centering
	\includegraphics[width=0.47\textwidth]{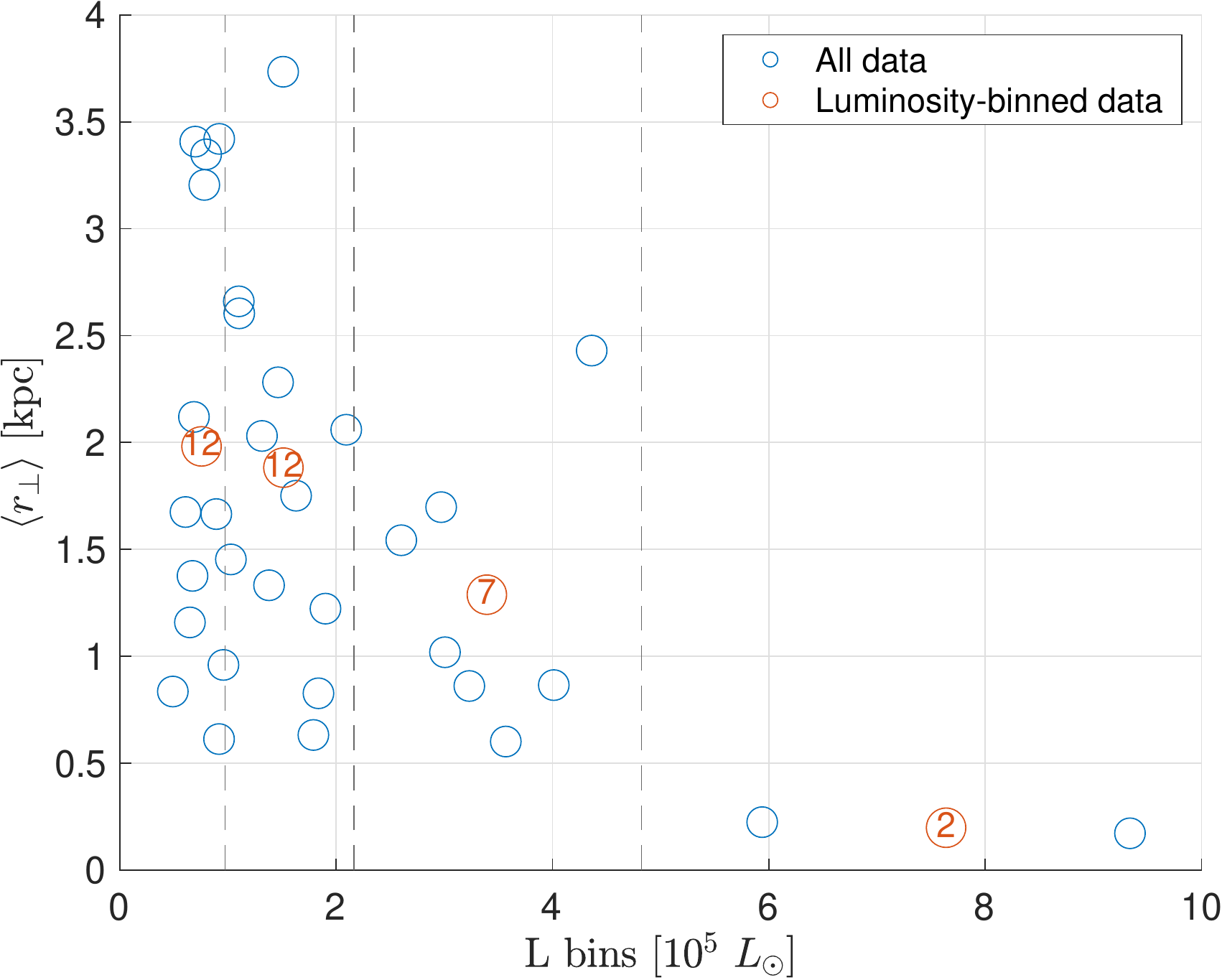}	
	\caption{Red circles with numbers: circle position along the $y$-axis shows the average projected distance $ \left<r_\perp\right> $, for GCs belonging to luminosity bins ($x$-axis) marked by vertical dashed lines. Numbers indicate the number of GCs per bin. The luminosity bins are equi-spaced in log scale. %We assume $ M/L_V = 1.6~M_\odot/L_\odot $ (adopting the derived value of \cite{muller2020spec}) and
	We assume distance $ D = 26.5 $~Mpc. Blue circles show the unbinned data, comprised of the $m_V<25.0$~mag objects inside $r<2r_{\rm eff}$ in \reffig{udg1_hst}. Fainter objects (with significant background contamination) are not shown here; in \refapp{faintset} we find that the faint object population yields $\left<r_\perp\right> \sim 2\div 2.5$~kpc, consistent with the pattern of the $m_V<25.0$~mag sample.
	%We have not presented here another lower luminosity sample of fainter objects, 
	}\label{fig:introData}
\end{figure}

%\textbf{nb, edit here}
It is noteworthy that most of the brighter GCs in the right panel of Fig.~\ref{fig:udg1_hst} are concentrated in the region $r<r_{\rm eff}$. 
%In particular, we utilize the apparent mass segregation of the GCs in UDG1.
% We believe that this population of GCs shows evidence for mass segregation. 
% To make this more manifest, in \reffig{introData} we show a scatter plot of the sample with GC projected distance (relative to the center of the stellar body) vs. GC luminosity. The data shows a clear trend: more luminous GCs are on average closer to the center of the galaxy. This luminosity or mass segregation calls for a quantitative dynamical explanation, which we attempt to provide in this work.
To explore this further, 
%including more detailed GC luminosity information, we focus on a high-quality sample of $33$ GCs, comprised of candidate GCs within $r<2r_{\rm eff}$ with $m_V<25$~mag. Spectrocscopical information is available for $11$ of the GCs in the high luminosity end \citep{muller2020spec}. Contamination can be estimated by comparison to a nearby field, and is expected to be very low, about $1$ object likely near the low luminosity end of the sample \citep{Danieli2021}. 
in \reffig{introData} we show the luminosity of this sample of 
GCs vs. their projected distance from the center of the galaxy. The data shows a clear trend: more luminous GCs are on average closer to the center of the galaxy. We estimate a p-value of about $1\%$ for the hypothesis that the data is a chance fluctuation and that there is no mass segregation (see \refapp{robust}). This luminosity or mass segregation calls for a quantitative dynamical explanation. %and our goal in the present paper is to provide this explanation. 

In this paper we show that this explanation can be naturally provided by dynamical friction. 
%
%We demonstrate in this paper that the effect responsible for the trend in \reffig{introData} is likely dynamical friction. 
The deceleration experienced by a GC due to dynamical friction in a galactic halo is roughly proportional to the GC mass $\MGC$. Therefore, more massive GCs inspiral closer to the center of the galaxy, resulting in mass segregation. This simple picture can be expected to hold over an intermediate duration of time: long enough to enable dynamical friction to act, but short enough so that GC mergers do not convert a large fraction of the total mass in GCs into a nuclear cluster. As we will demonstrate, using more detailed analytic estimates as well as a suite of numerical simulations, \UDG\ as we view it today may indeed be in this intermediate stage.

The paper is organized as follows. In \refsec{roughtheory} we discuss dynamical effects that shape the GC population in \UDG\ and similar galaxies. In \refsec{udgdatamodels} we recapitulate observational studies of \UDG, and define benchmark mass models. In \refsec{simulations} we set up and study N-body simulations, in which some dynamical effects (notably dynamical friction and GC mass loss) are modeled semi-analytically. In \refsec{discussion} we discuss the results. We conclude in \refsec{conclusions}. 

We reserve some details to Appendices. In \refapp{robust} we show the sensitivity of the results to GC selection criteria and the significance of the mass segregation trend. In \refapp{twobodyrelax} we derive two-body relaxation in an external potential. In \refapp{projection} we discuss projection effects. In \refapp{faintset} we present a preliminary analysis of the faintest GC candidates in \UDG. In \refapp{othersims} we present a number of convergence and stability tests of the simulations.

%%%%%%%%%%%%%%%%%%%%%%%%%%%%%%% Section 2 %%%%%%%%%%%%%%%%%%%%%%%%%%%%%%% 
\section{Back of the envelope analytic  estimates}\label{sec:roughtheory}
%The GC population of \UDG\ calls for a careful analysis in the context of galactic dynamics. 
%We start by applying analytic considerations to the study of the GC distribution. 

%%%%%%%%%%%%%%%%%%%%%%%%%%%%%%%%%%%%%%%%%%%%%%
%%%%%%%%%%%%%%%%%%%%%%%%%%%%%%%%%%%%%%%%%%%%%%
%%%%%%%%%%%%%%%%%%%%%%%%%%%%%%%%%%%%%%%%%%%%%%
%\subsection{Dynamical friction}\label{sec:dfrough}
The orbits of GCs traversing a background medium are processed by dynamical friction (DF; \citealt{Chandra43}). A convenient expression for the time-scale of dynamical friction is presented in \cite{Hui2017}, %(and further applied and analysed in \cite{Bar:2021jff})
\be\label{eq:tauDF}
\tau_{\rm DF} &\equiv & \frac{v^3}{4\pi G^2\rho \MGC C } \\ &\approx & 2\left(\frac{v}{10\frac{{\rm km}}{{\rm sec}}}\right)^3\frac{3\times 10^6\frac{M_\odot}{{\rm kpc}^3}}{\rho}\frac{3\times 10^{5}~M_\odot}{\MGC}\frac{2}{C}\,{\rm Gyr}\;. \nonumber
\ee
where $ \rho $ is the density of the medium inducing the DF, $ \MGC $ is the GC mass, and $ C $ is a dimensionless factor encoding the details of the velocity dispersion of the medium and a Coulomb logarithm. For some dwarf galaxies and UDGs, $ \tau_{\rm DF} \lesssim  10$~Gyr, meaning that DF should be effective over the life of the galaxy. This was noticed long ago for the Fornax dwarf spheroidal satellite galaxy \citep{Tremaine1976a} and more recently for NGC1052-DF2 \citep{Nusser2018a,Dutta_Chowdhury_2019}\footnote{See also \cite{Lotz:2001gz} for a survey of GCs in dwarf elliptical galaxies in the Virgo cluster and \cite{SanchezSalcedo2022} for an analysis of GCs in dwarf spheroidal and dwarf irregular galaxies.}.%; a survey of dwarf ellipticals was conducted in Ref.~\cite{Lotz:2001gz}\footnote{\textbf{Unclear to me yet what stops us from analysing that data. Are galaxies there simply much more massive? probably yes. Actually seems they barely find evidence for DF.}} {\textbf{more refs on original literature}}.

\UDG\ with its unusually large population of GCs is likely another system where DF is effective. %In fact, the DF effect may provide a test of mass models and formation models of \UDG\ and similar systems. 
The mass segregation observed in \reffig{introData} can be interpreted as a natural outcome of DF, because of the dependence $ \tau_{\rm DF}\propto 1/\MGC $ in \refeq{tauDF} (neglecting logarithmic dependence on $\MGC$, sequestered in $ C $). 
To illustrate how the $\MGC$ scaling leads to mass segregation, consider a cored halo, for which $ \tau_{\rm DF}$ is independent of radial position to leading order \citep{Bar:2021jff}. In such a system, a GC on a circular orbit which starts its life at radius $ r_0 $, migrates during time $t$ to a lower radius $ r \approx r_0 \exp\left(-t/2\tau_{\rm DF}\right) $ \citep{Bar:2021jff}. Accounting for projection and averaging over a population of GC orbits (see \refeq{rprojvsrDist}), one finds
\be\label{eq:corelogmass}
\ln \left< r_\perp\right>_{\rm core} = \ln \left< r_{0,\perp}\right>_{\rm core} - \frac{\Delta t}{2\tau_{\rm core}^{(0)}}\frac{\MGC}{\MGC^{(0)}} \; ,
\ee
where angle brackets denote population average. 

Assuming that the orbit distributions of GCs of different masses start with the same average initial radius, the simple model in \refeq{corelogmass} can be compared to data, with two free parameters: (i) the initial average projected radius, $\left<r_{0,\perp}\right>_{\rm core}$, and (ii) the core DF time measured in units of the age of the system, $\tau_{\rm core}^{(0)}/\Delta t$, computed for a reference GC mass $\MGC^{(0)}$. 

\begin{figure}[htbp!]
	\centering
	\includegraphics[width=0.45\textwidth]{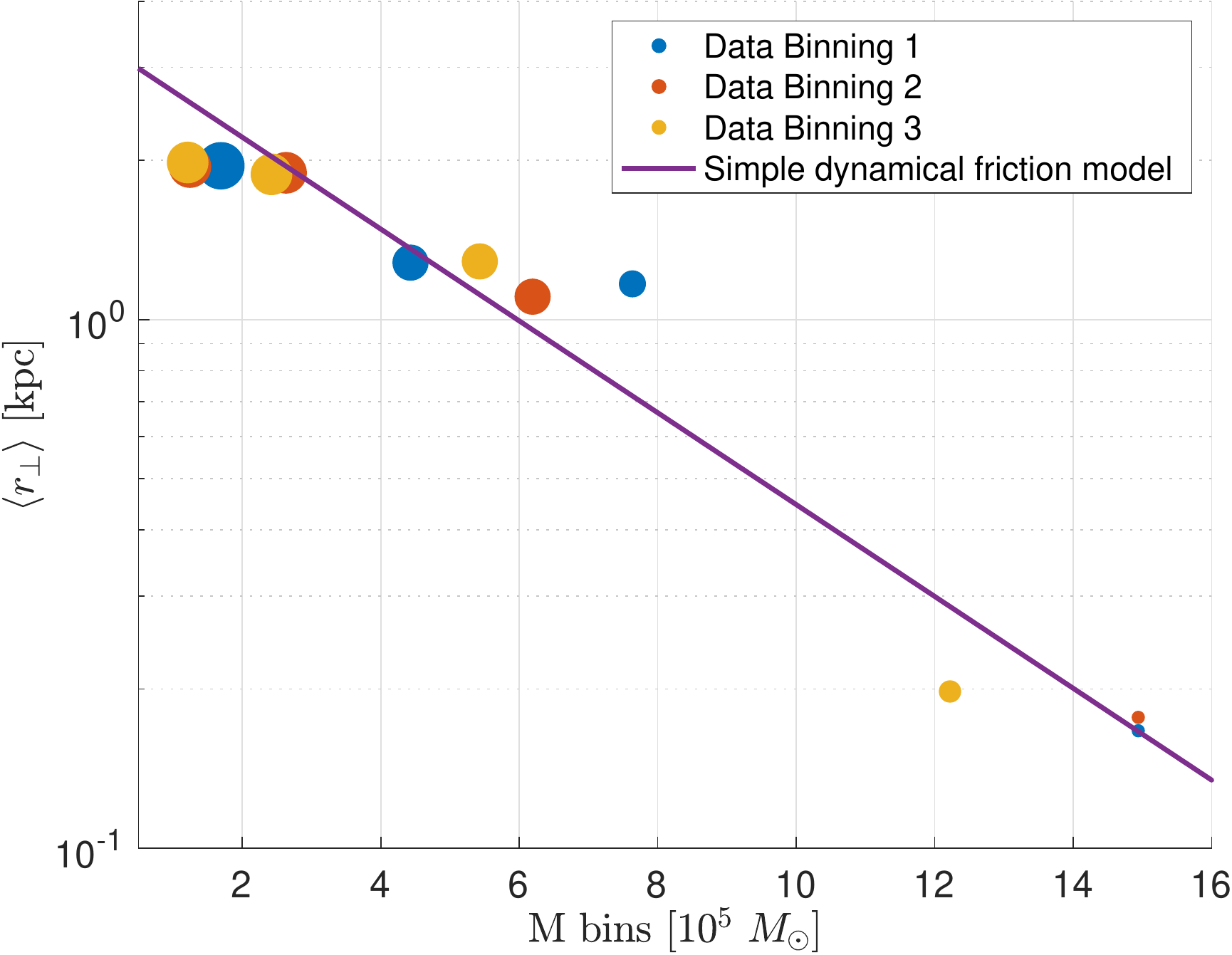}	
	\caption{The data from \reffig{introData}, overlaid with the simple dynamical friction model captured by Eq.~\eqref{eq:corelogmass}. The results are shown with three different choices of binning in GC mass. Point size indicates the amount of GCs per bin. %-- all logarithmically spaced with varying spacings and shifts. 
	(The last two points on the right should overlap; their position was slightly displaced for clarity.)}\label{fig:DataAndSimpleModel}
\end{figure}

In \reffig{DataAndSimpleModel} we compare this model to the data from \UDG. We set $ \left<r_{0,\perp}\right>_{\rm core}= 3 $~kpc, somewhat larger than the observed stellar average projected radius $ \approx 2.1 $~kpc; and $\tau_{\rm core}^{(0)}/\Delta t=(5~{\rm Gyr})/(10~{\rm Gyr})$, with $\MGC^{(0)}=5\times10^5$~M$_\odot$, amounting to
\be \label{eq:fittedtauDF}
\tau_{\rm DF}\approx 5~ \frac{5\times 10^{5}~M_{\odot}}{\MGC}~{\rm Gyr} \; .
\ee 
To convert from GC luminosity to mass, we assume a mass-to-light ratio $\MGC=1.6\left(\LGC/{\rm L}_\odot\right)$~M$_\odot$, following \cite{muller2020spec}. 
%
%Assuming that the orbit distribution of GCs of different masses start with the same average initial radius, one can compare \refeq{corelogmass} with \reffig{introData}. We plot such a comparison in \reffig{DataAndSimpleModel}. The model has two parameters, essentially: (i) the initial radius $ \left<r_{0,\perp}\right>\approx 3 $~kpc -- somewhat larger than the stars' observed average projected radius, $ \approx 2.1 $~kpc; (ii) the dynamical friction time 
%\be \label{eq:fittedtauDF}
%\tau_{\rm DF}\approx 5~ \frac{5\times 10^{5}~M_{\odot}}{\MGC}~{\rm Gyr} \; .
%\ee 
%
%The model result is overlaid  in \reffig{DataAndSimpleModel} where 
The data in \reffig{DataAndSimpleModel} is shown for three different choices of binning in  $\MGC$\footnote{The bin settings are: (1) $M/(10^5~{\rm M}_\odot)=[0.7,3,6,12,25]$,  (2) $M/(10^5~{\rm M}_\odot)={\rm exp}[\ln0.7:0.8:\ln30]$, and (3) $M/(10^5~{\rm M}_\odot)={\rm exp}[\ln0.7:0.9:\ln30]$.}. %As will be noted in \refsec{udgdatamodels}, this modeling  interpretation supports the possibility that the gravitational potential of \UDG\ is dark matter-dominated. 

In the rest of this section we discuss a number of additional effects that are, to some extent, intertwined with DF. These include gravitational GC-GC interactions; GC mergers; deformation of the background stellar and dark matter halo and dynamical heating by GCs; and GC mass loss. Some of these effects are interesting and could, under specific circumstances, modify the simple DF analysis. We will include a treatment of all of these effects in the numerical simulations described in Sec.~\ref{sec:simulations}.

%%%%%%%%%%%%%%%%%%%%%%%%%%%%%%%%%%%%%%%%%%%%%%
%%%%%%%%%%%%%%%%%%%%%%%%%%%%%%%%%%%%%%%%%%%%%%
%%%%%%%%%%%%%%%%%%%%%%%%%%%%%%%%%%%%%%%%%%%%%%
\subsection{Mergers of GCs}\label{sec:mergers}
A large density of GCs could lead to a high rate of GC-GC mergers. A crude estimate of the merger rate per GC is
\be
\Gamma&\sim& n_{\rm GC}\sigma_I v \\
&\sim& \frac{0.1}{10~{\rm Gyr}} \frac{n_{\rm GC}}{\left[\frac{20}{\frac{4\pi}{3}(2~{\rm kpc})^3}\right]}\frac{\sigma_I}{\pi (20~{\rm pc})^2} \frac{v}{10~\frac{{\rm km}}{\rm sec}} \;,\nonumber
\ee
where $ n_{\rm GC} $ is the number density of GCs and $ \sigma_I $ is the merger cross-section. Multiplying by the currently observed number of GCs yields $ \sim 3 $ mergers in \UDG. 

The crude estimate above can be compared with results of numerical simulations performed in \cite{Dutta_Chowdhury_2020} for a different galaxy, NGC1052-DF2 (DF2). DF2 hosts a stellar core comparable to that of \UDG, but has only about a third of the number of GCs. Performing a simulation with ``live'' GCs (i.e. made of a collection of stars rather than a single object, so that GC collisions can be resolved in some detail), \cite{Dutta_Chowdhury_2020} found an average number of GC mergers in DF2, over $ 10 $~Gyr, of about $ 0.3$ (at a rate of $0.03\,\mathrm{Gyr}^{-1}$). Since the number of mergers $ \propto n_{\rm GC}^2 $, this would be consistent with $ \sim 3 $ mergers in \UDG\ over $ 10 $~Gyr. We note however that \cite{Dutta_Chowdhury_2020} used the observed present-day distribution of GCs in DF2 as initial conditions for their simulations. Therefore, they simulated the future of the GC system, and since DF causes the GC distribution to converge inwards with time, it is likely that the (already small) reported merger efficiency was even smaller over a similar time scale in the past. In our work we will attempt to trace the history of the GC system in \UDG, so the initial conditions we select correspond to a GC system that is less dense than currently observed. Thus, the rate of mergers we find is indeed generically small.

%We will not have much more analytic insight to add on mergers (see discussion in \cite{Dutta_Chowdhury_2020}). 
In \refsec{simulations} we will discuss how we implement GC mergers in our simulations. We agree with the conclusions of \cite{Dutta_Chowdhury_2020} that DF enhances the GC merger rate, as an outcome of the increase of GC density with time. 
%As such, it should also be noted that mergers typically become more effective over time, meaning mergers depend to some extent on the age of the system. This can be demonstrated within the toy model of the previous section. Taking $ n_{\rm GC} \propto r^{-3}\propto e^{3t/2\tau_{\rm DF}} $, we see that the density grows exponentially in a constant $ \tau_{\rm DF} $ model, therefore so does the merger rate.

%Mergers can play a role in modifying the GC mass function. For example, a GC which is seemingly too massive to originate in initial conditions, may be the result of GC mergers. 

It is tempting to speculate that the  most luminous (and most centrally-located) GCs of \UDG\ (see \reffig{introData}) could be the result of DF-induced mergers. A closely related hypothesis was brought up long ago in the context of nuclear clusters in other galaxies \citep{Tremaine1975,CapuzzoD1993,OstrikerGnedin1997,CapuzzoD1997,Gnedin2014,ArcaSedda2014}. %There, nuclear stellar clusters were hypothesized to form due to a similar effect. 
We return to this possibility later on.%, finding that it may indeed be the case in some models (\refsec{resultsAlmostObs}), subject to several caveats (\refsec{caveats}). 

\subsection{Mass loss of GCs}\label{sec:massloss}

Old massive ($\gtrsim 10^5~M_\odot$) GCs are expected to have lost a part of their mass over their life due to stellar evolution and dynamical processes (for a recent review, see \cite{Krumholz2019}). We estimate the importance of this effect for our analysis, adopting a phenomenological approach. We treat GCs as point masses, losing mass at a prescribed rate without modeling the ``microphysics'' of the process. As a benchmark, we adopt the mass loss rate from \cite{Shao:2020tsl} (c.f. Fig.~C1 there).\footnote{Strictly speaking, the results there are reported around $2\times 10^5~M_\odot$ whereas our analysis extends to larger GC masses, which are expected to lose a smaller mass fraction. We neglect this complication in the following. }
%in application to Fornax-like dwarf galaxies within the E-MOSAICS simulations. 
With this prescription, GCs lose $ \sim 30~\% $ of their initial mass over a short $ \sim 0.5~{\rm Gyr} $ interval in an early phase, followed by a $ \sim {\rm Gyr} $ intermediate phase of $ \sim 20\% $ mass loss. The remainder $ \sim 10~{\rm Gyr} $ is characterized by a slower steady mass loss of about $30\%$ of the GC mass (compared to the beginning of that last phase). 

Assuming that dynamical relaxation time scales are longer than $\sim0.5$~Gyr (although see \refsec{selfRelax} for possible exceptions), it is a reasonable approximation to simply consider the initial GC distribution to be defined after the first brief mass loss episode.
%, as it occurs rather quickly compared to DF time-scales. 
Therefore, it seems reasonable to assume a mass loss rate of $ \dot{\MGC} \sim -(\MGC^{(0)}/3)/10~{\rm Gyr} $, i.e. $ \MGC(t) = \MGC^{(0)}[1-\delta \times t/t_0] $ with $ \delta = 1/3 $, $ t_0 = 10 $~Gyr. In a simplified model like that leading to Eq.~\eqref{eq:corelogmass}, this can be roughly incorporated by using an ``effective'' GC mass $ \MGC^{(\rm eff)} \approx \MGC^{(\rm obs)}(1+\delta/2)\approx 1.2 \MGC^{(\rm obs)} $, where $ \MGC^{(\rm obs)} $ is the currently observed GC mass (neglecting mergers). This amounts to an effective $ \tau_{\rm DF} $ that is $ \sim 20\% $ shorter compared to a naive expectation based on the currently observed GC masses. 
%To summarize, accounting for mass loss is expected to enhance DF somewhat compared to na\"ive expectations without mass loss. 
We thus expect that mass loss is not a crucial factor in the dynamics of \UDG. Nevertheless, for completeness, when we set up simulations in \refsec{simulations} we take this effect into account.

%%%%%%%%%%%%%%%%%%%%%%%%%%%%%%%%%%%%%%%%%%%%%%
%%%%%%%%%%%%%%%%%%%%%%%%%%%%%%%%%%%%%%%%%%%%%%
%%%%%%%%%%%%%%%%%%%%%%%%%%%%%%%%%%%%%%%%%%%%%%
\subsection{Relaxation of GCs between themselves}\label{sec:selfRelax}
Two-body relaxation between stars or star clusters is typically thought to be unimportant on the scales of galaxies \citep{BinneyTremaine2},\footnote{Except very near galactic centers or for some candidates of dark matter \citep{Hernandez:2004bm,Hui2017,Bar-Or:2018pxz}.} but diffuse galaxies with a rich GC population may present a counter-example. Assume $ N $ GCs of equal masses spread over a radial scale $ R $ with velocity scale $ v $, comprising a fraction $ f $ of the total mass within $ R $ (i.e. $ f \equiv M_{\rm GCs}/M $). It is straightforward to extend classic arguments \citep{BinneyTremaine2} to derive a two-body relaxation time scale (for details, see \refapp{twobodyrelax})
\be\label{eq:relaxtime}
t_{\rm relax} &\sim & \frac{0.1N}{\ln \frac{N}{f}}\frac{1}{f^2} t_{\rm cross} \\  &\sim & 10 \frac{N}{30}\frac{R}{2~{\rm kpc}}\frac{10~{\rm km/sec}}{v}\left(\frac{0.1}{f}\right)^2~{\rm Gyr}  \; ,\nonumber
\ee
where $ t_{\rm cross}\sim R/v $. Here, the reference value chosen for $v$ represents a somewhat extreme scenario in which the gravitational potential of \UDG\ is dominated by the stellar mass. 
We note that the spectroscopic study of \cite{Forbes2021} reported a line-of-sight velocity dispersion $17\pm 2~$km/sec, suggesting a dark matter-dominated halo and yielding a long two-body relaxation time-scale for GCs in \UDG.

Two-body relaxation of GCs assists DF in inducing mass segregation. 
In \refapp{twobodyrelax} we present an N-body simulation in a smooth external potential that demonstrates this effect. 

As briefly reviewed in \refsec{massloss}, GCs are expected to lose $\mathcal{O}(1)$ of their mass over their life. With this in mind, \cite{Danieli2021} pointed out that the GC population may have initially comprised an $ \mathcal{O}(1) $ fraction of the stellar mass in \UDG. This scenario could make two-body relaxation surprisingly efficient, if, in addition, the total halo mass of (and therefore velocity dispersion in) \UDG\ is small. To see this, note that inserting $f\approx0.5$ (\`a la the GC-dominance hypothesis of \cite{Danieli2021}) along with $v\approx10$~km/sec (low mass / no dark matter hypothesis) into Eq. \eqref{eq:relaxtime} yields $t_{\rm relax}\sim0.5$~Gyr, a short relaxation time that could in principle affect the GC distribution at a noticeable level even during the brief initial mass loss phase of the GCs.\footnote{This scenario requires that most of the GCs were formed nearly at the same time, and no more than a few 100 Myr apart.} If the halo is dark matter dominated (as supported by the spectroscopic study of \cite{Forbes2021}), then $f\approx0.5$ along with $v\approx20$~km/sec gives $t_{\rm relax}\sim5$~Gyr, making two-body relaxation relatively unimportant.
%
%Assuming a dark-matter dominated halo, we could estimate 
%\be\label{eq:relaxtimeEarlyDM}t_{\rm relax} &\sim & 5 \frac{N}{30}\frac{R}{2~{\rm kpc}}\frac{20~{\rm km/sec}}{v}\left(\frac{0.1}{f}\right)^2~{\rm Gyr} \; .\ee
%Here, since the initial GC mass loss time scale is significantly shorter than $t_{\rm relax}$ (see \refsec{massloss}), this relaxation channel is not effective in this scenario. A second scenario, taking the more extreme assumption that dark matter was negligible for the dynamics of GCs when the galaxy was very young, we find
%\be\label{eq:relaxtimeEarlyBaryon}t_{\rm relax} \sim 0.5 \frac{N}{30}\frac{R}{2~{\rm kpc}}\frac{10~{\rm km/sec}}{v}\left(\frac{0.5}{f}\right)^2~{\rm Gyr} \; ,\ee
%This is a strikingly short relaxation time, potentially competitive with the time-scale of the early phase of GC mass loss. For \UDG, it would hint that an early period in the life of the galaxy could have contributed to mass segregation of GCs, had the galaxy been dominated by the GCs themselves. We discuss this point within concrete models in \refsec{discussion}.

Put in a wider scope, these estimates suggest that there may be regions in ``parameter space'' of ultra-diffuse galaxies where two-body relaxation of GCs could be important.

Lastly, although we focused on the implications of two-body relaxation on the scale of an entire galaxy, the effect can manifest in part of a galaxy. Consider the possibility that $\mathcal{O}(10)$ GCs are driven by DF close the galactic center and stall there, e.g. due to core stalling \citep{Read:2006fq}. Re-purposing \refeq{relaxtime} for this case, we find
\be
t_{\rm relax} \sim 0.3 \frac{N}{10}\frac{R}{1~{\rm kpc}}\frac{20~{\rm km/sec}}{v}\left(\frac{0.2}{f}\right)^2~{\rm Gyr} \; .
\ee
Here values are motivated by \UDG, c.f. \refsec{udgdatamodels}. The short relaxation time that we find suggests that even if DF becomes ineffective due to core stalling, mass segregation of GCs may proceed due to their N-body interaction, potentially allowing the formation of a nucleus.

%%%%%%%%%%%%%%%%%%%%%%%%%%%%%%% Section 3-epsilon %%%%%%%%%%%%%%%%%%%%%%%%%%%%%%% 
%\section{Data of \UDG}\label{sec:udgdata}
\section{Observational constraints and mass models of \UDG}\label{sec:udgdatamodels}
In this section we summarize observational constraints on \UDG, and describe halo mass models that we will use in numerical simulations.

% The distance $D$ to \UDG\ can be derived in several techniques. 
% By using surface brightness fluctuations, \cite{Danieli2021} derived $D \sim 21\pm 5$~Mpc. 
Similar to \cite{Danieli2021}, we adopt a distance of $D = 26.5\pm 0.8$~Mpc to \UDG, based on the distance to the NGC5846 group, reported in \citet{KourkchiTully2017:dis}. 
% By assuming that \UDG\ is a satellite of the NGC5846 galaxy (and not only a satellite of the galaxy group \cite{muller2020spec}), one obtains $D = 26.5\pm 0.8$~Mpc, which we adopt throughout the paper. 
Note that the association of \UDG\ with the NGC5846 galaxy is not guaranteed and may perhaps be disfavoured from kinematical measurements; \cite{Forbes2021} reports a radial velocity $2167\pm 2$~km/sec to \UDG, whereas NGC5846 galaxy was measured at $1712\pm 5$~km/sec \citep{Cappellari11}. A radial-velocity difference of $\approx 455$~km/sec is rather high for a satellite. We note that the trend seen in \reffig{introData} is qualitatively insensitive to the distance estimate, although detailed constraints on the galaxy halo could be affected. For example, since $r_\perp \propto D$ and  $\MGC\propto D^2$, within the scope of a simple analysis as in \reffig{DataAndSimpleModel} we can estimate that a $10\%$ ($20\%$) uncertainty on $D$ yields an $\approx 20\% $ ($40\%$) uncertainty on the DF time-scale $\tau_{\rm core}^{(0)}$ (with larger distances implying smaller DF).

%We adopt a distance of $26.5$~Mpc (see discussion in \cite{Danieli2021}). 
The stellar luminosity was found to be well-described by a \sersic\ profile with index $n=0.61$, half-light radius $R_e =1.9$~kpc and luminosity $ L_V = 0.6\times 10^{8}~L_\odot $ \citep{Danieli2021}.

\UDG\ was noted for its large GC content \citep{muller21udg,muller2020spec,Forbes2021}. The highest quality photometric data of these GCs was obtained using 2-orbit WFC3/UVIS observations of the \textit{Hubble Space Telescope} \citep{Danieli2021}. The compact objects catalog was processed into different populations corresponding to different selection criteria of magnitude, angular size, position and color. Here we primarily adopt the two sets in the magnitude range $21 \lesssim m_V \lesssim 25$ due to their high quality and low contamination (about $1$ expected object out of $33$ in $r<2R_e$, based on a nearby background field), c.f. right panel of \reffig{udg1_hst}. We note that our results are insensitive to the photometric selection criteria, as discussed at \refapp{robust}.

At the faint end, $25 \lesssim m_V \lesssim 26.5$, the catalog suffers larger contamination (about $24$ expected objects out of $43$). We refrain from using it in our main analysis, but note that it can be useful in principle within a more comprehensive statistical analysis. We show a preliminary analysis in \refapp{faintset}.

\begin{figure*}[hbtp!]
	\centering
	\includegraphics[width=0.485\textwidth]{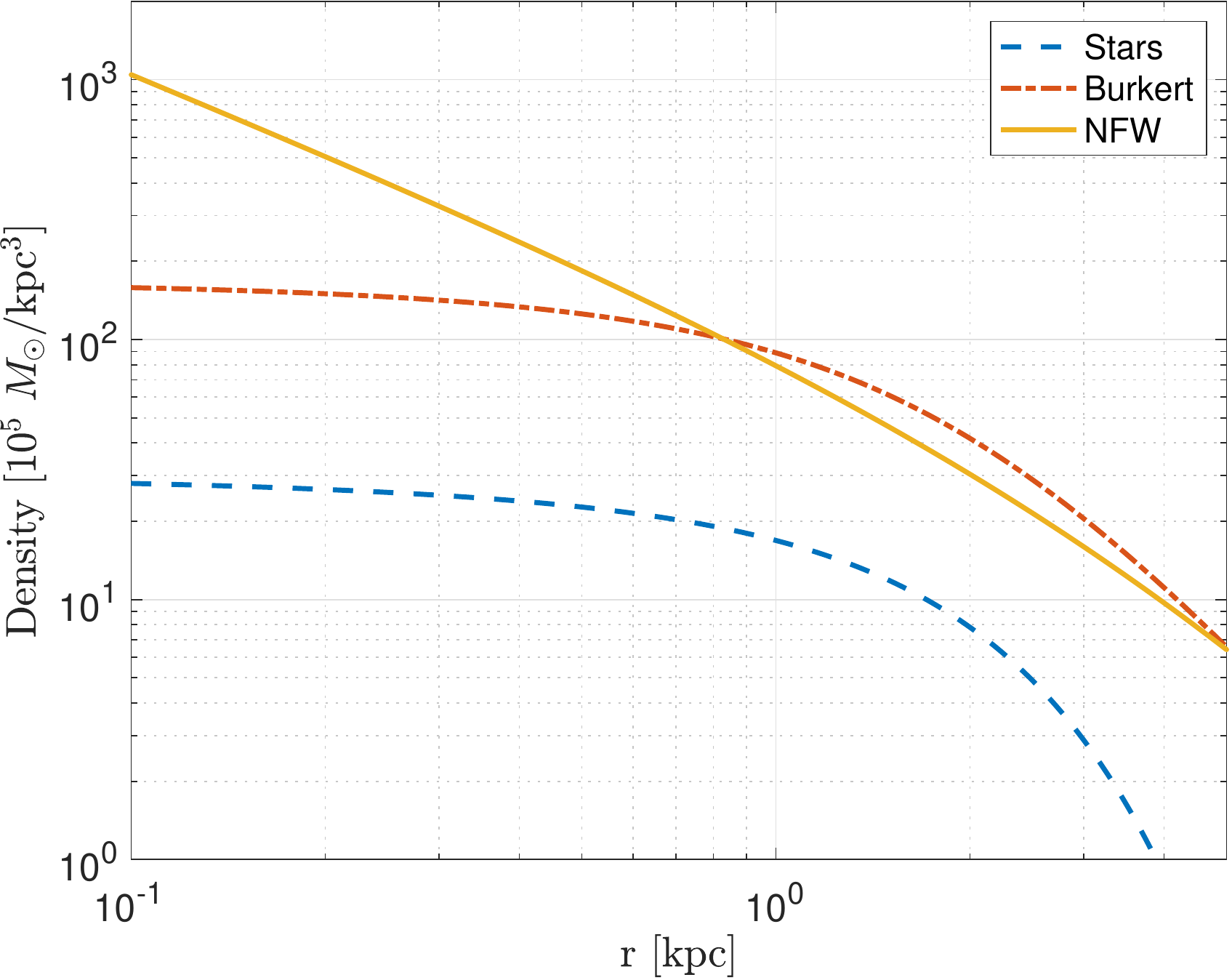}	\includegraphics[width=0.485\textwidth]{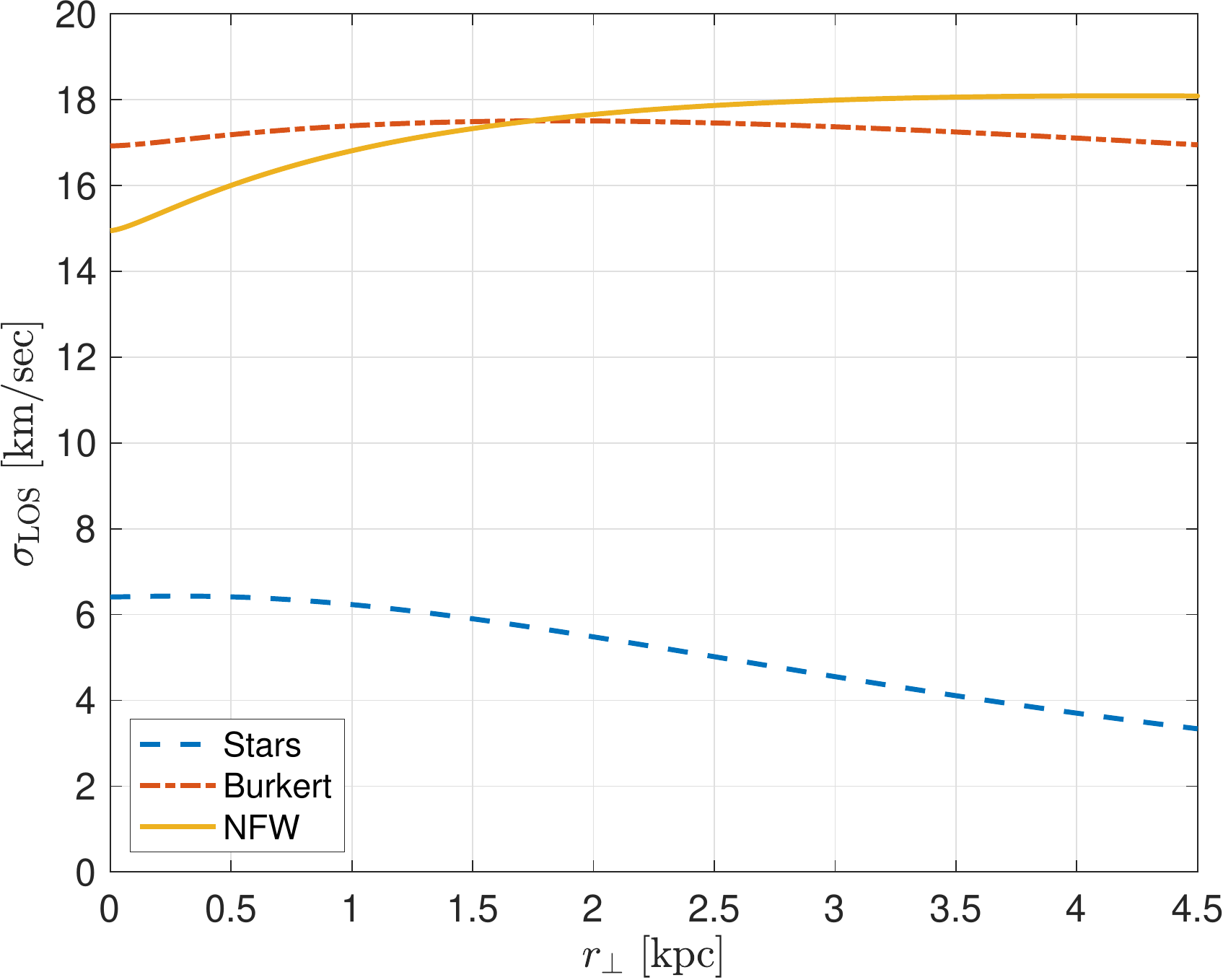}
	\includegraphics[width=0.485\textwidth]{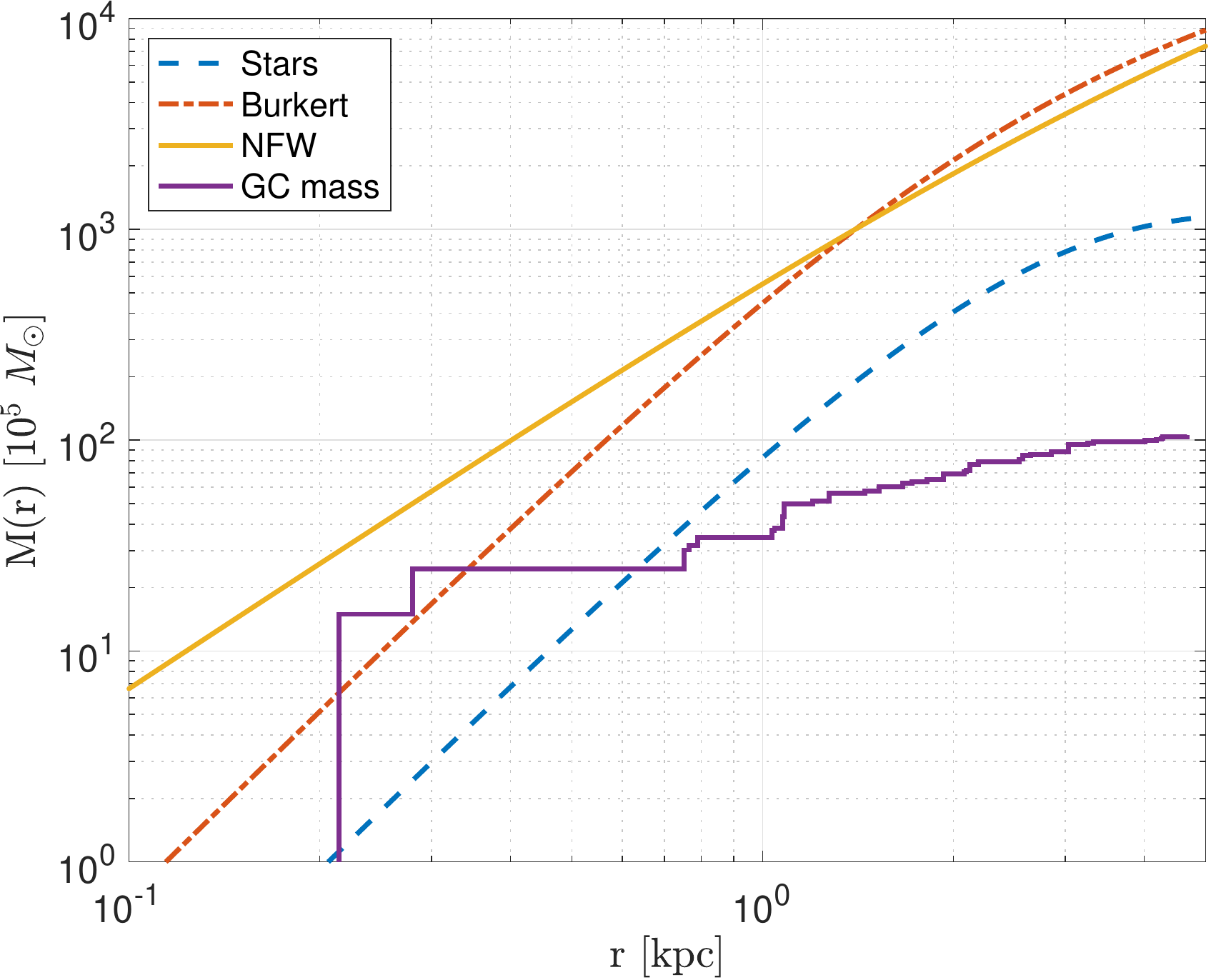}	\includegraphics[width=0.485\textwidth]{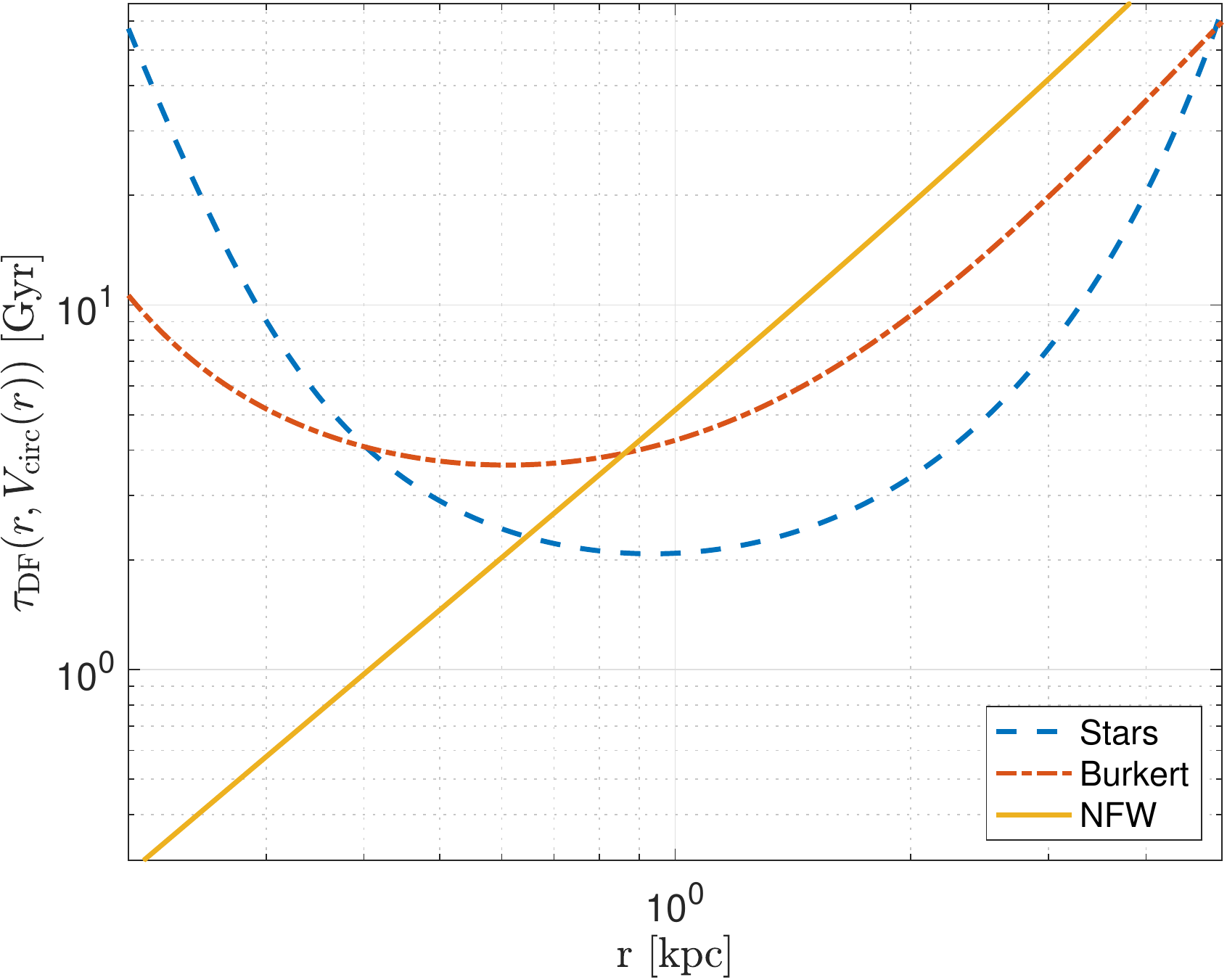}
	\caption{Properties of mass models that we use. \textbf{Top left}: mass density as a function of radius. \textbf{Top right}: Expected line-of-sight velocity dispersion of the observed distribution of stars \cite{Danieli2021} assuming isotropic velocity dispersion. \textbf{Bottom right:} the dynamical friction time evaluated with the velocity of a circular orbit for $ \MGC = 5\times 10^5~M_\odot $, with Coulomb logarithms as described in \refsec{simoutline}. \textbf{Bottom left:} the enclosed mass of different models with a rough comparison to GCs mass profile, deprojected via the approximation $ r = 1.25\times r_\perp $ (see \refeq{rperpvsr}).}\label{fig:modelsProp}
\end{figure*}

Spectroscopic studies of \UDG\ have confirmed the membership of $11$ GCs in the galaxy \citep{muller2020spec,Forbes2021}. Furthermore, \cite{Forbes2021} reported a GC line-of-sight velocity dispersion $\sigma_{\rm LOS} = 17\pm 2~$km/sec, implying that dynamics are dominated by dark matter. In what follows we describe two dark matter-dominated mass models that roughly saturate the reported $\sigma_{\rm LOS}$ from \cite{Forbes2021}, and one mass model that neglects dark matter altogether. 
%
%%%%%%%%%%%%%%%%%%%%%%%%%%%%%%% Section 3 %%%%%%%%%%%%%%%%%%%%%%%%%%%%%%% 
%\section{Mass-models of \UDG}\label{sec:udgmodels}
%
%To set-up concrete modeling of dynamics in \UDG, we define three mass models:
\begin{enumerate}
	\item[\stars] A mass model following the observed stellar luminosity, adopting $ M/L_V = 2~M_\odot/L_\odot $ following \cite{muller2020spec}.
	
	\item[\Burk] A mass model following \cite{Burkert1995}, $ \rho = \rho_0 r_0^3/[(r+r_0)(r^2+r_0^2)] $. We set $ r_0 = 2 $~kpc and $ \rho_0 = 1.66\times 10^{7}~M_{\odot}/{\rm kpc}^3 $. The value we adopt for $\rho_0$ allows us to test a profile that is significantly more massive than the \stars\ profile, yet sufficiently dilute to have $ \tau_{\rm DF}\lesssim 10 $~Gyr. In addition, this value is consistent with a velocity dispersion $17\pm 2 $~km/sec reported in \cite{Forbes2021} (saturating the $ 1 \sigma$ upper bound of \cite{muller2020spec}).
	%We adopt this value of $\rho_0$ for two reasons: (i) We want to test a model appreciably more dense than the \stars-only model yet sufficiently dilute to have $ \tau_{\rm DF}\lesssim 10 $~Gyr; (ii) We would like to compare with \cite{Forbes2021}, who reported a line-of-sight velocity dispersion of $ \sim 17\pm 2 $~km/sec (or, similarly, saturating the $ 1 \sigma$ upper bound of \cite{muller2020spec}). The density $ \rho_0 \approx 1.66\times 10^{7}~M_{\odot}/{\rm kpc}^3 $ satisfies both requirements.
	
	\item[\NFW] A mass model following \cite{Navarro:1996gj}, $ \rho = \rho_c\delta_c/[(r/R_s)(1+r/R_s)^2] $. We set $ R_s = 6 $~kpc and $ c=6 $, defined in the usual way in $ \delta_c $ \citep{Navarro:1996gj}. The predicted stellar kinematics in this model is comparable to that in the \Burk\ model. For reference, the virial mass of this model is $M_{200} = 200\frac{4\pi}{3}\rho_c c^3R_s^3 \approx 6\times 10^9~M_\odot$.
\end{enumerate}
In \reffig{modelsProp} we show the density, line-of-sight velocity dispersion, DF time scale, and enclosed mass of the different mass models. 

We can compare \reffig{DataAndSimpleModel} to the DF time scale in the bottom-right of \reffig{modelsProp}. None of the models have constant $ \tau_{\rm DF}$ and $ \rho$ as a function of radius, as assumed in the toy model of Sec.~\ref{sec:roughtheory}, but the cored models (\stars\ and \Burk) come close. All of the models predict significant DF over a 10~Gyr time scale, for GC orbits entering within a few kpc from the galaxy center.
%Evidently, the \Burk\ model fits \refeq{fittedtauDF} better than the \stars-only model, which na\"ively predicts too much DF. The DF time-scale of the NFW model behaves different radially, but it is in the correct ballpark to induce an appreciable DF effect.

%%%%%%%%%%%%%%%%%%%%%%%%%%%%%%% Section 4 %%%%%%%%%%%%%%%%%%%%%%%%%%%%%%% 
\section{Numerical simulations}\label{sec:simulations}
Our goal in the simulations is twofold. First, we aim to test the possibility that the apparent mass segregation in \UDG\ is due to DF. A central unknown in the problem is the initial distribution of GCs; our first task is to explore a range of initial conditions and see if a reasonable starting distribution can naturally evolve into the observed one. Second, provided that we can indeed identify reasonable initial conditions for the distribution of GCs, consistent with current observations, we also aim to examine whether the GC data can discriminate between different halo models.

Predicting the long term dynamics of a GC population in a galaxy like \UDG\ would optimally involve direct integration of live GCs \`a la \cite{Dutta_Chowdhury_2020} with baryonic effects \`a la \cite{Shao:2020tsl} in high resolution models of galaxies \`a la \cite{Meadows20}. 
%That would be a worthwhile endeavor which we leave for future works. 
Here, we take a more modest approach, 
%designed to direct the investigation on the unprecedented galaxy, \UDG, and hopefully to prepare the ground for future discovered galaxies of a similar kind. We 
simulating the N-body dynamics of GCs in a smooth external gravitational potential, adding dynamical friction using a semi-analytic prescription \citep{Petts2015,Bar:2021jff}.

\subsection{Simulation description}

\subsubsection{Outline}\label{sec:simoutline}
\textbf{Dynamics.} We set-up an N-body simulation, where each body represents a GC with a Plummer softening of $ \epsilon= 7 $~pc, i.e., the gravitational potential due to GC $i$ is $ \Phi_i(\mathbf{r}) = -GM_i/\sqrt{(\mathbf{r}-\mathbf{r}_i)^2+\epsilon^2} $. The background halos of stars and dark matter are modeled by a smooth and static profiles corresponding to the models in \refsec{udgdatamodels}, such that the background gravitational acceleration is modeled as $ -GM(r)\hat{r}/r^2 $. 

DF is implemented using a deceleration term $ -\Vvec/\tau_{\rm DF} $. Specifically, we assume $C = \ln\Lambda\left[{\rm erf}(X)-\exp(-X^2)2X/\sqrt{\pi}\right]$ (corresponding to a Maxwellian velocity distribution), where $X = V/\sqrt{2}\sigma$ and $\sigma$ is the local velocity dispersion of the medium.  %
The Coulomb logarithm $\ln\Lambda$ is modeled following \cite{Bar:2021jff}, where in the \stars\ and \Burk\ cases we select 
%the isothermal (``ISO'') expression,
$\Lambda_{\rm ISO} = 2V^2r/(G\MGC)$, and for \NFW\ $\Lambda_{\rm NFW} = b_{\rm max}^2\sigma^2/(G\MGC)$ with $b_{\rm max} = 0.5~$kpc. In both case we use in practice $\ln \Lambda\to \frac{1}{2}\ln(1+\Lambda^2)$ to regulate the logarithm when it becomes small and the treatment breaks down. As discussed in \cite{Bar:2021jff}, we can expect that the semi-analytic procedure captures the correct numerical value of $\tau_{\rm DF}$ to a factor of 2 or so.

As GCs inspiral inwards due to DF, some of their orbital energy is transferred to and heats the background medium. Accurate modeling of this effect requires simulations that resolve the particles of the medium, which is beyond the scope of this work.
%Since the gravitational potentials of galaxies typically vary by $\mathcal{O}(1)$ throughout the stellar body, an inspiral due to DF can cause massive objects to gain $ \mathcal{O}(1) $ binding energy with respect to the galaxy. Since energy is conserved, the friction medium is ``heated up'' and loses binding energy. Properly accounting for this effect requires  simulations that resolve the background halo particles, which is beyond the scope of our analysis. In \refsec{simulations}, when we employ simple simulations with a semi-analytic implementation of DF, we treat this issue in a crude way by ``turning off" DF below the radius where GCs have a comparable mass to the enclosed mass of the medium. 
%
%Following the discussion in \refsec{dfheating}, 
Instead, to roughly model this effect, we limit DF to radii where $ M_{\rm halo}(r)-M_{\rm GC,enclosed}(r)/2 > 0 $. At radii $r< r_{\rm DF,crit}$, where $r_{\rm DF,crit}$ is defined by $M_{\rm halo}(r)=M_{\rm GC,enclosed}(r)/2$, we turn off the DF deceleration term. We update $r_{\rm DF,crit}$ every $ 0.1 $~Gyr. For cored profiles (\stars\ and \Burk) we turn DF off at $ r< 0.3 R_e $ to mimic core stalling \citep{Read:2006fq,Kaur2018,Meadows20,Dutta_Chowdhury_2019}.

We implement GC mergers using an effective merger criterion. The merger criterion we choose is the simultaneous fulfillment of the conditions $ E_{12} \equiv \mu \mathbf{V}_{12}^2/2+U_{12}(\mathbf{r}_{12})<0 $ and $ r < r_{\rm merger} = 20~{\rm pc}\sim  {\rm few}\times \epsilon $~pc (c.f. \cite{Dutta_Chowdhury_2020}). Here, $ \mu = m_1m_2/(m_1+m_2) $ is the reduced mass of the GC pair, $ \mathbf{V}_{12} \equiv \dot{\mathbf{r}}_{12} =  \dot{\mathbf{r}}_2-\dot{\mathbf{r}}_1 $ is the relative velocity, and $ U_{12}(r_{12}) $ is the relative potential. For point-like objects, $ U_{12}(r_{12}) = -Gm_1m_2/r_{12} $. In general, $ U_{12} = \frac{1}{2}\int(\rho_1\Phi_2+\rho_2\Phi_1)d^3x $. For Plummer spheres with softening parameter $ \epsilon $ we find that $ U_{12}\approx -Gm_1m_2/({r_{12}^{2.11}+(1.7\epsilon)^{2.11}})^{1/2.11} $ provides a very good approximation, which we adopt in the simulations.

Upon a merger, we assign the new combined GC mass $ M = m_1+m_2 $, velocity $ \mathbf{V} = (m_1\mathbf{V}_1+m_2\mathbf{V}_2)/(m_1+m_2) $, and location $ \mathbf{R} = (\mathbf{r}_1+\mathbf{r}_2)/2 $. 
This corresponds to a linear momentum-conserving ``sticking'' of GCs. Energy is not conserved in this process: we neglect mass loss during the merger, thus energy must be transferred to the internal dynamics of the GC, which we do not model.

We approximate the process of continuous mass loss by decreasing GC masses in time steps of $ 0.1 $~Gyr, following the mass loss trend described in \refsec{massloss}. 

\textbf{Initial conditions.} Simulated GCs start in random positions in an (on average) isotropic distribution. We test initial GC radial distributions which start-off as a \sersic\ profile with different values of $R_e$. For simplicity, throughout we retain $ n=0.61 $, similar to the stellar distribution.

%We assume ergodicity of the initial GC phase space distribution function, and use this to define the initial velocity distribution \citep{BinneyTremaine2}.%\footnote{Ergodicity is assumed with respect to the halo, not the GCs themselves, therefore different-mass GCs have the same velocity distribution.}
We initiate the GC velocity distribution such that the radial distribution would remain stationary in the absence of DF, mass loss, and mergers \citep{BinneyTremaine2}.
Defining $ \mathcal{E} = \Psi(r)-v^2/2 $ (where $\ \Psi(r)\equiv -\Phi(r) $ taken conventionally to asymptote to zero at $ r\to \infty $), the goal is to derive the distribution function $ f(\mathcal{E}) $ based on the number-density profile of GCs $ n_{\rm GC} $, under a spherically-symmetric external potential $\Phi$, satisfying the Poisson equation for the halo's mass density $ \nabla^2\Phi = 4\pi G\rho_{\rm halo} $. We adopt a numerically-convenient expression for $ f(\mathcal{E}) $ \citep{magnithesis}
\be
f(\mathcal{E}) &=& \frac{1}{\sqrt{8}\pi^2}\Bigg[\frac{1}{\sqrt{\mathcal{E}}}\left(\frac{dn_{\rm GC}}{d\Psi}\right)\Big|_{\Psi=0}+2\sqrt{\mathcal{E}}\left(\frac{d^2n_{\rm GC}}{d\Psi^2}\right)\Big|_{\Psi=0}\nonumber\\&+&2\int\limits_0^{\mathcal{E}}d\Psi \sqrt{\mathcal{E}-\Psi}\frac{d^3n_{\rm GC}}{d\Psi^3}\Bigg] \; .
\ee
For the density profiles that we use, the first two terms vanish, leaving
\be
f(\mathcal{E}) \approx \frac{1}{\sqrt{2}\pi^2}\int\limits_0^{\mathcal{E}}d\Psi \sqrt{\mathcal{E}-\Psi}\frac{d^3n_{\rm GC}}{d\Psi^3} \; .
\ee

We track the system during a $ 10 $~Gyr time period.

As discussed in the next subsection, we test different possibilities for the GC initial mass distribution. 
%The GC initial mass distribution can either be selected to match the observed GC mass function, or can be sampled from a different distribution. We elaborate on our strategy in the next subsection. 
As a rule, we aim for an initial GC mass function that approximately matches the current GC mass function, accounting for mass loss. %Therefore, we correct the initial GC mass function to would-be pre-mass loss values.

%We define two procedures of initializing the GC mass distribution
%\begin{enumerate}
%	\item Adopting the observed GC luminosities with $ M/L_V = 1.6 $, based on a selected sample of $ 31 $ GCs from \cite{Danieli2021}. 
	
%	\item Sampling $ N_{\rm GC} \sim 40 $ GCs from a distribution $ f(\ln M)\propto \exp\left[-\frac{1}{2}\left(\frac{\ln M-\mu_M}{\sigma}\right)^2\right] $. The expected mass is $ \left<M\right> = \int Mf(\ln M)d\ln M/\int f(\ln M)d\ln M = M_0 \exp(\sigma^2/2) \sim M_0 $ (in the limit $ \sigma \ll 1 $), where $ \mu_M \equiv \ln M_0 $. The standard deviation is $ \Delta M \equiv \sqrt{\left<M^2\right>-\left<M\right>^2} = M_0 e^{\sigma^2/2}\sqrt{e^{\sigma^2}-1} \sim \sigma M_0 $ (in the limit $ \sigma\ll 1 $). We fix the parameters $ \mu $, $ \sigma $ such that the expected mass roughly follows the observed GC sample above (amounting to $ \sim 105~M_{\odot} $ with $ M/L_V = 1.6 $). This approach employs the simplifying assumption that GCs do not lose mass through mergers. 
%\end{enumerate}
%In both cases we correct the initial GC masses to pre-mass loss values.

\textbf{Implementation.} We implement the code in MATLAB using the \textsf{ode45} solver, partially based on \cite{MATLABnbody}. We use kpc-Gyr-$ 10^5M_{\odot} $ units, for which $ G = 0.449~ {\rm kpc}^3/10^5~M_\odot/{\rm Gyr}^2 $ and kpc/Gyr$\  = 0.979$~km/sec. We use a constant time-step of $dt= 2\times 10^{-5}\times 2\pi R/v $ ($ R $ and $ v $ being characteristic radius and velocity scales of the system). This amounts to $ dt\sim 10^{-5} $~Gyr.

\textbf{Convergence.} When DF, mergers, and mass loss are turned off, we find that energy is conserved to better than $ 1\% $ over $ 10~$Gyr.

\textbf{Sensitivity to parameters.} We tested the sensitivity of our results with respect to several parameters of the simulations, rerunning with (i) different Plummer softenings, $\epsilon = 4$ and $12$~pc; (ii) different merger radii $r_{\rm merger} =10$ and $35$~pc; (iii) unrestricted DF, i.e. without turning off DF for $r< 0.3R_e$ and $M_{\rm halo}-M_{\rm GCs}/2<0 $; (iv) a higher central concentration initial GC distribution -- with a \sersic\ index of $2$. In every case we retained all other parameters constant. We show the results in \refapp{othersims}. In general, we find that the choices of $\epsilon$, $r_{\rm merger}$ and DF near the center do not appreciably impact the radial distribution of GCs. Mergers, however, do depend relatively strongly on the choice of these parameters -- but still restricted to no more than few merger events per simulation ($\lesssim 0.03$ mergers/GCs). In the simulation run with an initially higher central concentration GC distribution we find more mergers ($\lesssim 0.1 $/GC), without significantly altering our main results.

\subsubsection{Method for comparing with observations}
%To compare between data and simulations we propose a moment analysis, as follows. 
For each halo and initial GC distribution model, we run 40 simulation realizations. We then compute the average projected distance $ \left<r_\perp\right> $ and the number of GCs $ \left<N_{\rm GC}\right> $ in the final state, 
%\footnote{One can compute higher moments such as $\left<r_\perp^2\right>$, but we do not find that it adds valuable information.}
splitting the GC sample into mass bins. We report 68\% confidence intervals for $ \left<r_\perp\right> $ and $ \left<N_{\rm GC}\right> $; these confidence intervals are dominated by the intrinsic randomness of the finite number of GCs per mass bin (we have verified that the averages and their confidence intervals are stable w.r.t. increasing the number of realizations per model). The predicted moments $ \left<r_\perp\right> $ and $ \left<N_{\rm GC}\right> $ can then be compared to the observed moments in the data. As noted in \refsec{udgdatamodels}, we only use GC candidates at $r_\perp < 2R_e = 3.8$~kpc, in order to minimize the background contamination from non-GCs sources. %Therefore, any simulation moments are also computed within that region.

%The moments $ \left<r_\perp\right> $ and $ \left<N_{\rm GC}\right> $ in different mass bins can be expected to exhibit statistical correlation, which complicates the analysis. 
%This correlation is physical, and can be estimated from the simulations as follows. Defining $ N_{\rm bins} $ the number of mass bins ($ \sim 4 $) and $ N_{\rm batch} $ the number of simulation runs per batch ($ \sim 40 $), we can compute the cross-correlation matrix $ C_{ij} = \mathbb{E}\left[(\left<r_\perp\right>_i-\mathbb{E}(\left<r_\perp\right>_i))(\left<r_\perp\right>_j-\mathbb{E}(\left<r_\perp\right>_j)/\sigma_i\sigma_j)\right] $, 
%where $ \left<r_\perp\right>_i $, $ i\in [1,N_{\rm bins}] $, and the expectation values are done over the $N_{\rm batch} $ runs. We find $ |C_{i\neq j}| \lesssim 0.2 $, consistent with low or no correlation within the level of our simulations.

%not significantly larger than a random noise. \textcolor{blue}{\bf (KB: if I compute the $\chi^2$ summing over the bins ignoring the covariance, how much do I miss the correct $\chi^2$ including the covariance?)}

We comment that when GC mergers are not important (average number of mergers per GC during 10~Gyr much smaller than unity), the variable $N_{\rm GC}$ simply reflects the observed current number of GCs per mass bin, and does not contain any additional information on the dynamics (apart, of course, from demonstrating the fact that mergers are not important). %The exception to this %When mergers are important, $N_{\rm GC}$ can test the model, to the extent that we can constrain the   
%We note that when mergers are important, $ N_{\rm GC} $ is sensitive directly to the halo through the induced velocity dispersion and indirectly through the induced dynamical friction since it increases the GC density -- both of which affect the merger rate (see \refsec{mergers}). When mergers do not take place, $ N_{\rm GC} $ must be taken as initial conditions and does not hold information on dynamics. An exception to this statement arises when one simulates a galaxy, but only compares to data in a region on the sky. Then, some GCs in the simulations are not selected in the field of view. Indeed, this is the case, as we shall show in the next subsection.

%%%%%%%%%%%%%%%%%%%%%%%%%%%%%%% Section 5 %%%%%%%%%%%%%%%%%%%%%%%%%%%%%%% 
\section{Results and Discussion}
\label{sec:discussion}

\subsection{Results: observed GC mass function as initial condition}
\label{sec:simresultsobs}

\begin{figure*}[htbp!]
	\centering
	\includegraphics[width=0.32\textwidth]{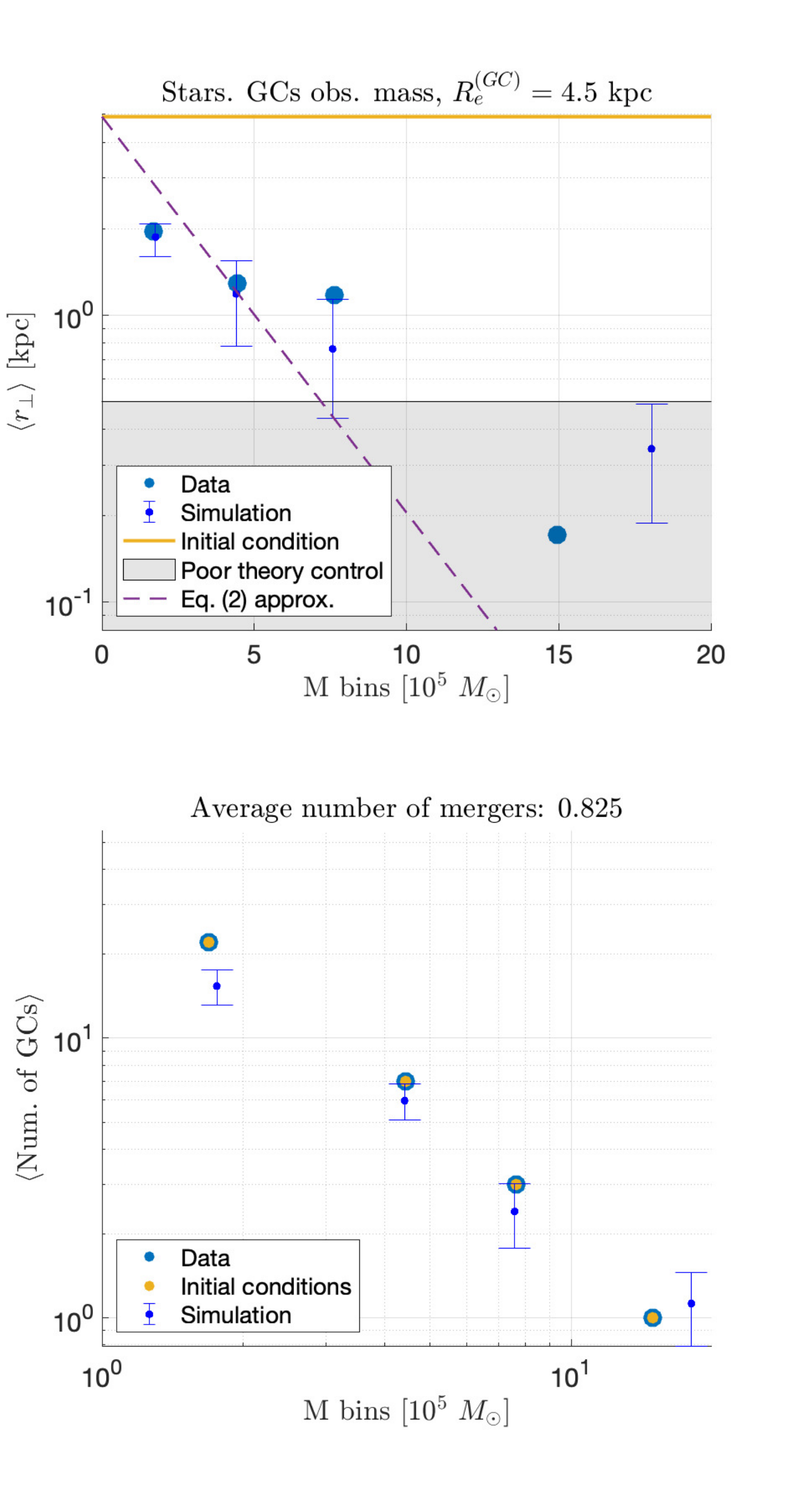}
	\includegraphics[width=0.32\textwidth]{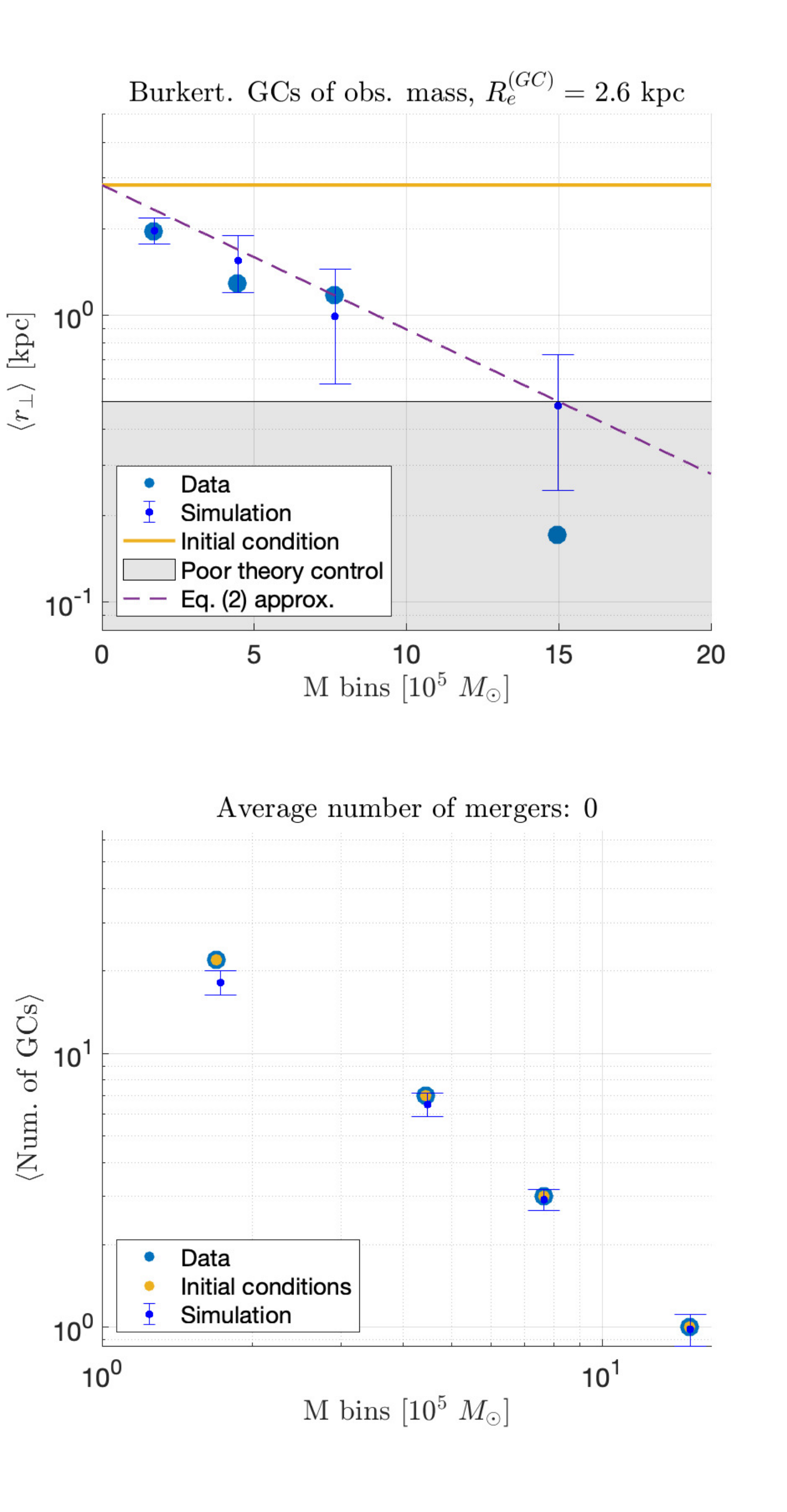}
	\includegraphics[width=0.32\textwidth]{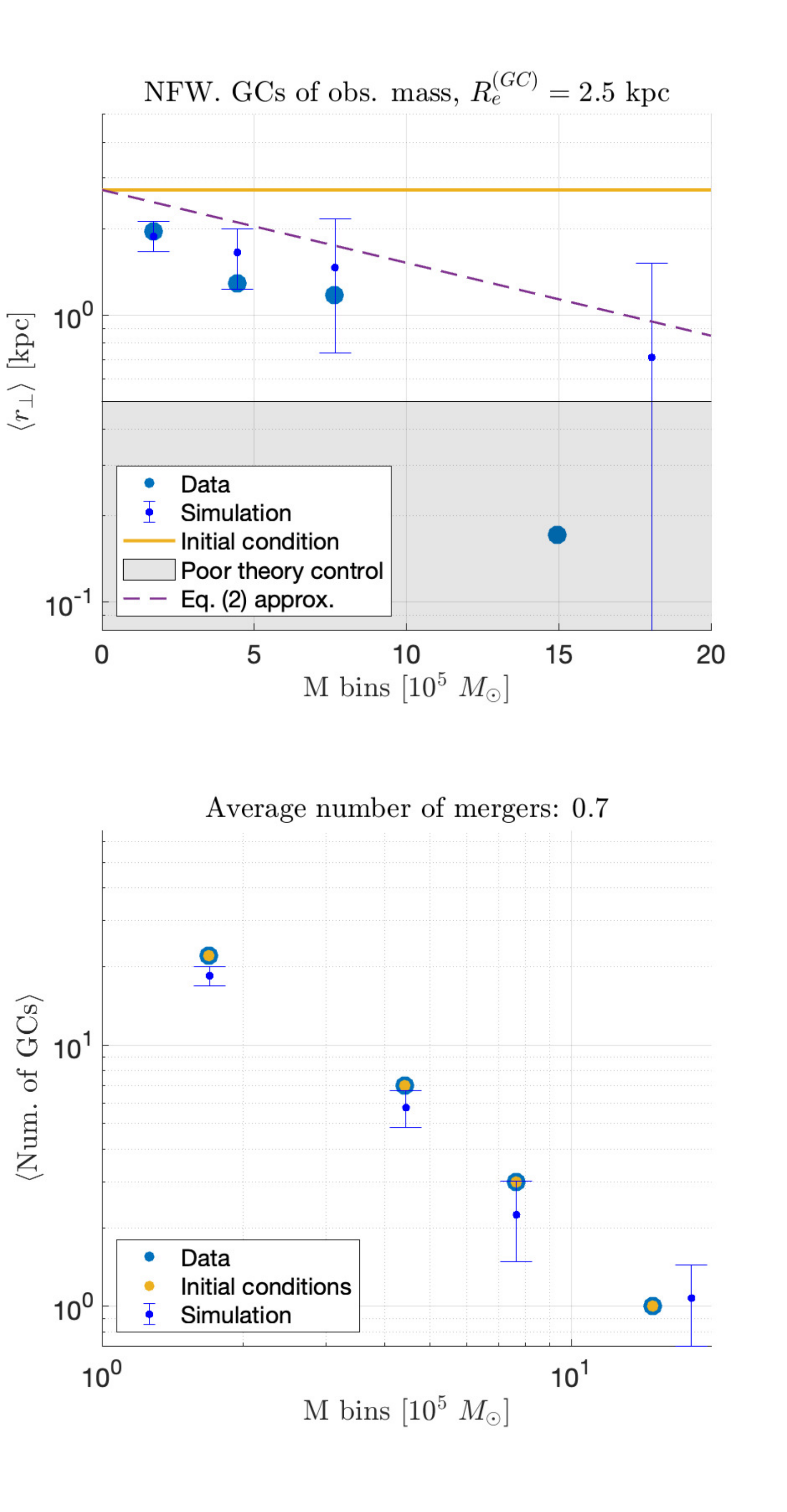}	
	\caption{ Each column of panels represents a batch of $ 40 $ realizations of a halo model and initial conditions. The initial set of GC masses is taken to match the observed set of $ 33 $ GCs \cite{Danieli2021}, correcting only for continuous mass loss. The initial radial distribution is \sersic\ with $ n=0.61 $, with half-light radius that is different for each halo model. \textbf{Left column:} \stars\ model. \textbf{Middle column:} \Burk\ model. \textbf{Right column:} \NFW\ model. \textbf{Top row:} The mean projected radius of GCs in different mass bins. The filled points are data from \cite{Danieli2021} (see discussion in \ref{sec:udgdatamodels} and \reffig{introData}). Points with error bars are simulation results including $ 68\% $ confidence intervals. Horizontal orange lines show the $ \left<r_\perp\right> $ of the initial GC distribution (common to all mass bins). The corresponding initial \sersic\ radius is $R_e= 4.5 $, $ 2.6 $ and $ 2.5 $~kpc for \stars, \Burk\ and \NFW, respectively. For comparison, the stellar body of \UDG\ is best-fit with $R_e \approx 1.9$~kpc, such that $\left<r_\perp\right>\approx 2.1$~kpc. Shaded gray region at $r_\perp < 0.5 $~kpc is to remind the reader that the predictions of the simulations are least robust in this area, notably because the mass in GCs may not be negligible w.r.t. to the halo enclosed mass (see text for more details). Dashed purple line is an adaptation of \refeq{corelogmass} using $\tau_{\rm core}^{(0)} = \tau_{\rm DF}(r,V_{\rm circ}(r))|_{r = R_e}$, $\left<r_{0,\perp}\right>_{\rm core}$ calculated using the initial GC distribution. \textbf{Bottom row:} The average number of GCs per mass bin, with big blue points representing the data, medium orange points the initial condition corrected for GC mass loss (by construction for this run, the initial condition is aligned with data). The small blue points with error bars represent simulation results with symmetric error estimate $ \sigma_N = \sqrt{\left<N_{\rm GC}^2\right>-\left<N_{\rm GC}\right>^2} $.  }\label{fig:simResultsObs}
\end{figure*}

In this subsection we run the simulations with the observed GC mass distribution as an initial condition: that is, the initial set of GCs is chosen to be identical to the observed set (c.f. \reffig{introData}, correcting only for mass loss). Thus, apart from mergers and mass loss, the only difference between the initial set of GCs and the currently observed set, is the radial distribution.
%function of the high quality sample, as described in \refsec{udgdata}. 
We seek approximate ``best-fit'' results for each halo model from \refsec{udgdatamodels}, scanning the GCs initial radius and determining agreement with the data by eye. (A more elaborate optimization procedure does not change the results appreciably.) We show the results in \reffig{simResultsObs}. In all three models we find reasonable fits to the current GC radial distribution (top row).

The main result of the analysis are the  \sersic\ radii of the initial GC distribution. These values are indicated in the top row of \reffig{simResultsObs} in the title and the horizontal orange line. For the \stars\ model, we find $R_e^{(GC)} = 4.5$~kpc, significantly more extended than the current distribution of the stellar body. For \Burk\ and \NFW\ models we find $R_e^{(GC)} = 2.6$ and $2.5$~kpc, respectively, just mildly more extended than the stellar body of \UDG. 

We also plot in the top row of \reffig{simResultsObs} an adaptation of \refeq{corelogmass} corresponding to  different models and initial conditions. We set the parameter $\left<r_{0,\perp}\right>_{\rm core}$ as the initial average projected distance and the DF time parameter $\tau_{\rm core}^{(0)} = \tau_{\rm DF}(r,V_{\rm circ}(r))|_{R_e=1.9~{\rm kpc},\MGC^{(0)}=5\times 10^5~M_\odot}$ at a reference mass and the half-light radius of the stellar body (recall, the sample of GCs we work with is restricted to $r<2R_e$). We see that \refeq{corelogmass} is a useful approximation, in reasonable agreement with the simulations.

In the bottom row of \reffig{simResultsObs} we plot the average number of GCs per mass bin. We find a small number of mergers in all cases $\lesssim 1$ (indicated in the title of the figures). One may notice that even in the absence of mergers (in the \Burk\ case), the simulation prediction for the number of GCs can be smaller than the initial one. This is the result of masking out GCs at $r_\perp > 3.8$~kpc when converting simulation results for comparison with observational data (we present ``initial conditions'' number without this cut).

%We note that in the NFW case (right panel of \reffig{simResultsObs}) we find an interesting behaviour in the largest mass bin. Our simulations predict a large variation in the projected distance of that bin. This originates from the strong $r$ dependence of $\tau_{\rm DF}$ in the NFW case -- about $\propto r^2$ for circular orbits (see \reffig{modelsProp} and \cite{Bar:2021jff}). This results 

%\begin{figure*}[htbp!]
%	\centering
%	\includegraphics[width=0.85\textwidth]{figures/BarMLS1s2A.eps}	\\
%	\includegraphics[width=0.85\textwidth]{figures/ISOMLS1s2A.eps}	\\
%	\includegraphics[width=0.85\textwidth]{figures/NFWMLS1s2A.eps}	
%	\caption{ As \reffig{simResultsObs} but with GC masses sampled out of a Gaussian distribution, as explained in the text.  }\label{fig:simResultsLow}
%\end{figure*}

\subsection{Results: almost-as-observed GC mass function as initial condition}\label{sec:resultsAlmostObs}
%\subsubsection{GC mass function: almost as observed}

In this subsection we consider an initial set of GC masses that is slightly different than the observed set. This allows us to demonstrate two points. The first point is that a small deficit in the predicted number of GCs in the low-$\MGC$ bins, as can be noted in the bottom row of Fig.~\ref{fig:simResultsObs}, can easily be compensated for by a small increase in the assumed initial number of low mass GCs. 
The second point concerns the possibility that GC mergers -- rather than pure DF -- are the origin of the most massive few GCs in the observed set. We find that, within the limitations of our simulations, this formation channel for the single most massive GC may be feasible, although assessing its likelihood in detail is somewhat beyond the expected domain of validity of our method.
%Motivated by the previous section's results, where some configurations had some GC mergers and the number of light GCs was underpredicted, let us slightly modify the observed GC mass spectrum. 

We perform this exploration using the NFW halo model. The initial set of GCs is chosen as follows. We break the most massive GC ($ \MGC\approx 1.5 \times 10^6~M_{\odot} $) into $ 3 $ GCs, one with $\MGC=7.5\times10^5~M_\odot$, and $ 2 $ with $\MGC=3.75\times10^5~M_\odot$. We also add $3$ light GCs with $M= \{1,1.75,2.5\}\times 10^5~M_\odot$. 

\begin{figure}
	\centering
	\includegraphics[width=0.48\textwidth]{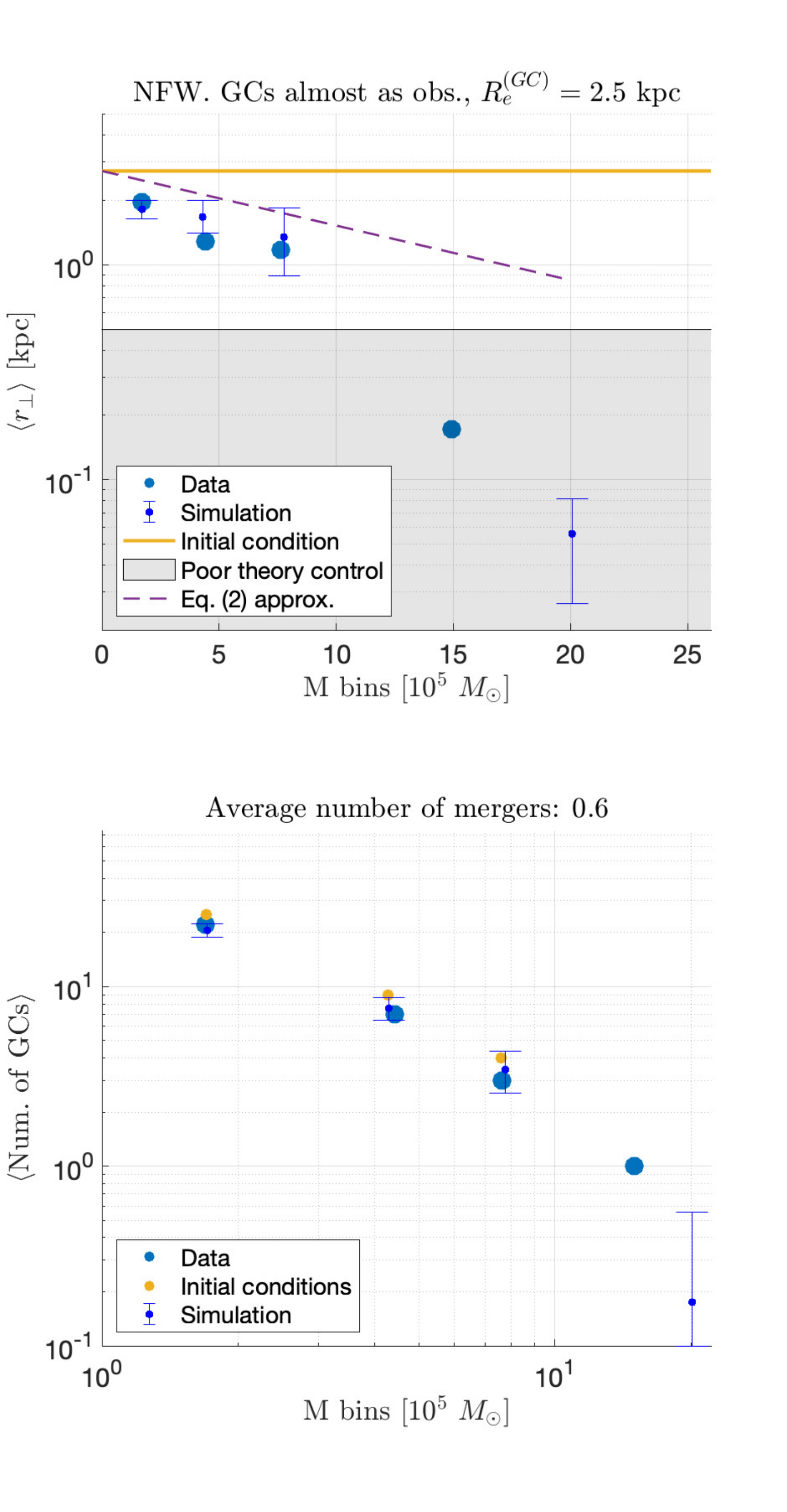}
	\caption{ An example where the observed GC mass function is slightly modified: the most massive GC, of mass $ \sim 15\times 10^{5}~M_\odot $ is traded in the initial GC mass function by $ 1 $ GC with half the mass, and $ 2 $ GCs with $ 1/4 $ of the mass. This shows how this massive GC may have been the result of mergers. Additionally, three light GCs with $\MGC\sim 1.75 \times 10^5~M_\odot$ were added to compensate for loss due to selection of data $r_\perp<3.8$~kpc. Indeed with this addition the simulation is a better fit to the data in the first data point.}\label{fig:simResultsModifiedSpec}
\end{figure}

The result of this exercise is shown in \reffig{simResultsModifiedSpec}. We find that: (i) The added low-mass GCs in the initial set bring the final set to perfect agreement with observations. (ii) The simulations do sometimes yield a sufficiently massive most-massive-GC, roughly consistent with observations, but this is not common, and happens in only about 20\% of the runs. When the most-massive-GC is produced by a merger, then this merger is essentially always taking place between the second-most-massive GC in the initial state and one of the intermediate $\MGC$ GCs. 

Both (i) and (ii) above are consequences of the paucity of mergers observed in our simulations. Regarding the low mass GCs, we expect that the result is quite robust.
%, given that low mass GCs typically occupy the outer regions of the halo where our simulation method should hold reasonably well. 
Regarding the more massive GCs, since these migrate into the inner halo, where our treatment of DF becomes less trustworthy, it is plausible that our simulations underestimate the massive GC merger rate to some extent. A refined treatment of the dynamics in the inner few hundred pc of the system would be needed to clarify this issue. 

\subsection{Discussion}
The main lesson that we draw from our analysis, analytical and numerical, is that DF provides a natural explanation for the apparent mass segregation of GCs in \UDG. The projected radial distribution obtained in the simulations is consistent with the trend $\ln\langle r_\perp\rangle\propto-\MGC$ (see Eq.~\eqref{eq:corelogmass} and  Fig.~\ref{fig:DataAndSimpleModel}). The slope and intercept of this trend are compatible with simple analytic estimates based on DF theory. The dark matter-dominated halo models are also consistent with independent constraints on the stellar velocity dispersion \citep{Forbes2021}, and suggest a reasonable initial condition for GCs that is only slightly more extended than the current observed stellar body. %More detailed numerical simulations confirm that DF can quantitatively reproduce the observed GC distribution, for initial conditions that appear reasonable, provided that the gravitational potential of \UDG\ is dark matter-dominated (see Fig.~\ref{fig:simResultsObs}).

%We would like to reiterate a few points:
%\begin{enumerate}
	%\item GC mergers are rare:
	We find a low merger rate, $ \lesssim 1$ per 10~Gyr (i.e. $\lesssim3\%$ per GC per 10~Gyr), for all of our halo models. Factors that may, in principle, relax this result include: (i) we ``turn off" DF in the inner halo, below a radius defined as to guarantee $ M_{\rm halo}\gtrsim M_{\rm GCs}/2 $. Dissipative effects below this radius ($ \sim 0.7 $~kpc) are not modeled, and could induce more mergers; (ii) our merger criteria may be too strict. We test both of these factors in \refapp{othersims}, finding that the results remain fairly robust (especially for  DM-dominated halo models) even when DF is kept ``on" throughout the halo and when the merger criteria are varied.  
	
	%If mergers are not effective, it is not very valuable to start with a generic GC mass distribution (which is currently imperfect, in any case). 
	
	The few mergers that are observed in the simulations often involve the most massive GCs, as these both have a larger intrinsic collision cross section and also settle into the inner halo such that their density is increased. Thus, the occasional mergers observed in the \stars\ and \NFW\ halo models can slightly skew the initial mass distribution, and with a probability of the order of 20\% could even account for the formation of the most massive GC in the sample through merger. 
	
	As a guide for numerical studies, it is possible to analytically estimate the mass of the nuclear cluster $m_{\rm nuc}$ resulting from GC mergers in an NFW halo. Following \cite{Bar:2021jff}, we approximate $\tau_{\rm DF}(r;\MGC)\approx \bar\tau(\MGC^{(0)}/\MGC)(r/\bar{r})^\beta$, applicable for circular orbits with $\beta\approx 2$, and define $r_{\rm cr}$, the radius below which GCs on circular orbits decay to the galactic center after time $\Delta t$, $r_{\rm cr}(\MGC) \approx \bar{r}(4 \Delta t(\MGC/\MGC^{(0)})/3\bar\tau)^{1/2}$. The nuclear cluster mass is then
	\be
	m_{\rm nuc} \approx \sum_i \MGC^{(i)}\int\limits_0^{r_{\rm cr}(\MGC^{(i)})} n_{i,0}(r)d^3r \; ,
	\ee
	where $i$ runs on GC mass bins, $n_{i,0}$ is the initial GC number density. Using this expression, we find $m_{\rm nuc} \approx 15\times 10^5~M_\odot$ for the NFW simulation in \reffig{simResultsObs}, consistent with the numerical results.

\subsection{Caveats and questionable simplifications}\label{sec:caveats}
Before we conclude, we would like to highlight a few possible caveats in our analysis. 
\begin{enumerate}
\item \textbf{Use of semi-analytic description of dynamical friction.} A semi-analytic description of DF has been shown to achieve reasonable agreement with dedicated simulations for a cuspy halo profile \citep{ArcaSeddaDFCuspy2014,Bar:2021jff}. However, the procedure may be less accurate for cored profiles. It is generally agreed that DF is suppressed near the center of a cored halo \citep{Read:2006fq,Cole2012,Kaur2018,Meadows20}. However, studies that employ direct numerical simulations do not agree on some details: \cite{Cole2012,Banik21DF} report stalling and buoyancy effects at about the core radius, whereas \cite{Meadows20} reports continued DF, consistent with constant $ \tau_{\rm DF} $, and broadly consistent with semi-analytic expectations \citep{Petts2015,Bar:2021jff}. 
%To decisively test core profiles using DF arguments, simulations of DF in cored profiles will have to converge to a definitive answer. 
%Additionally, a short phase of ``super-Chandrasekhar'' friction \cite{Petts2015} -- at about $ r\sim R_e $ -- is not well described by the semi-analytic treatment \cite{Petts2015,Bar:2021jff}.

\item \textbf{Merger prescription and tidal disruption.} We neglected the internal dynamics of GCs in the treatment of mergers, and ignored GC tidal disruption by other GCs and by the host halo. Some support for this approximation comes from \cite{Dutta_Chowdhury_2020}, that argued that in a similar setting (NGC1052-DF2) the tidal capture of GCs is not a dominant effect. It would be useful to repeat our calculations using simulations of live GCs, to resolve internal GC dynamics.

We note that disruption of GCs by the halo is unlikely to be important. For circular orbits near the center of a \NFW\ profile, where the effect is most significant, the tidal radius is $r_{\rm tidal}\approx (\MGC/M_{\rm 200})^{\frac{1}{3}} r^{\frac{1}{3}} R_s^{\frac{2}{3}}$ (\citealt{renaud2011,orkney2019}). We find for our \NFW\ model $r_{\rm tidal} \approx 40\,(r/0.1~{\rm kpc})^{\frac{1}{3}}(\MGC/10^5~M_\odot)^{\frac{1}{3}}~{\rm pc}$, which is much larger than GCs sizes, even when evaluated at a very small radius, $0.1$~kpc.

%\item System age. The age of \UDG\ was estimated to be $ 11.2^{+1.8}_{-0.8} $~Gyr, with $ 11 $ spectroscopically-confirmed GCs in the same ballpark (with few Gyr uncertainties) \cite{muller2020spec}. %We integrated to $ 10 $~Gyr. 

%Mergers, if active at all, happen towards the end of the simulation. It happens simply because GC density grows larger due to DF. It raises the question of whether our results depend on the assumed age of the system.

%\item System morphology. We assumed spherical symmetry. \UDG\ does not exhibit significant deviations from that assumption, and GCs appear to be distributed isotropically across the galaxy \cite{Forbes2021,muller21udg,Danieli2021}. 

\item \textbf{Mass to light ratio.} We assumed a GC mass-to-light (M/L) ratio of $ M/L_V = 1.6~M_\odot/L_\odot $, motivated by \cite{muller2020spec} who reported this value, derived from the stacked measured spectra of $11$ GCs. 
%Several questions arise from this procedure. (i) 
Stacking aside, a GC-by-GC analysis suggests a small spread in the mass-to-light ratio ($0.2~M_\odot/L_\odot$), laregly consistent with the $1\sigma$ uncertainty of the mass-to-light ratio for the stacked spectra ($1.6^{+0.3}_{-0.1}~M_\odot/L_\odot$).  %\textcolor{blue}{\bf (KB: how much? and BTW, why is the stellar $M/L=2$ different to the GC $M/L$? is that commonly assumed/reasonable?)} 
Moreover, most of the GCs in the high quality sample that we analysed ($22$ out of $33$) have no spectroscopic data, and thus no direct $M/L$ estimate. Nevertheless, their relatively uniform color \citep{Danieli2021} suggests they are probably similar to the rest of GC sample. 
%We think this is not important, as our analysis is statistical rather than individual-GC-based. 

A systematic uncertainty in the mass-to-light ratio (either GC-by-GC, or overall) would affect DF estimates, since $ \tau_{\rm DF}\propto 1/\MGC $. 
%We think that $\sim 20\%$ uncertainty would not significantly change our results, but if for some reason there is $\mathcal{O}(1)$ uncertainty, it would probably make the analysis far less discriminative. 

\item \textbf{Initial conditions.} Our exploration of possible initial conditions for the GC sample was rudimentary, and could be made more systematic. 
%This was enough to establish DF as a plausible origin for the observed GC mass segregation, but a more systematic analysis could be useful to draw firmer conclusions about halo model discrimination. 
An essential part of the basic preference we find for DF obviously stems from our choice to initiate the GC sample with the same initial radial distribution across GC mass bins; one could, if one wanted, entertain the possibility that DF is ineffective in the system (some models of dark matter, for example, can effectively quench DF \citep{Hui2017,Bar:2021jff}, and that for some reason, more massive GCs are preferentially formed deeper into the host halo compared to less massive GCs, in a formation pattern that mimics the natural expectations from DF. Indeed, galaxy formation simulations may indicate that more massive GCs form closer to the center of galaxies \citep{RenaCampos2021}.

Another concrete example for a mechanism that could also induce GC mass segregation was briefly discussed in \refsec{selfRelax}: if the total mass of \UDG\ was dominated by GCs during a brief ($\mathcal{O}(100~{\rm Myr})$) early epoch before substantial GC mass loss took place, and under the (perhaps highly simplified) assumption that GCs formed at the same time, two-body relaxation of the GCs could have contributed to the mass segregation. This would essentially amount to mass segregation in initial conditions, since we do not attempt to model this epoch within our simulations.

We stress that the DF within our models is an irreducible effect. It should contribute to mass segregation also in the scenario of initial mass segregation. We note however that this expectation may spoil in the case of a high merger rate. 

%{\bf We stress that DF is an irreducible effect in cold dark matter. Therefore, }
%This would lead to a somewhat larger initial GC radius $R_e^{(GC)}$, primarily for the \stars\ model (as \Burk\ and \NFW\ are not expected to have ever been dominated by GCs). 
%Since we do not attempt to model the very early period in the life of the galaxy, such a process, if indeed effective, would have to be encoded in initial conditions within the scope of our analysis. We leave this investigation for future work.

%Although we try to remain agnostic about the initial conditions of GCs, we make a few assumptions about them. We assume GCs start with an ergodic distribution function. {\textbf{Is it motivated by any formation mechanism?}} \textbf{Comment on formation mechanisms that predict rotation in initial conditions? \cite{Forbes2021} comments on this.} Additionally, we assume that all GCs for a given halo start with the same \sersic\ profile. The only parameter we explore in that respect is the \sersic\ radius (keeping $ n=0.61 $, the same index of the \stars\ model, see \refsec{udgdata}). This treatment artificially precludes the possibility of GC formation as a function of radius depends also on GC mass. 

%It would be interesting to explore initial conditions characterized by some initial mass segregation.

\item \textbf{Galaxy mergers.} We did not consider the possibility that \UDG\ has undergone mergers with other galaxies. 

\item \textbf{Tidal stripping.} Likely a member of NGC5846 galaxy group \citep{muller2020spec, Danieli2021}, \UDG\ may have been affected by tidal forces. Taking as a benchmark the near galaxy NGC5846 and the distance assumption of $26.5$~Mpc, the two galaxies are separated by projected distance $r_{\rm gal} = 164$~kpc. Taking this as an estimate for the true distance between the galaxies, the tidal radius is
%\footnote{Projection would hint that the current true radius is larger by $\mathcal{O}(20\%)$, but orbital kinematics hint an opposite sign effect, since it is more likely to find an object near the apocenter rather than near the pericenter.},
\be
r_{\rm tidal} &\approx & r_{\rm gal} \left(\frac{M_{\rm \UDG}}{2M_{\rm NGC5846}}\right)^{1/3} \\ &=& 6\left(\frac{M_{\rm \UDG}}{10^8~M_\odot}\right)^{\frac{1}{3}}\left(\frac{10^{12}~M_\odot}{M_{\rm NGC5846}}\right)^{\frac{1}{3}}~{\rm kpc}\;. \nonumber
\ee
Here, the reference value for the mass of \UDG\ is on the low side, neglecting any contribution from dark matter; thus, it is relevant for the \stars\ model. For the mass of NGC5846, we a used $ 10^{12}~M_\odot$. We see that the GC initial condition found for the \stars\ model (half-mass radius of $4.5$~kpc) may indicate some level of inconsistency with the tidal radius estimate. The dark matter-dominated halo models (\NFW\ and \Burk) seem conveniently compatible with the tidal radius -- although the halos themselves may be somewhat affected by tidal stripping.

%It should be noted that the assumption that \UDG\ is a bound satellite of NGC5846 is rather preliminary. In particular, the difference of $\approx 450$~km/sec in radial velocity may pose a challenge to this hypothesis.

\end{enumerate}

%%%%%%%%%%%%%%%%%%%%%%%%%%%%%%% Section 5 %%%%%%%%%%%%%%%%%%%%%%%%%%%%%%% 
\section{Summary}
\label{sec:conclusions}

We have shown that the observed radial distribution of globular clusters (GCs) in NGC5846-UDG1 (\UDG) is suggestive of mass segregation. The mass segregation pattern can naturally be explained by dynamical friction (DF). While the basic imprint of DF appears clear, uncertainties on the initial distribution of GCs at formation complicate the task of drawing robust constraints on the dark matter content of the  halo. Assuming that GCs form at a characteristic radius that is not widely different from that of the bulk of the stellar population (not in GCs), and that the characteristic GC formation radius does not depend on GC birth mass, the data provides dynamical support for a massive dark matter-dominated halo for \UDG. This demonstrates that dynamical arguments (and not only kinematics) can shed light on the distribution of dark matter in galaxies.

The dynamical preference for a massive halo can be further tested with kinematics data. Indeed, it is broadly consistent with existing kinematics results from \cite{Forbes2021}. 

Our study motivates, and can benefit from several technical improvements in the scope and detail of our numerical simulations, as noted in the main text. Input from the theory of the formation of GCs, their expected mass function at birth and their initial characteristic radial scale (especially in comparison with the main stellar system), could provide better-informed priors for the initial conditions, allowing the dynamics analysis to produce sharper constraints on the dark matter halo. Alternatively, a more detailed phenomenological scrutiny of GC initial conditions in \UDG\ may be useful to inform GC formation theory, especially if combined with kinematics constraints.

More observational studies of UDG1-like galaxies, where GCs at various masses can be reliably identified and characterized, will play an important role in future work of this kind. The large sky coverage and point-source depth of the upcoming Vera Rubin Observatory Legacy Survey of Space and Time \citep{LSST:2009} will enable mapping of GCs associated with low surface brightness galaxies in the nearby universe. Looking ahead, the Nancy Grace Roman Space Telescope \citep{Spergel2015:Roman} will open an unprecedented window into studies of extragalactic GCs. Its wide-field imaging, high spatial resolution, and sensitivity will allow probing GCs below the turn-over magnitude across the sky and out to larger distances.

\acknowledgments
We thank Diederik Kruijssen, Shaunak Modak, Aaron Romanowsky, Scott Tremaine, Sebastian Trujillo Gomez and Pieter van Dokkum for comments on the manuscript and Boaz Katz and Karamveer Kaur for useful discussions. We thank Oliver M\"uller for a useful discussion on the spectroscopy of \UDG. We are also grateful to the anonymous reviewer for useful suggestions that improved the manuscript. NB is grateful for the support of the Clore scholarship of the Clore Israel Foundation. KB and NB were supported by grant 1784/20 from the Israel Science Foundation. SD is supported by NASA through Hubble Fellowship grant HST-HF2-51454.001-A awarded by the Space Telescope Science Institute, which is operated by the Association of Universities for Research in Astronomy, Incorporated, under NASA contract NAS5-26555.

\begin{appendix}
\section{Robustness of the GC sample and mass segregation}\label{app:robust}
It is important to substantiate that our results are not sensitive to the photometric selection criteria of \cite{Danieli2021}. \reffig{udg1_hst} demonstrates that our sample ($m_V < 25.0$~mag) is expected to be contaminated by $1$ object at the low-luminosity end. Therefore contamination is very unlikely to change the results. 

\begin{figure}[htbp!]
	\centering
	\includegraphics[trim={6cm 2.2cm 6cm 0.5cm},clip,width=0.7\textwidth]{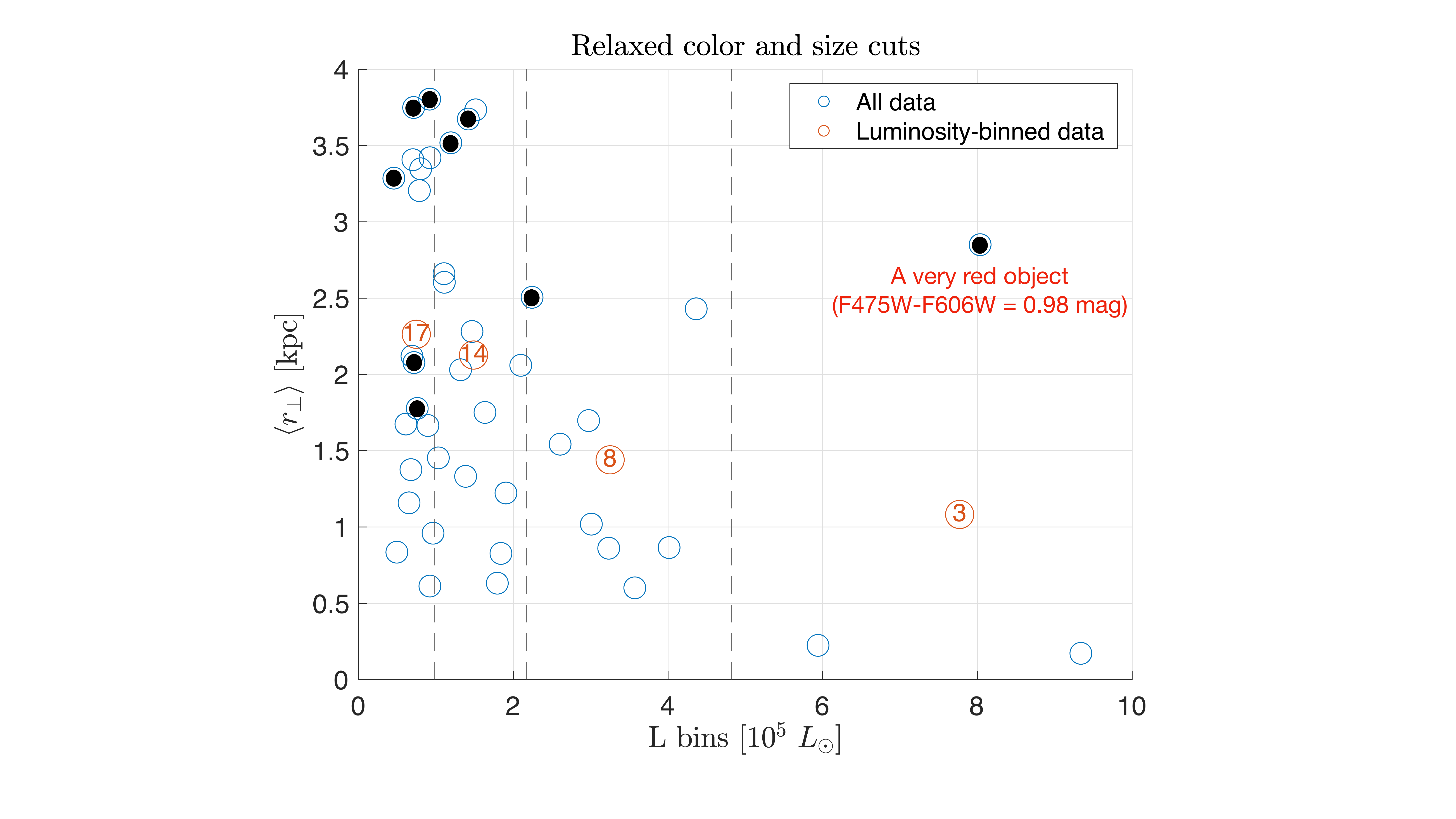}
	\caption{A modified GC sample with relaxed photometric selection criteria, as explained in the text. The ``very red object'' refers to an object whose color ($\mathrm{F475W-F606W} = 0.98$~mag) differs by about $20$ standard deviations than similar other spectroscopically confirmed GCs ($\mathrm{F475W-F606W} = 0.39 \pm \,0.03$~mag). }\label{fig:modifiedData}
\end{figure}

Another risk of using photometric selection criteria is the potential oversight of true GCs. To test this possibility, we re-examine the data presented in \cite{Danieli2021} by generously relaxing the selection criteria: changing the FWHM size and color from $2.4<{\rm FWHM} < 4.5$~pix ($2.1<{\rm FWHM} < 4.5$~pix) and $0.2 < \mathrm{F475W-F606W} < 0.6 $ ($0.08 < \mathrm{F475W-F606W} < 0.8 $) for $m_V < 24.5$~mag ($24.5<m_V<25$~mag) to $1.0<{\rm FWHM} < 10.0$~pix and $0 < \mathrm{F475W-F606W} < 1 $. This new sample is shown in \reffig{modifiedData}, after masking two bright objects whose spectroscopy suggests non-membership with \UDG \,\citep{muller2020spec}\footnote{We thank Oliver M\"uller for help on this point.}. Relaxing the selection criteria, $ 9 $ new objects are added (whose would-be luminosities are estimated assuming distance to \UDG): (i) $ 8 $ out of the $9$ are relatively faint objects at large radii, consistent with a contamination of $12$ objects by comparing to a nearby background field. (ii) A very red and bright object with a color of $\mathrm{F475W-F606W} = 0.98$~mag. It is suggestive to compare this to the $11$ spectroscopically-confirmed GCs (which have similar brightness) and are narrowly distributed at $\mathrm{F475W-F606W} = 0.39~$mag with a standard deviation of $0.03$~mag. Due to this large difference, we assume this object to be foreground. We note that it is difficult to substantiate spectroscopically as this object is very near on the sky to a very bright foreground star.

\begin{figure}[htbp!]
	\centering
	\includegraphics[height=0.40\textwidth]{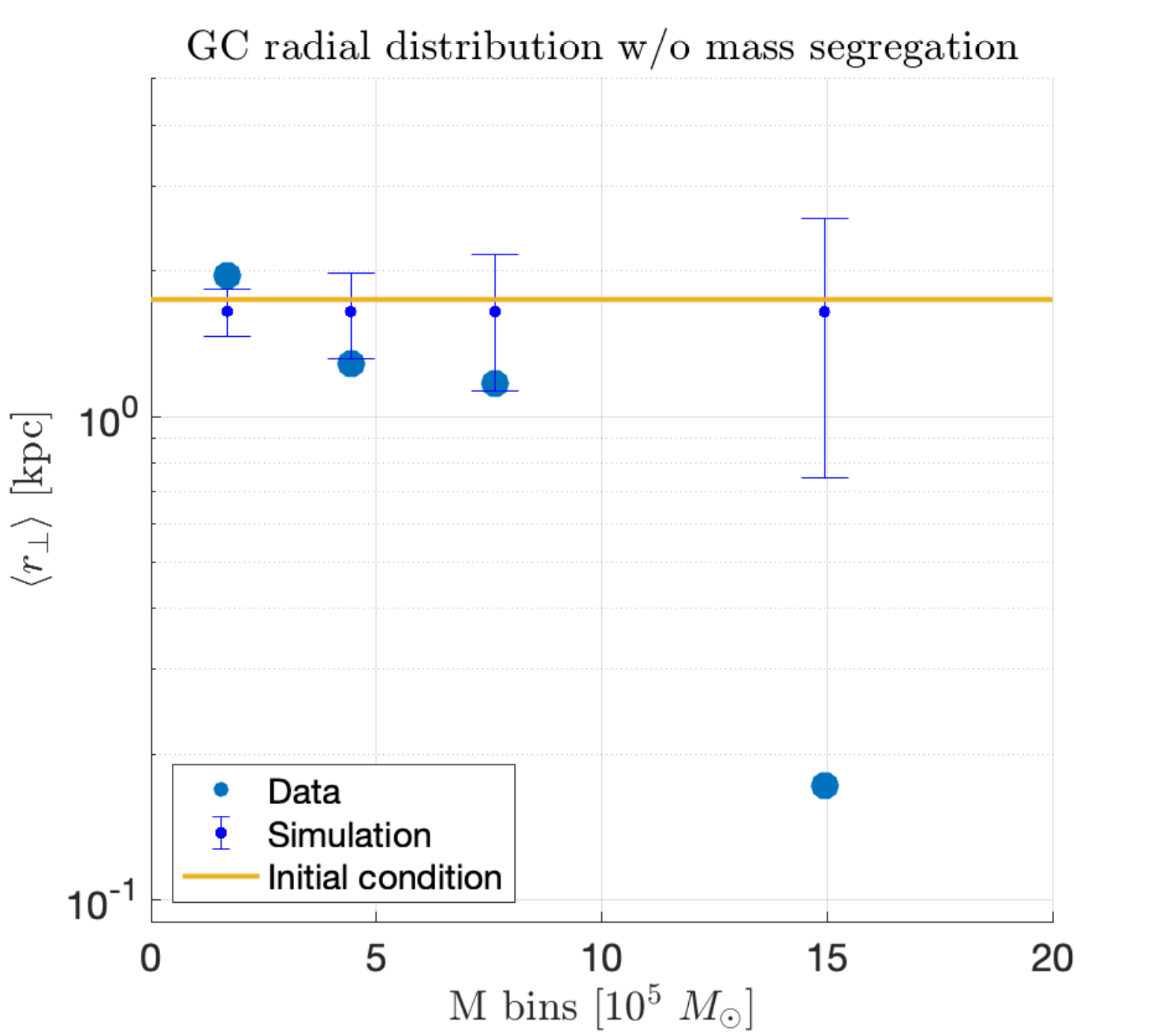}	\includegraphics[height=0.40\textwidth]{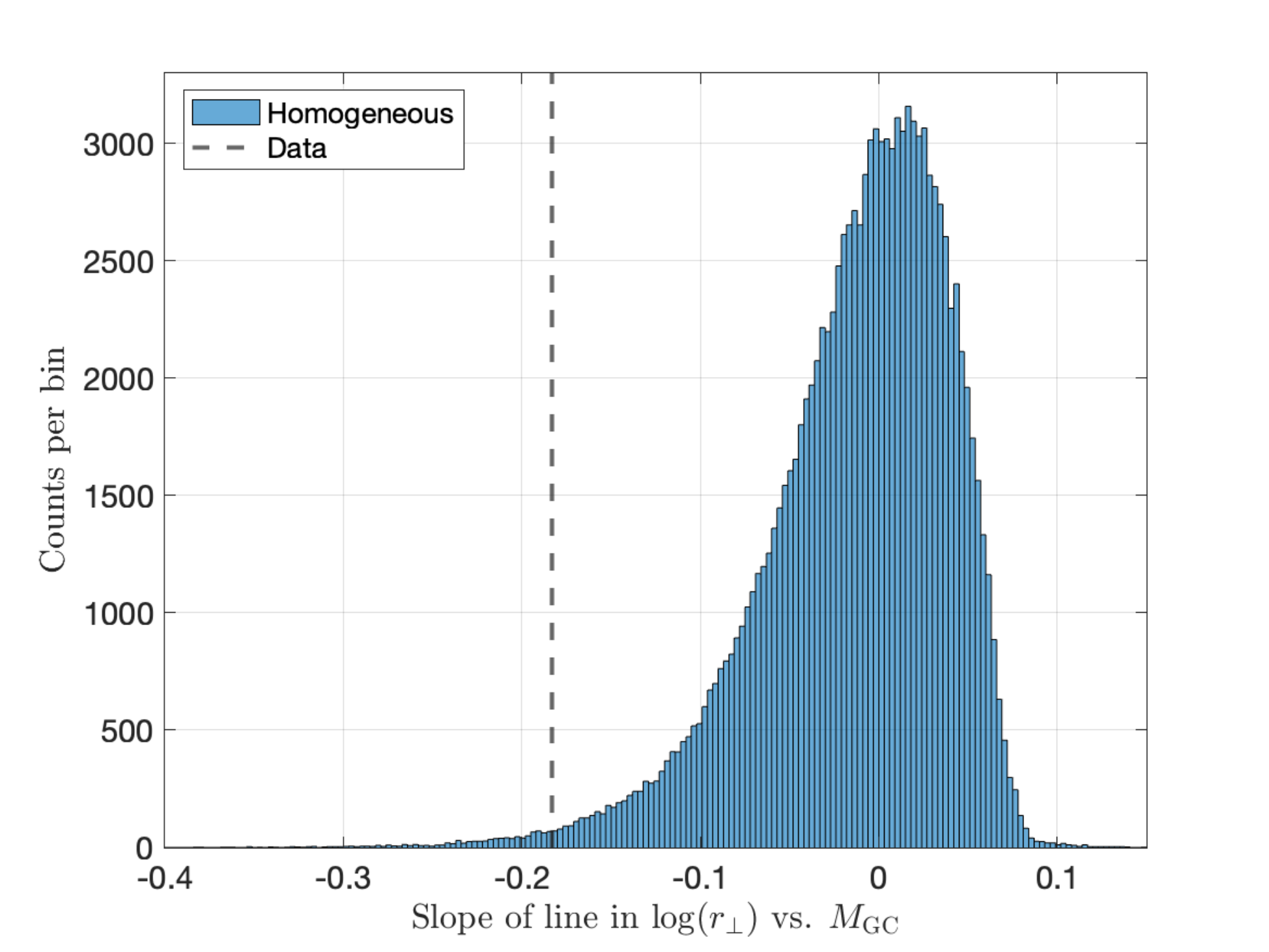}
	\caption{\textbf{Left:} Same as upper panels of \ref{fig:simResultsObs}, for a ``Simulation'' that is the mock GC data discussed in the text -- without any time integration -- such that it presents no mass segregation. \textbf{Right:} A histogram of the slope of line in $log(r_\perp)$ vs. $\MGC$ -- a test statistic $\mu$, essentially -- for mock data of GCs in \UDG\ all drawn from the same radial distribution (``Homogeneous''). The vertical dashed line shows this test statistic for the data ($\mu_{\rm data}$) of \UDG. The probability $p(\mu<\mu_{\rm data} = 0.008$.}\label{fig:noMassSegregation}
\end{figure}

It is also important to quantify the significance of the mass segregation trend, argued in \reffig{introData}. To that end, we generate a mock sample of GCs from a common radial distribution -- exactly like our simulations setup (\refsec{simulations}). We carry out two exercises with this mock data. First (\reffig{noMassSegregation}, left), we show an example of the radii of GCs vs. their mass, as in Fig. \ref{fig:simResultsObs}, showing that the no-mass-segregation hypothesis is not compatible with the data, in comparison to the models shown in the paper. Second, we define a test statistic that is a good proxy to mass segregation: the slope of the line in $\log(r_\perp)$ vs. $\MGC$. We show a histogram of this test statistic for the mock data. We estimate from this a p-value $\lesssim 1\%$ for the hypothesis that the data contains no mass segregation. These exercises support our claim of a mass segregation trend in the data of \UDG.

\section{Two-body relaxation of GCs in a background potential}\label{app:twobodyrelax}
The classic two-body relaxation time-scale of a self-gravitating system can be extended to the case of a N-body system in an external potential, following \cite{BinneyTremaine2}. Assume $ N $ GCs, each with a mass $ m $ spread along a characteristic radius $ R $ with a characteristic velocity $ v $ induced by a body of mass $ M_{\rm tot} = Rv^2/G  $. The mean square change in velocity per crossing time is then
\be
\Delta v^2 \approx 8N\left(\frac{Gm}{Rv}\right)^2\ln\Lambda = 8N \left(\frac{m}{M_{\rm tot}}\right)^2 v^2\ln\Lambda = \frac{8}{N} \left(\frac{Nm}{M_{\rm tot}}\right)^2 v^2\ln\Lambda\; .
\ee
Defining $ f \equiv Nm/M_{\rm tot} = M_{\rm GCs}/M_{\rm tot} $ (likely $ \sim 0.01\div 0.1 $ for the example of \UDG) and using $ \ln\Lambda = \ln R/b_{90} $, with $ b_{90} $ being the impact factor parameter where a mass is deflected by $ 90 $ degrees, i.e. $ b_{90} = 2Gm/v^2 $, one finds $ \ln \Lambda = \ln (N/f)  $. We therefore obtain 
\be
\Delta v^2 = \frac{8}{N} f^2 v^2\ln \frac{N}{f}\; .
\ee
The relaxation time is
\be
t_{\rm relax} \sim t_{\rm cross}N_{\rm cross} \approx  t_{\rm cross}\frac{v^2}{\Delta v^2} \approx \frac{0.1N}{\ln \frac{N}{f}}\frac{1}{f^2} t_{\rm cross} \; ,
\ee
where $ t_{\rm cross} \sim  R/v $.

For a UDG with $ v \sim 10 $~km/sec and $ r\sim 2 $~kpc, $ t_{\rm cross}\sim 0.2 $~Gyr, one finds
\be
t_{\rm relax} \sim 10 \left(\frac{N}{30}\right)\left(\frac{R}{2~{\rm kpc}}\right)\left(\frac{10~{\rm km/sec}}{v}\right)\left(\frac{0.1}{f}\right)^2~{\rm Gyr} \; .
\ee
(it is possible to replace $ f = Nm/M_{\rm tot} = Nm G/(Rv^2) $, which changes dependencies on variables.)

\begin{figure}[htbp!]
	\centering
	\includegraphics[width=0.6\textwidth]{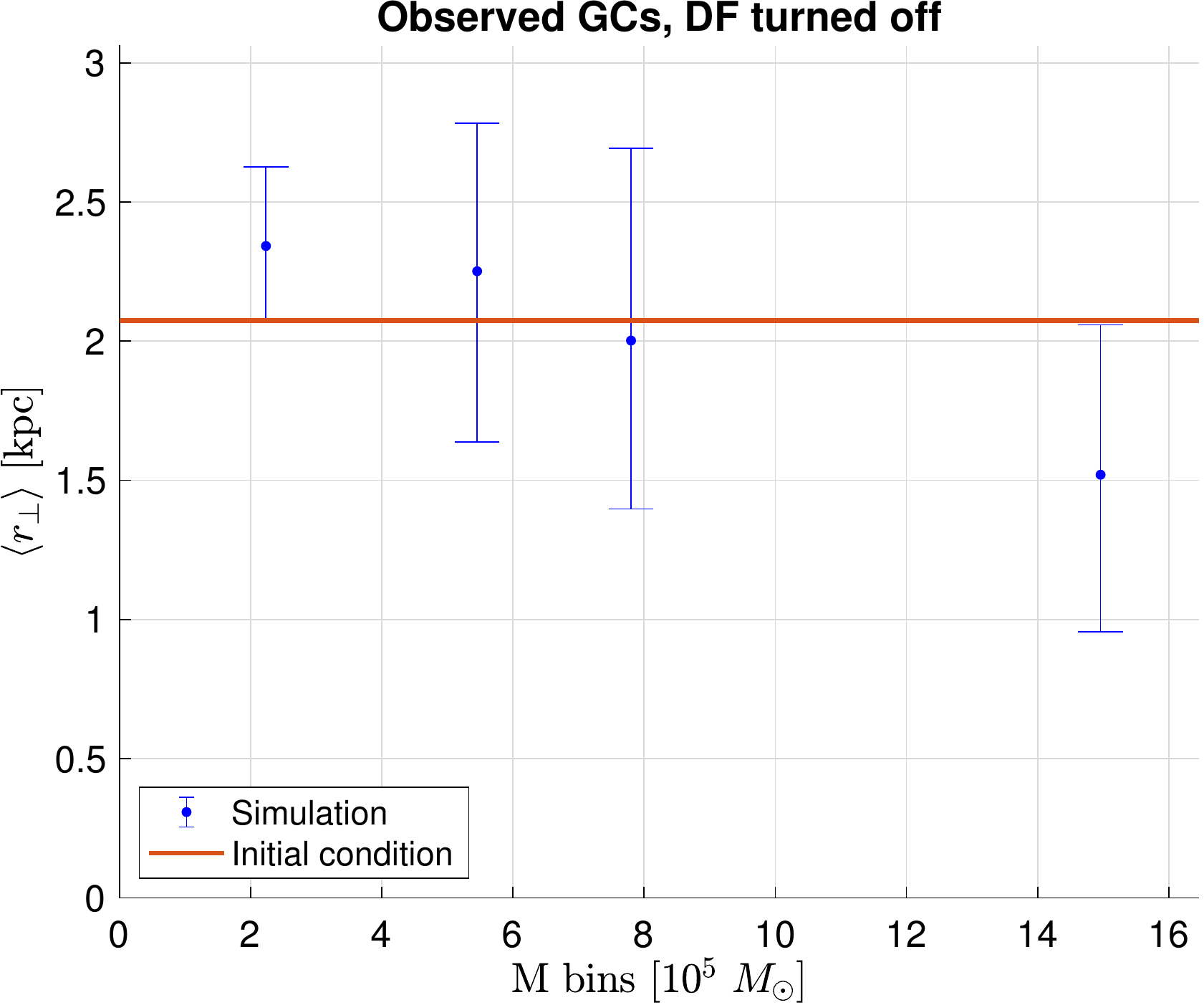}		
	\caption{A simulation batch where dynamical friction is turned off, using \stars\ model, as explained in \refsec{udgdatamodels}. The interaction between GCs causes the mass segregation. Of course, the segregation depends on the GC mass function, which we take to be the sample of $ 33 $ observed GCs \citep{Danieli2021}.}\label{fig:barNoDF}
\end{figure}

We find the interesting result that two-body relaxation of GCs is marginally effective in a galaxy such as NGC5846-UDG1, in the case where its dynamics is dominated by the observed stellar body. Indeed, we find support for these estimates in a N-body simulation where external dynamical friction is turned off, see \reffig{barNoDF}. Unsurprisingly, more massive GCs sink to small radii whereas lighter GCs experience a ``buoyancy''-like effect. This should be kept in mind: while lighter constituents (stars and dark matter) of the galaxy mostly operate as friction, the GC distribution works both as friction and as heating. Of course, this effect is only appreciable in light mass models such as the \stars\ model.

\section{Projection effects}\label{app:projection}

The expectation value of the projected distances can be re-expressed as
\be\label{eq:rprojvsrDist}
\left<r_\perp \right> = \int d^3r n(r) r_\perp = 2\pi \int dr r^2 n(r) \sin\theta \sin\theta d\theta  = \frac{\pi^2}{2}\int dr r^2 n(r)  =\frac{\pi}{4}\left<r\right>  \; ,
\ee
where we assumed spherical symmetry.

For a single GC, taking a circular orbit with radius $ r $ for simplicity, the time-average expected projected radius is
\be
\left<r_\perp\right>_t = \frac{1}{T}\int\limits_0^T r\sin (\omega t) dt = \frac{\pi}{2}r \; .
\ee
This is of course only a crude estimate. We may do better by using simulation data. Taking the last $ 1 $~Gyr of $ r(t) $ and $ r_\perp(t) $ of different simulations and mass models, we find (subscript $ t $ denotes averaging over time)
\be\label{eq:rperpvsr}
\left<r_\perp\right>_t \approx 0.8 \left<r\right>_t \; .
\ee
Indeed, this agrees with the estimates used in the literature, e.g. $ r_\perp \approx (\sqrt{3}/2)r $ and $ r_\perp \approx (3/4)r $ in \cite{Hui2017} and \cite{Meadows20}, respectively.

\section{Faintest objects set}\label{app:faintset}
\begin{figure}[htbp!]
	\centering
	\includegraphics[width=0.6\textwidth]{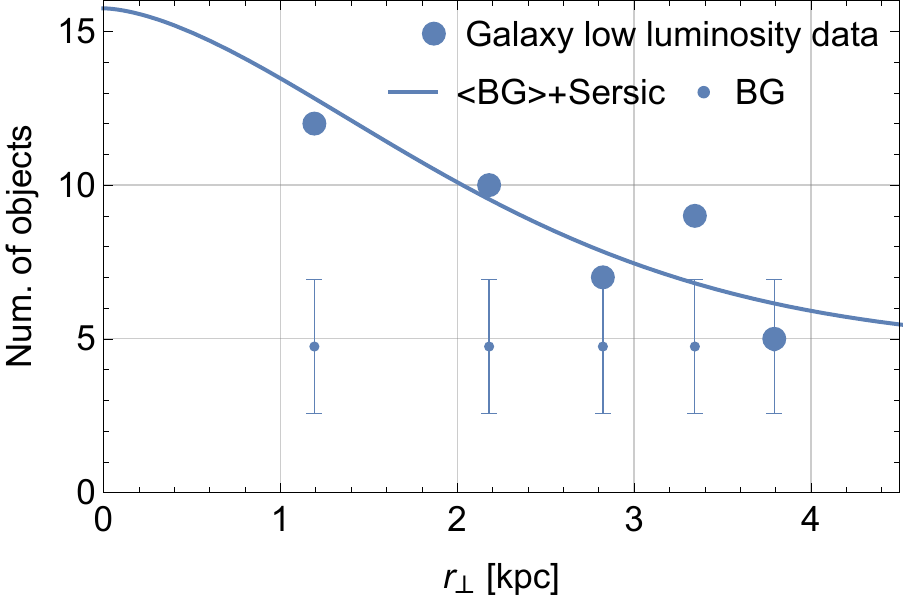}		
	\caption{Thick points are the number of objects in the lowest luminosity sample of \cite{Danieli2021} per radial bin. The projected radius is taken as $\left<r_\perp\right> = \int\limits_{r_1}^{r_2}\Sigma(r_\perp)r_\perp d^2r_\perp / \int\limits_{r_1}^{r_2}\Sigma(r_\perp) d^2r_\perp \approx (2/3)(r_2^3-r_1^3)/(r_2^2-r_1^2) $, taking $\Sigma \approx $~const. per bin with $r_1$ and $r_2$ inner and outer radii of each bin. The contamination (BG) is shown in points with error-bars $\pm $~standard deviation. The contamination is universal among different bins because we chose the bins to have equal area. The solid line is the sum of the expected contamination ($\left<{\rm BG}\right>$) and a \sersic\ profile with $n=0.61$ and $R_e=2.3$~kpc, slightly more extended than the \stars\ in \UDG\ (since bins are equal area, the \sersic\ surface density can be adopted immediately). }\label{fig:faintGCs}
\end{figure}

Below a luminosity threshold corresponding magnitude $ m_V \approx 25 $, there is considerable contamination of other sources in the field of view of the galaxy \citep{Danieli2021}. In the selection criteria of \cite{Danieli2021}, the lowest luminosity set contains $ 43 $ objects, whereas a nearby background sample with $ 6.5 $ times more area contains $ k_0 = 155 $ sources -- or about $ 24 $ objects per galactic-area, implying $\mathcal{O}(1)$ contamination in that low-luminosity set. In order to nevertheless extract information about the radial distribution of true GCs in this set, we first divide the galaxy into radial bins of area $S_{\rm bin}^{(i)}$. When the background area $S_{\rm bg} \gg S_{\rm bin}^{(i)}$, the expected background objects per bin is $\lambda_i\approx k_0\times S_{\rm bin}^{(i)}/S_{\rm bg} $. Assuming the background is Poisson-distributed, we can estimate the true number of objects in a given bin. In the current work, we refrain from a full statistical analysis, which would require a more careful treatment and modeling of the statistical distribution of true GCs. Instead, we plot in \reffig{faintGCs} the data along with contamination per radial bin and demonstrate that a \sersic\ profile in the ballpark of the \stars\ distribution is a good fit to the data.

%We assume Poisson statistics, i.e., assuming $ \sigma $ expected sources in an area corresponding the background ($ S_{\rm bg} $) and $ k $ observed objects, $ P(k|\sigma) = \sigma^k\exp(-\sigma)/k! $. Inverting assuming no prior on $ \sigma $, $ {\rm f}(\sigma|k) = {\rm P}(k|\sigma)/{\rm normalization} $. Inspecting an area of size $ S_{\rm bin} $, one expects a distribution function for the contamination $ k_{\rm bin} $
%\be
%{\rm P}(k_{\rm bin}|{\rm BG-sample}) &=& \int d\sigma_{\rm bin} {\rm P}\left(k_{\rm bin}|\sigma_{\rm bin} \right){\rm f}\left(\sigma_{\rm bin}|{\rm BG-sample} \right) \\ &\propto & \int d\sigma {\rm P}\left(k_{\rm bin}|\alpha\sigma \right){\rm f}\left(\sigma|{\rm BG-sample} \right) \propto \int d\sigma \frac{(\alpha\sigma)^{k_{\rm bin}}e^{-\alpha\sigma}}{k_{\rm bin}!}\frac{\sigma^ke^{-\sigma}}{k!} \\ & = & \frac{\Gamma(k+k_{\rm bin}+1)}{k!k_{\rm bin}!}\alpha^{k_{\rm bin}}(1+\alpha)^{-k-k_{\rm bin}-1} \; .
%\ee
%$ \alpha \equiv S_{\rm bin}/S_{\rm bg} $, relating $ \sigma_{\rm bin} = \alpha \sigma $. This gives
%\be
%\left<k_{\rm bin}\right> = \alpha(k+1) \; \; , \; \; \Delta k = \sqrt{\left<k_{\rm bin}^2\right>-\left<k_{\rm bin}\right>^2} = \sqrt{\alpha(\alpha+1)(k+1)}
%\ee

\section{Further simulations}\label{app:othersims}
\begin{figure*}[htbp!]
	\centering
	\includegraphics[width=0.32\textwidth]{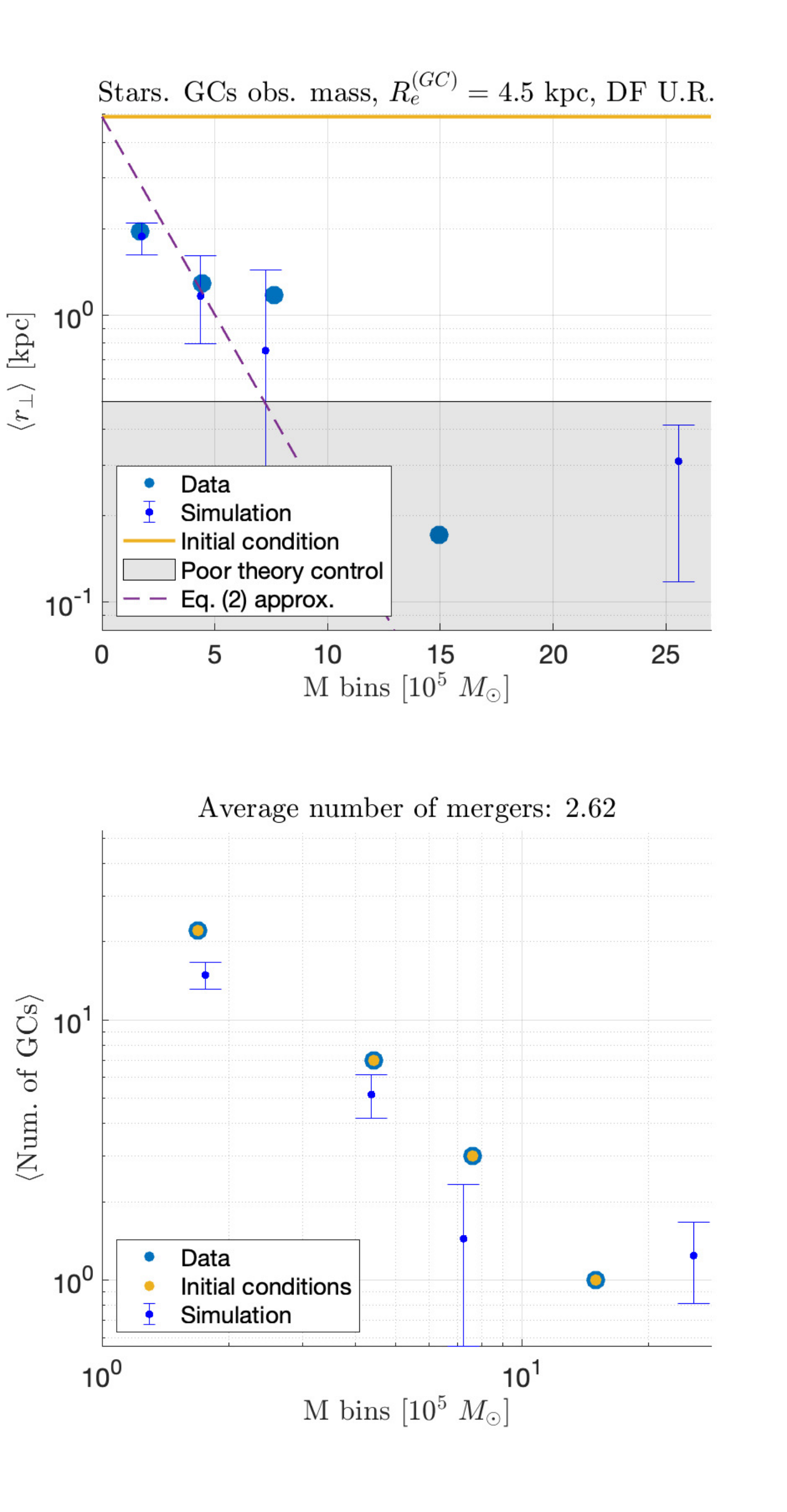}
	\includegraphics[width=0.32\textwidth]{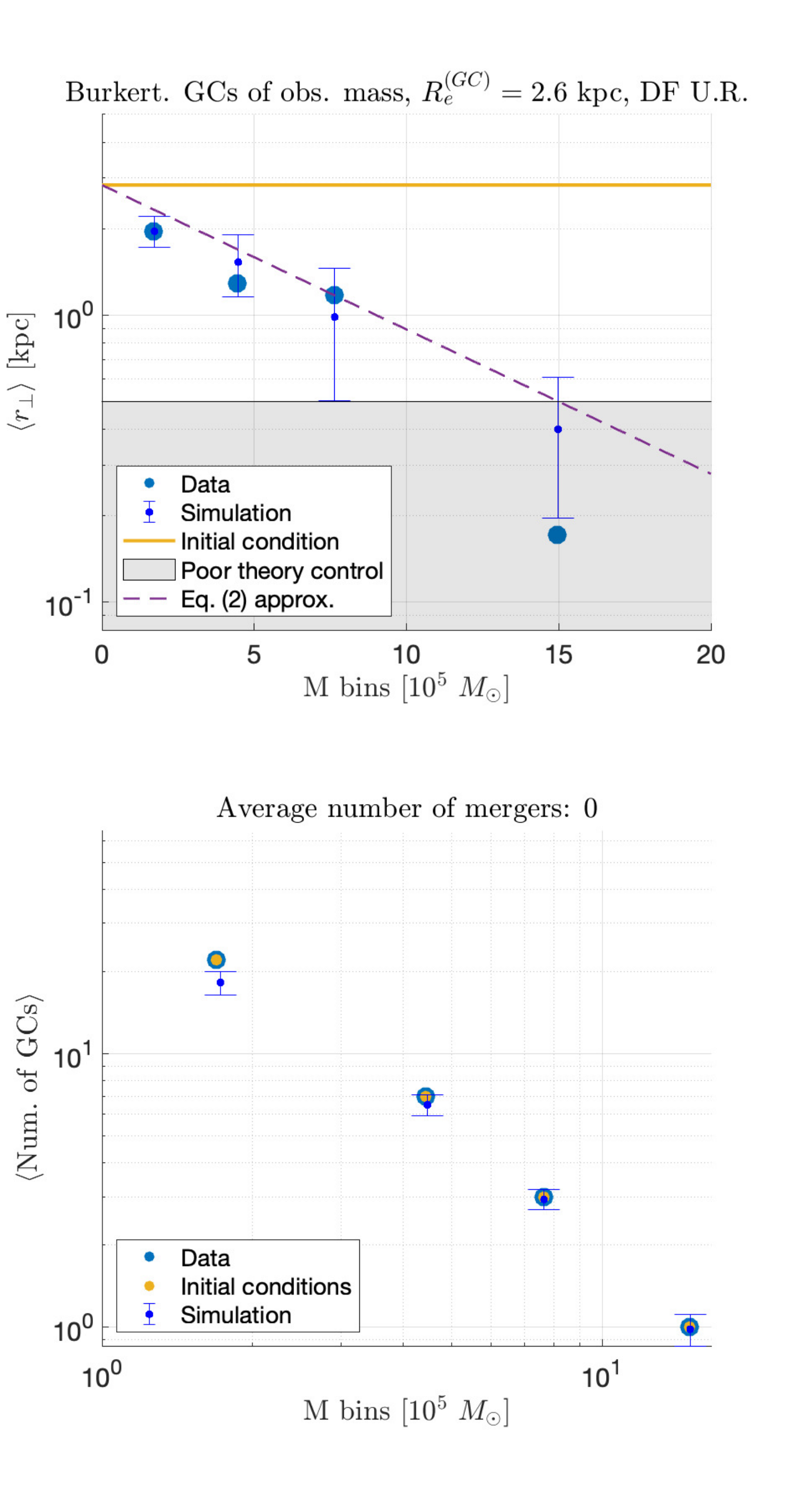}
	\includegraphics[width=0.32\textwidth]{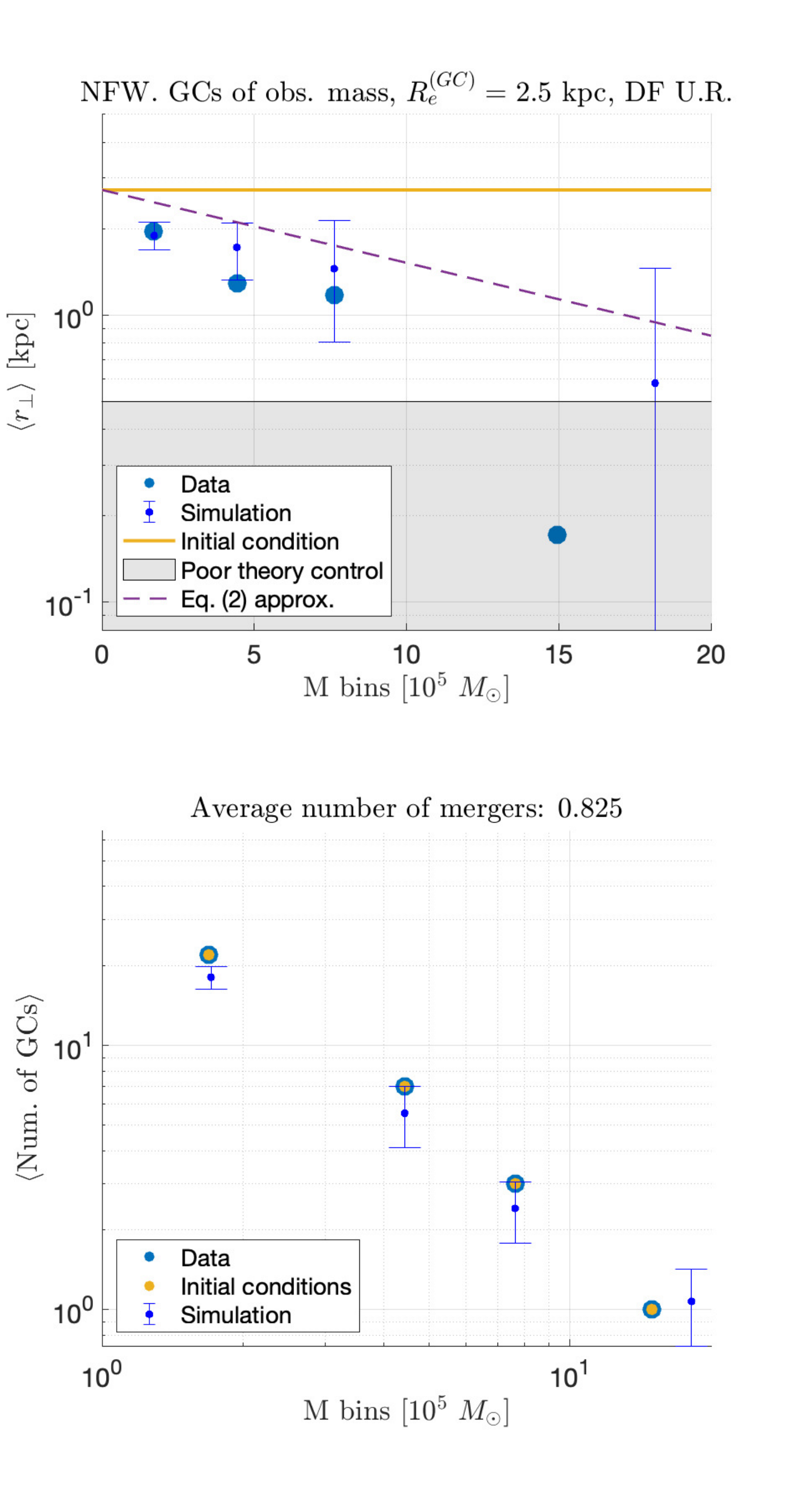}	
	\caption{Like \reffig{simResultsObs}, but with unregulated DF (``DF U.R.''), i.e. without turning off DF at any radius, either due to core stalling conditions $ r< 0.3R_e$ or the enclosed mass condition $M_{\rm halo} -M_{\rm GCs}/2<0$. }\label{fig:simResultsObsUR}
\end{figure*}

\begin{figure*}[htbp!]
	\centering
	\includegraphics[width=0.32\textwidth]{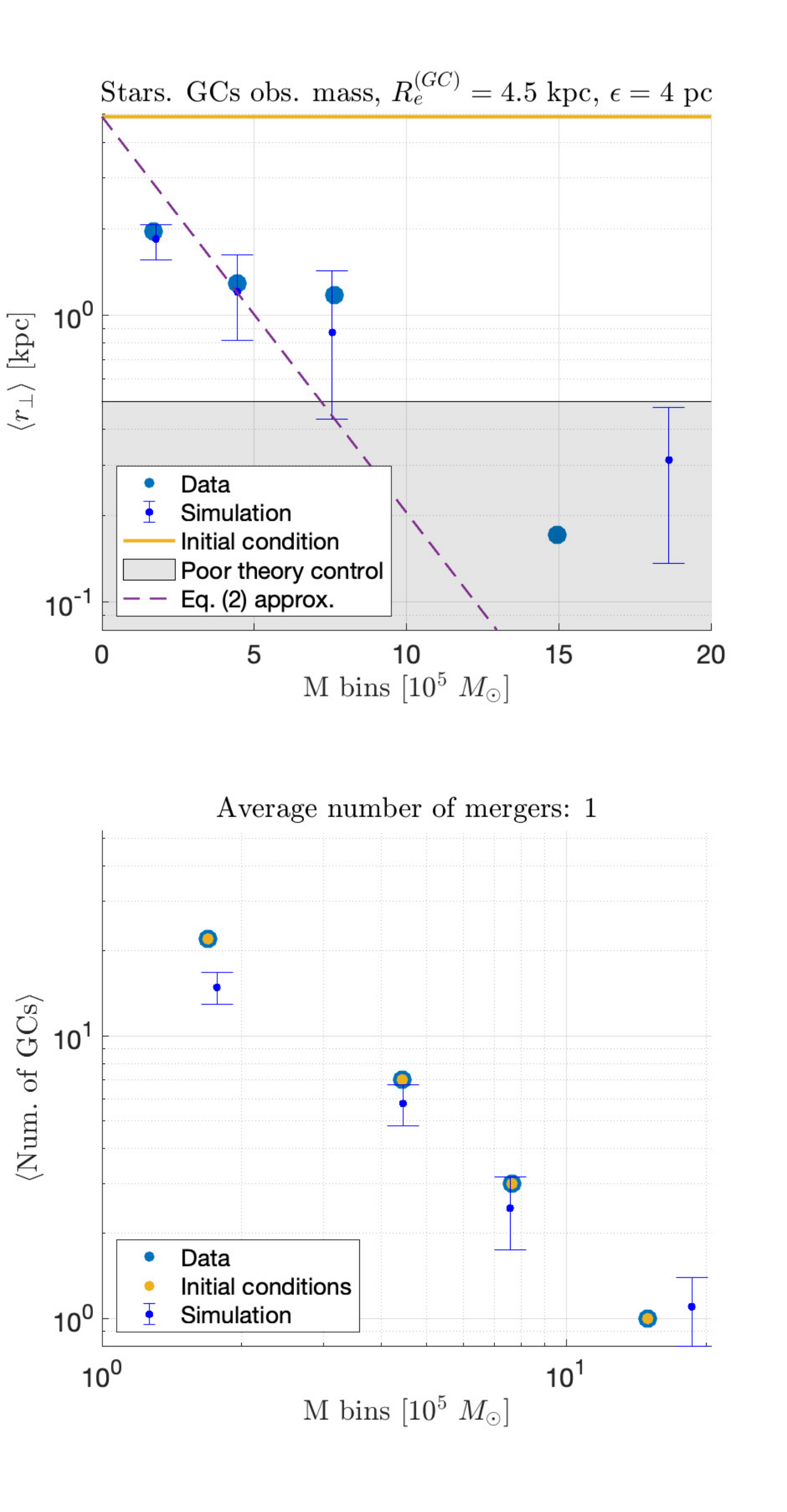}
	\includegraphics[width=0.32\textwidth]{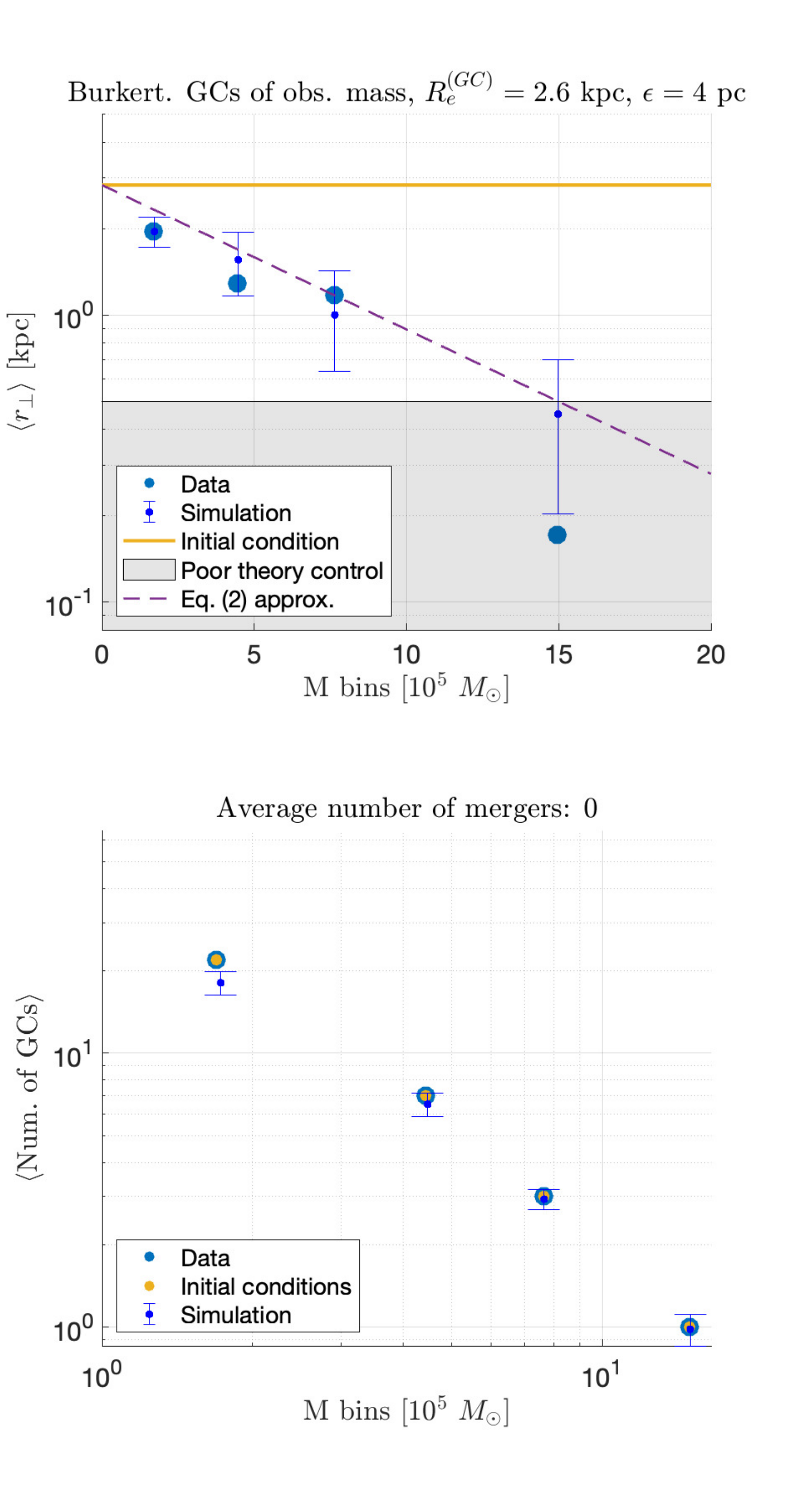}
	\includegraphics[width=0.32\textwidth]{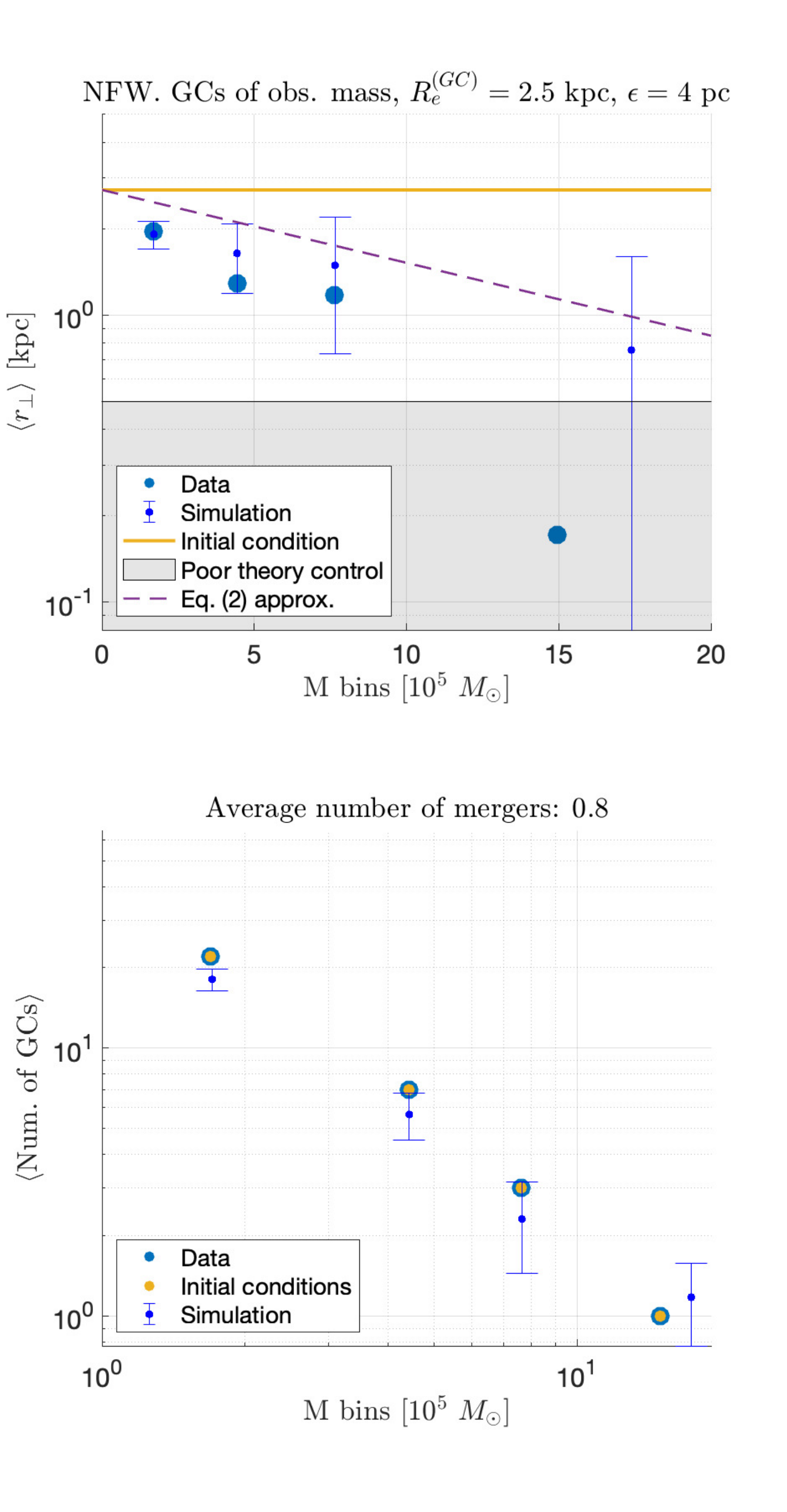}	
	\includegraphics[width=0.32\textwidth]{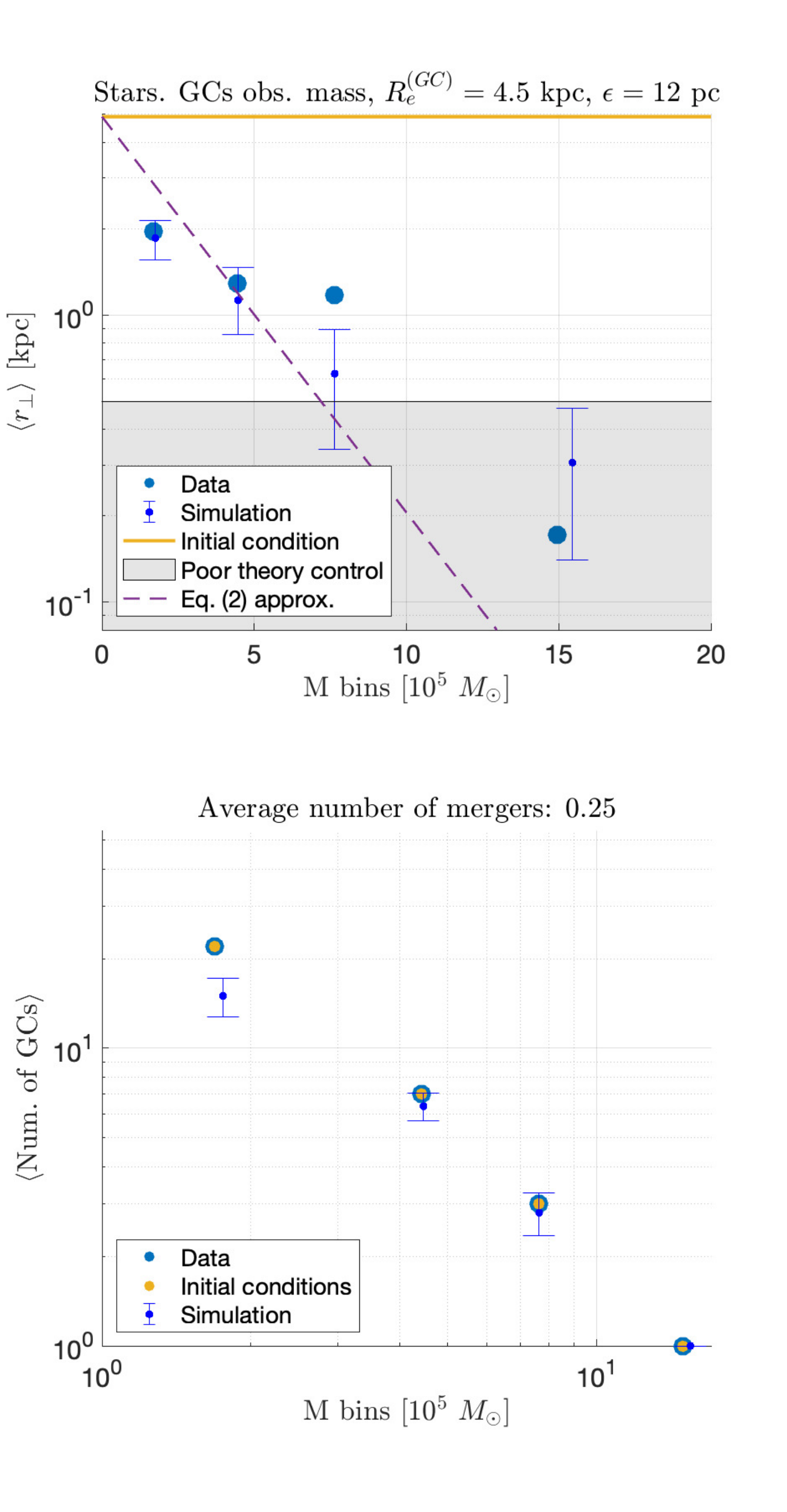}
	\includegraphics[width=0.32\textwidth]{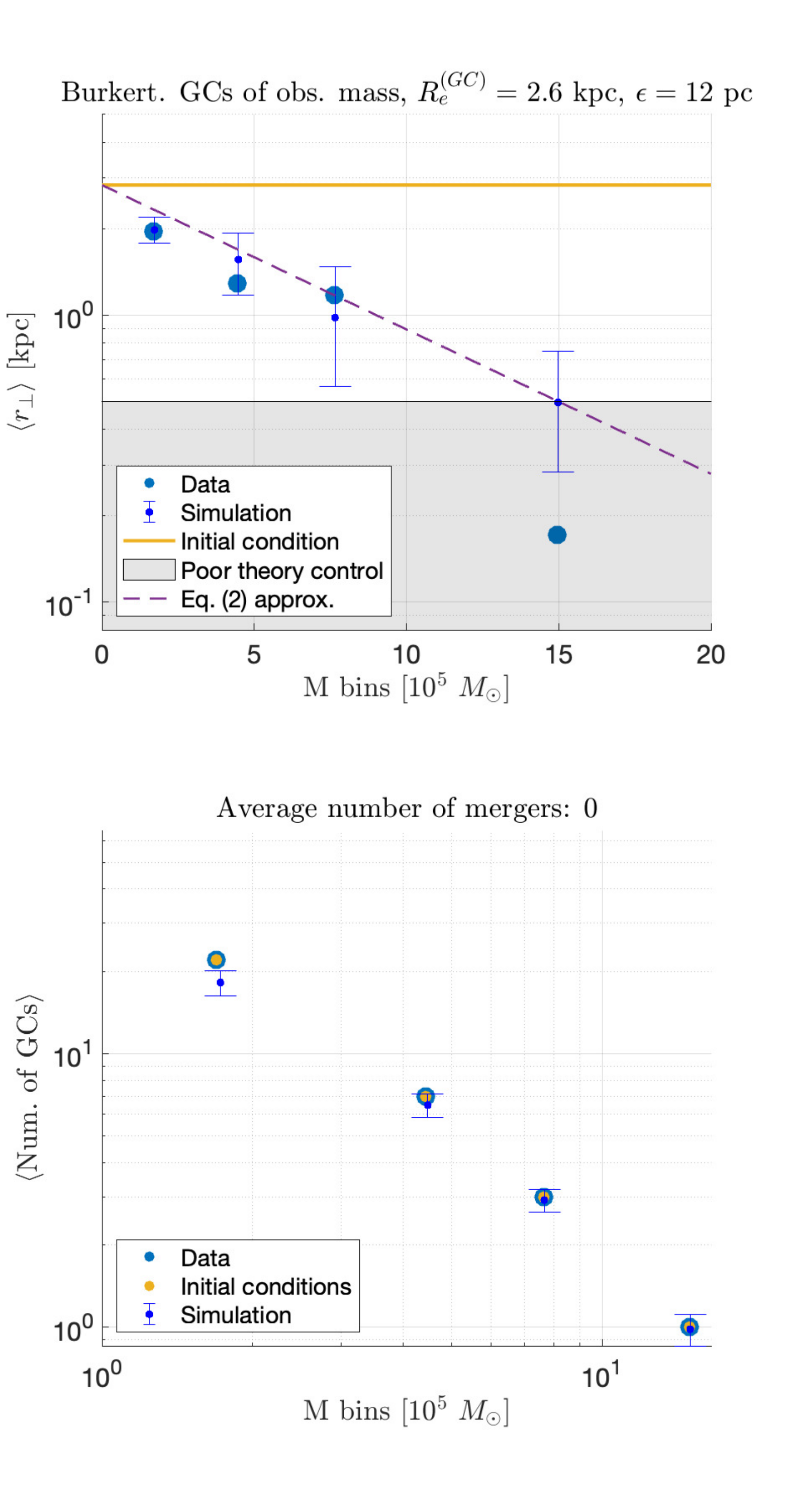}
	\includegraphics[width=0.32\textwidth]{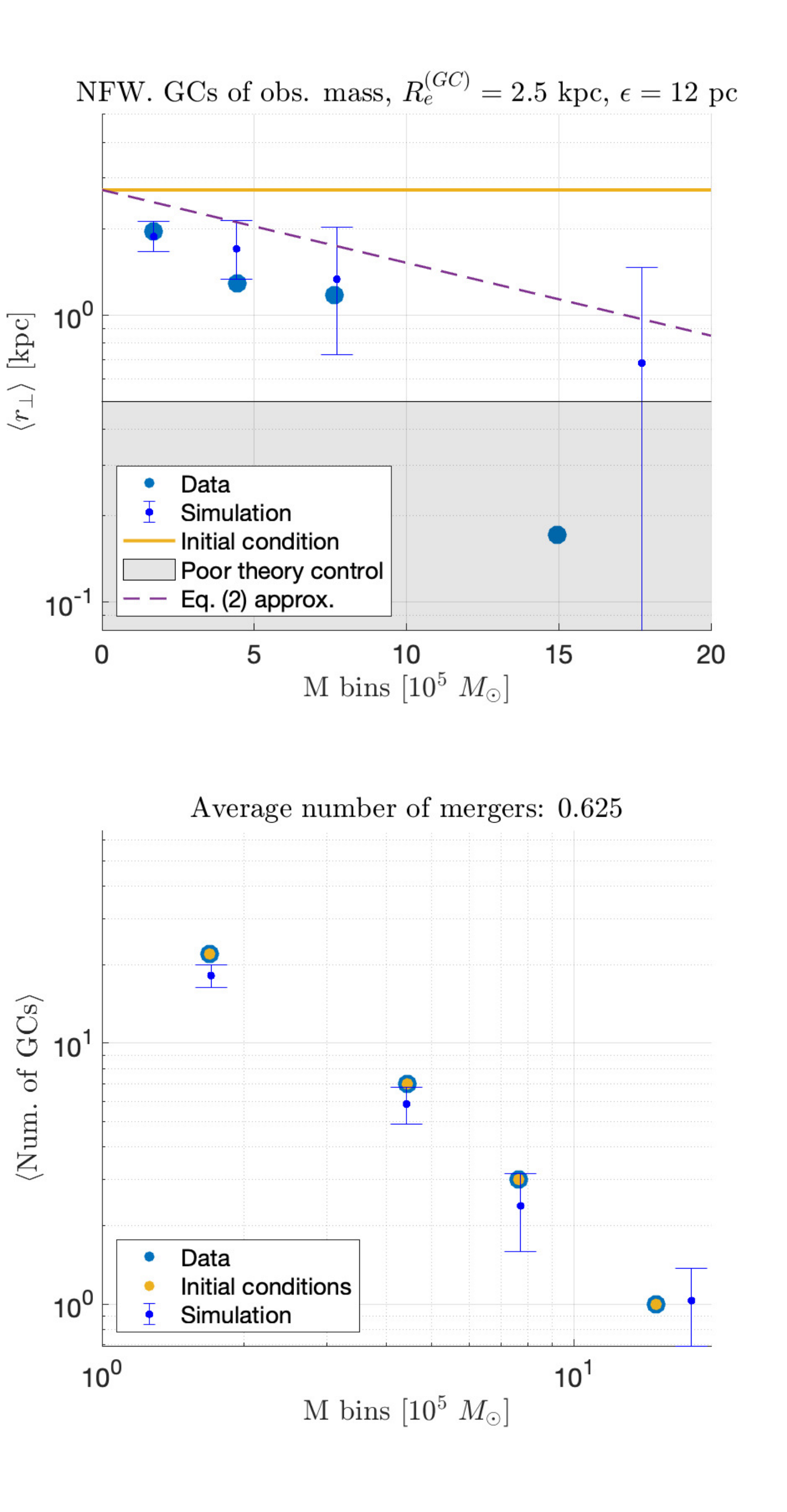}	
	\caption{Like \reffig{simResultsObs}, but $\epsilon = 4$~pc (top) and $\epsilon=12$~pc (bottom). }\label{fig:simResultsObsEps}
\end{figure*}

\begin{figure*}[htbp!]
	\centering
	\includegraphics[width=0.32\textwidth]{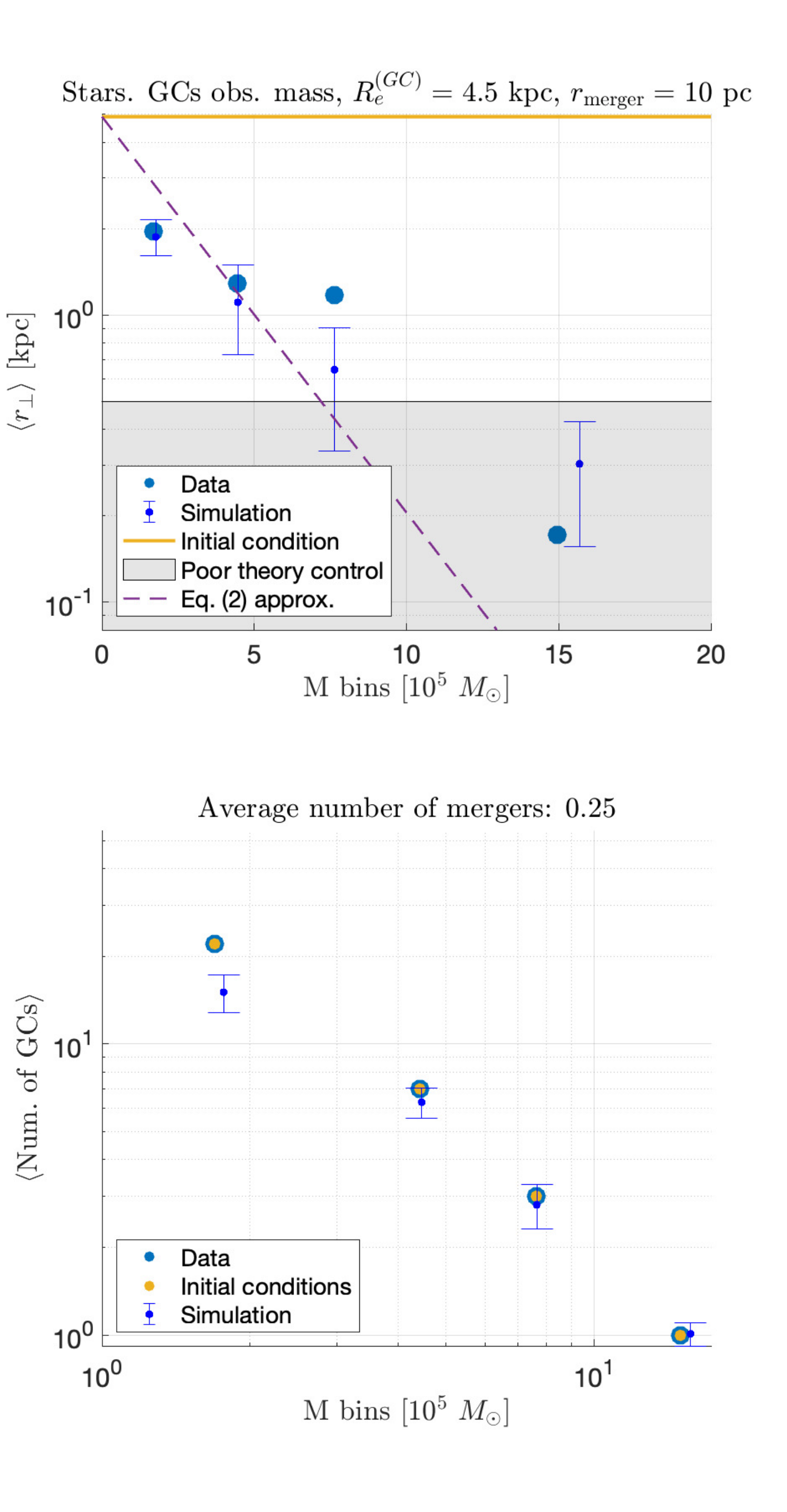}
	\includegraphics[width=0.32\textwidth]{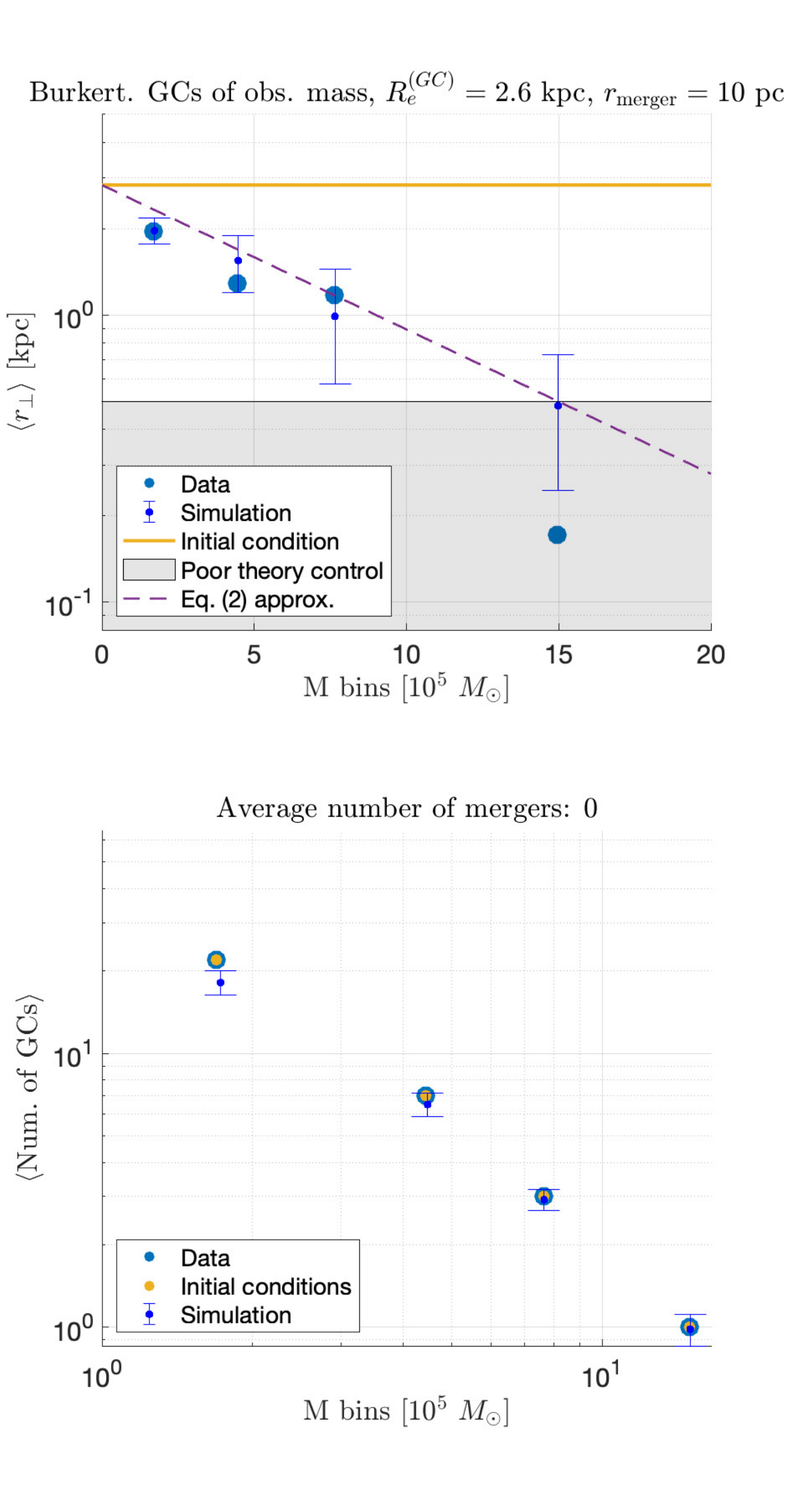}
	\includegraphics[width=0.32\textwidth]{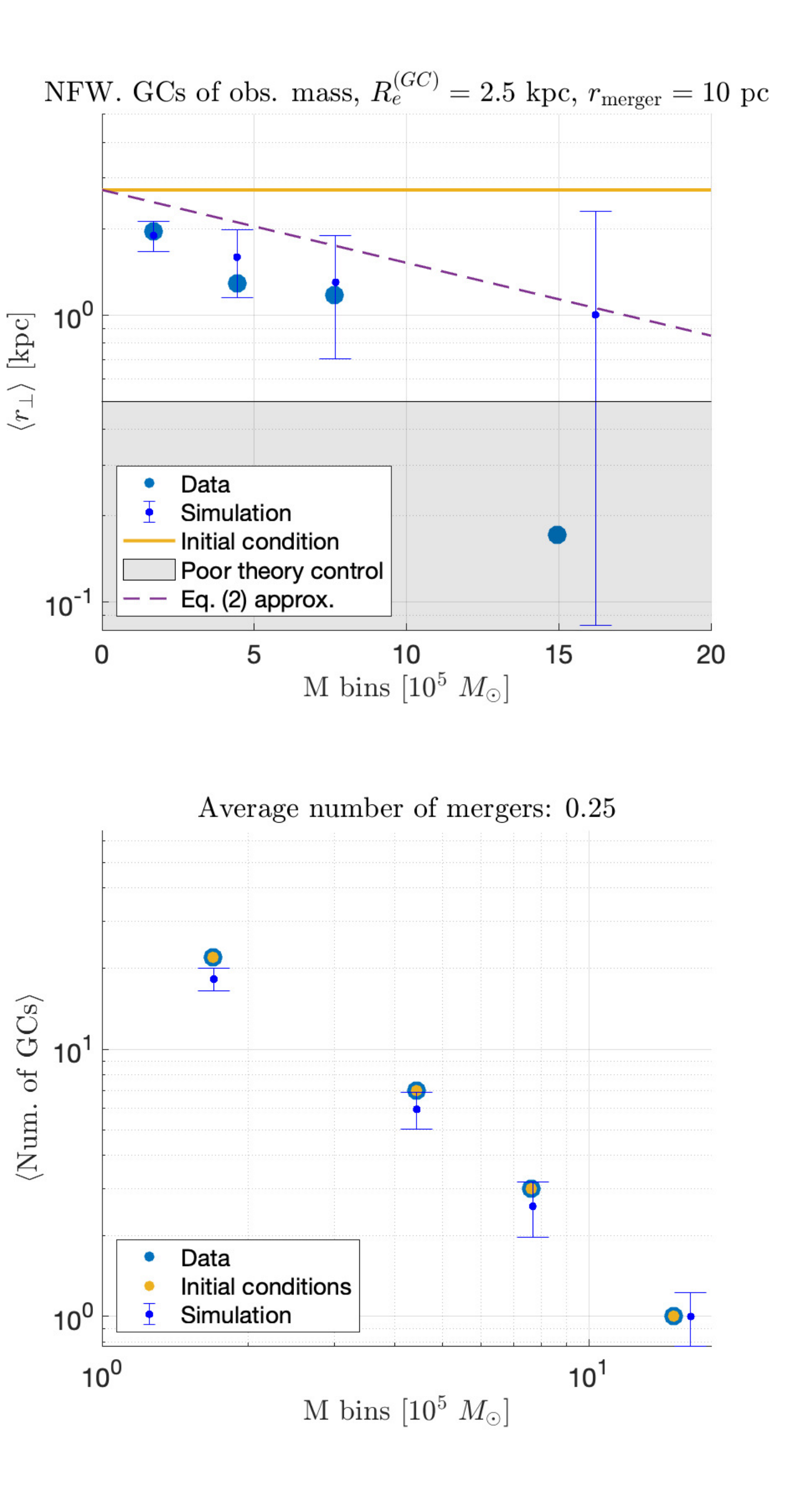}	
	\includegraphics[width=0.32\textwidth]{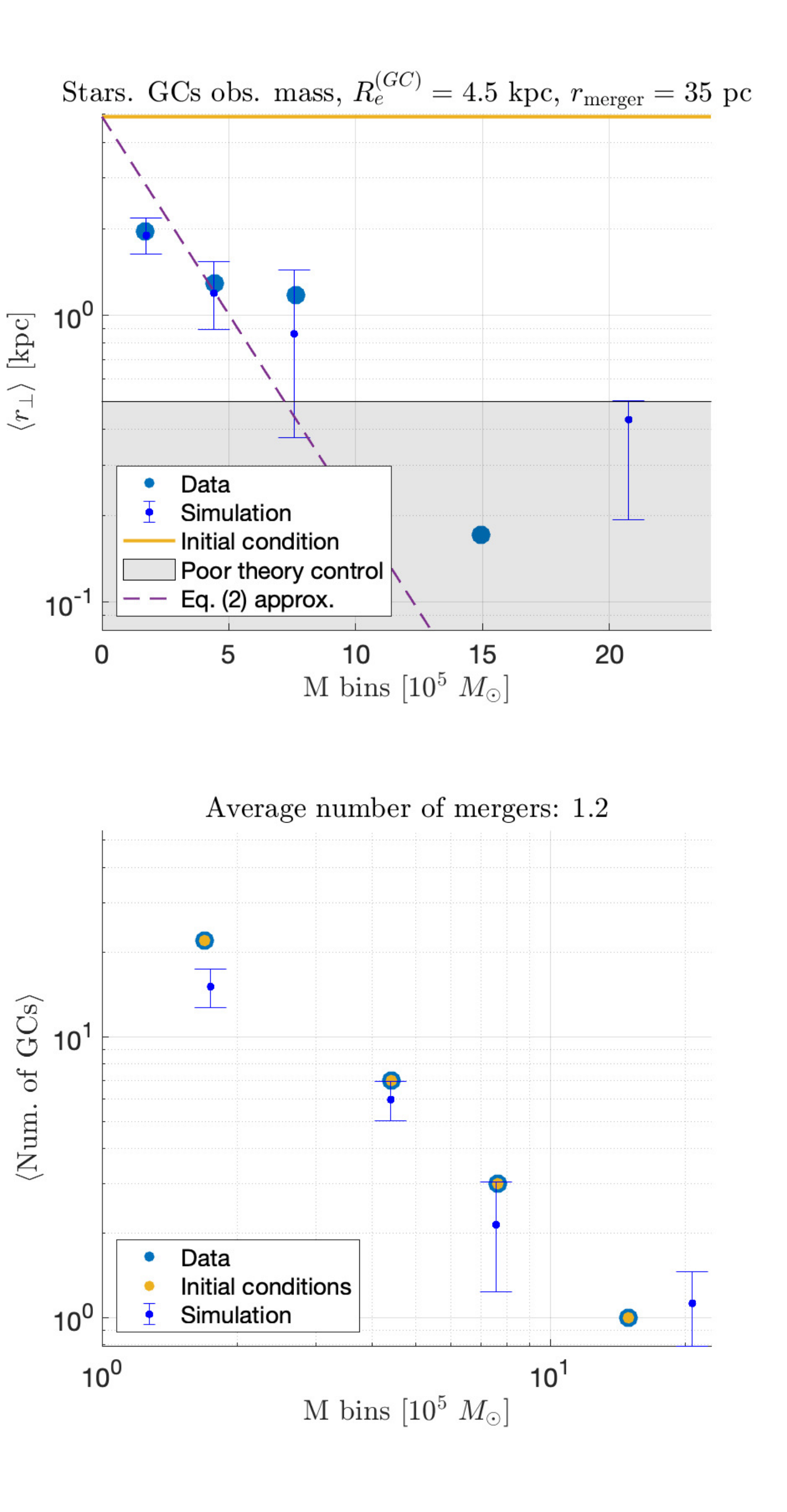}
	\includegraphics[width=0.32\textwidth]{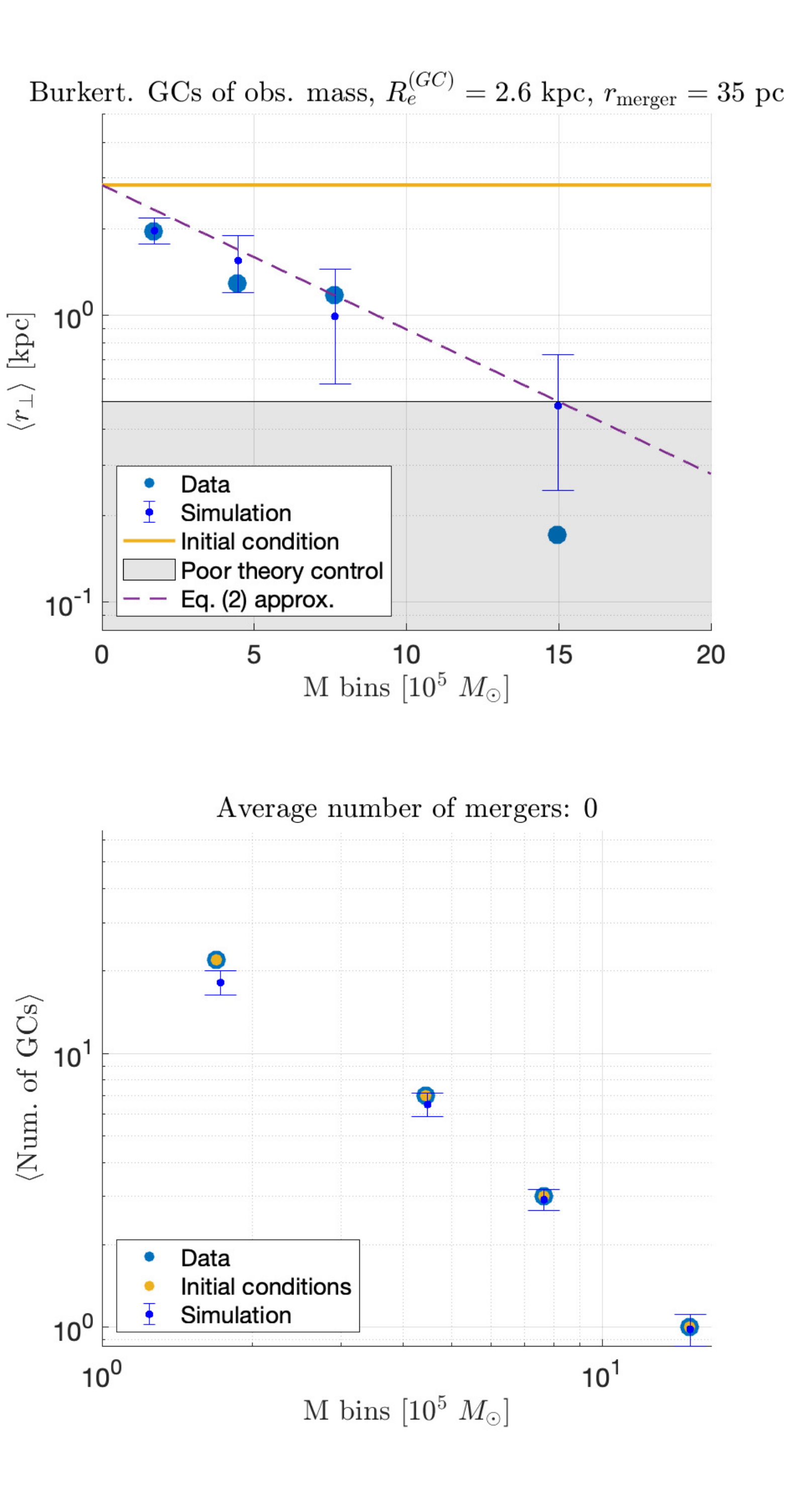}
	\includegraphics[width=0.32\textwidth]{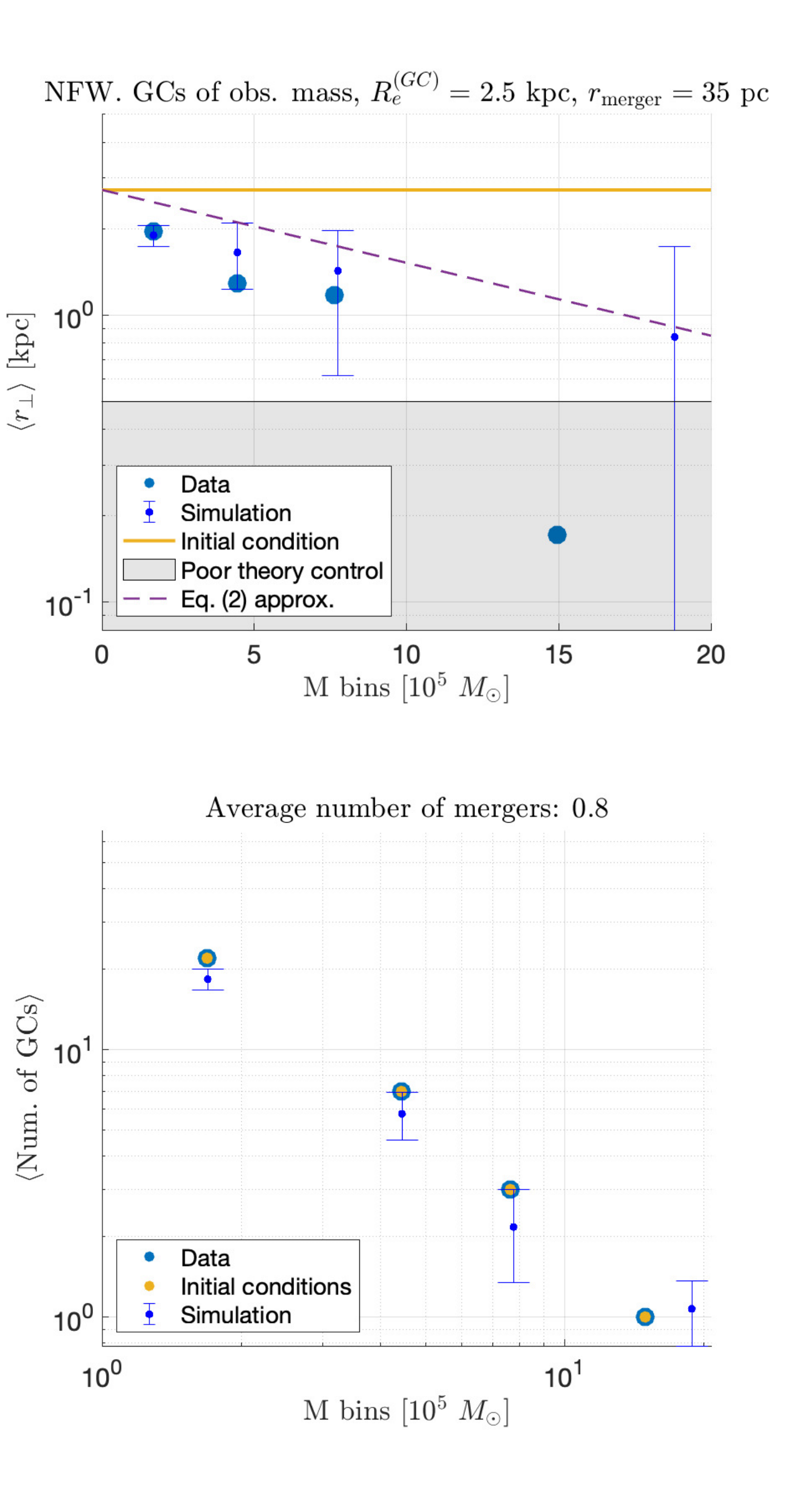}	
	\caption{Like \reffig{simResultsObs}, but $r_{\rm merger} = 10$~pc (top) and $r_{\rm merger}=35$~pc (bottom).}\label{fig:simResultsMrg}
\end{figure*}

\begin{figure*}[htbp!]
	\centering
	\includegraphics[width=0.32\textwidth]{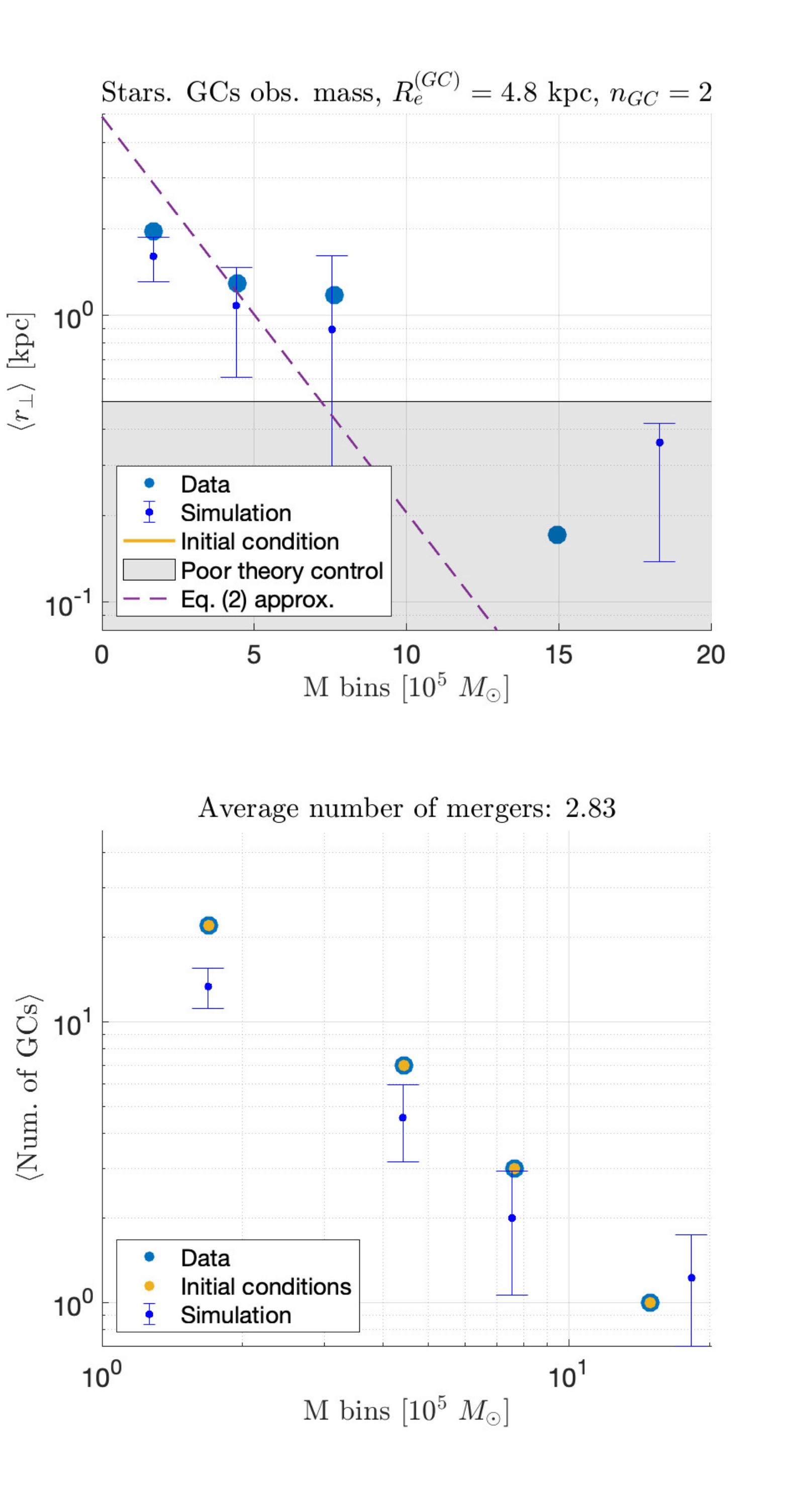}
	\includegraphics[width=0.32\textwidth]{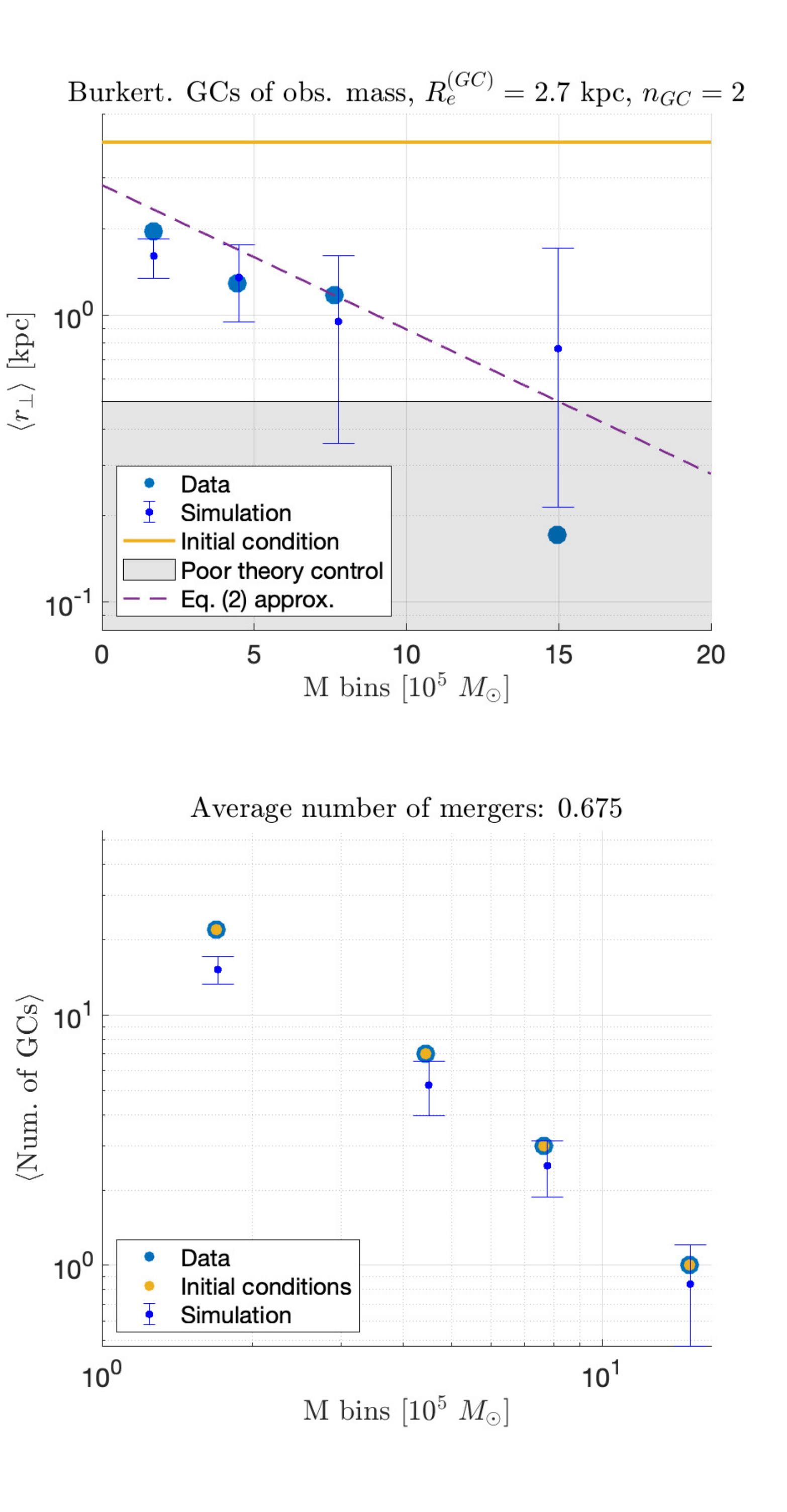}
	\includegraphics[width=0.32\textwidth]{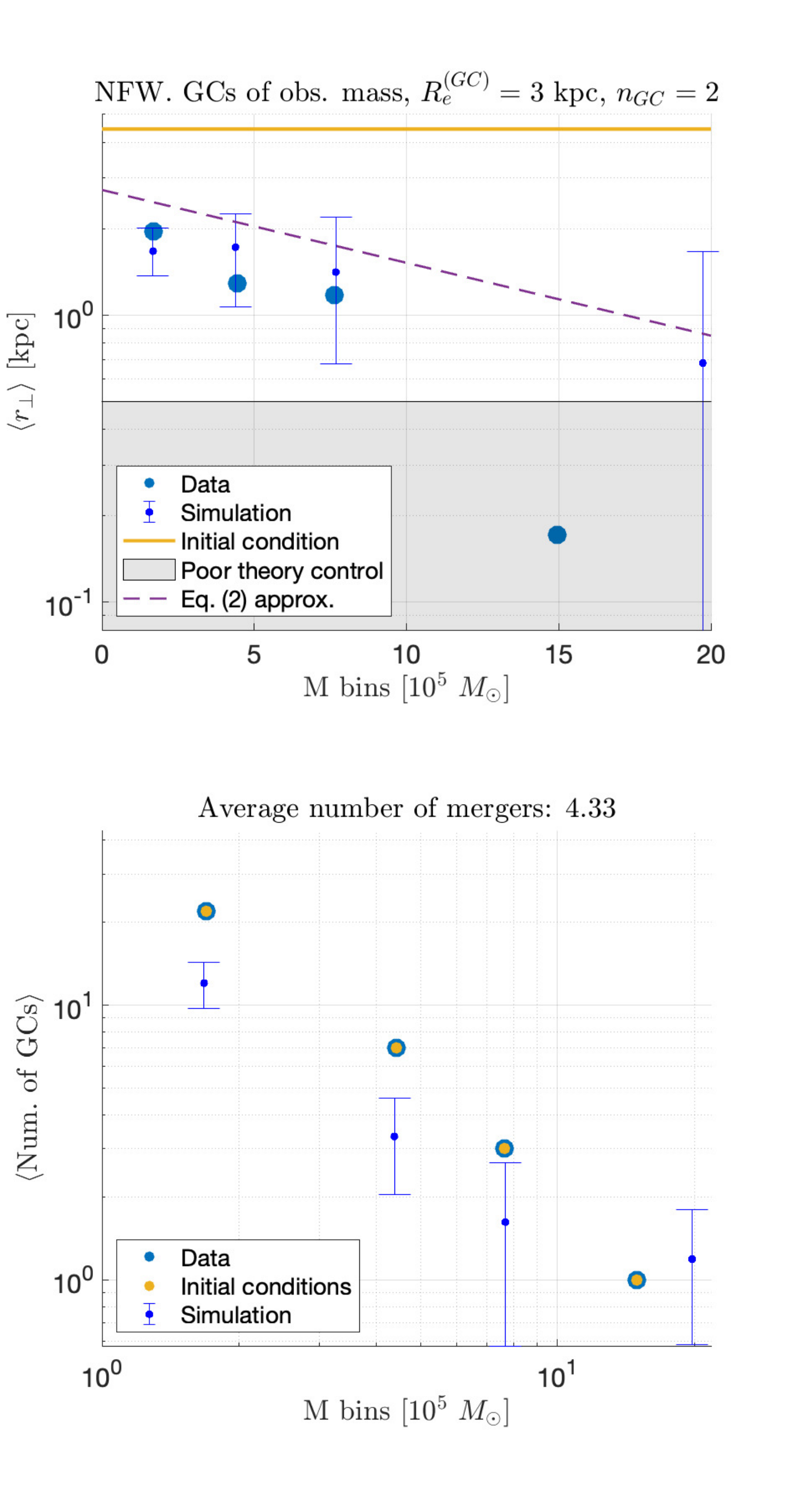}	
	\caption{Like \reffig{simResultsObs}, but with $n_{GC} = 2$, i.e., a relatively cuspy GC distribution.}\label{fig:simResultsCuspyGC}
\end{figure*}

In this Appendix we demonstrate the sensitivity of our results in \reffig{simResultsObs} under the change of several modelling choices that we made in \refsec{simulations}. In \reffig{simResultsObsUR} we show the simulations without the restriction of DF at $r<0.3R_e$ and $M_{\rm halo} -M_{\rm GCs}/2<0$. Namely, in producing this plot, we let DF remain active throughout the halo. In \reffig{simResultsObsEps} we show the results of simulations for different values of the GC Plummer softening parameter $\epsilon$. In \reffig{simResultsMrg} we show the results of simulations for different values of the critical merger radius $r_{\rm merger}$. In \reffig{simResultsCuspyGC} we show the results of simulations for a more centrally-concentrated initial distribution of GCs, using a \sersic\ index $n_{GC} = 2$.

\end{appendix}
%\vspace{6 pt}
%\newpage
%\clearpage

\bibliography{ref}

\end{document}